%% file: main.tex
\newif\ifabstract
\abstracttrue
\newif\iffull
\ifabstract \fullfalse \else \fulltrue \fi

\documentclass[11pt]{article}
\usepackage{amsfonts}
\usepackage{amssymb}
\usepackage{amstext}
\usepackage{amsmath}
\usepackage{xspace}
\usepackage{ntheorem}
\usepackage{graphicx}
\usepackage{url}
\usepackage{graphics}
\usepackage{colordvi}
\usepackage{hyperref}
\usepackage{subfigure}
\usepackage{framed}
\usepackage{enumitem}
\usepackage{algorithm2e}
\usepackage{float}
\usepackage{thmtools}
\usepackage{thm-restate}
\usepackage{cleveref}

\advance \oddsidemargin by -0.8in
\newcommand{\myparskip}{3pt}
\parskip \myparskip

\newenvironment{proofof}[1]{\noindent{\bf Proof of #1.}}%
        {\hspace*{\fill}$\Box$\par\vspace{4mm}}

\newcommand{\vbl}{\operatorname{vbl}}

\newcommand{\ceil}[1]{\ensuremath{\left\lceil#1\right\rceil}}
\newcommand{\floor}[1]{\ensuremath{\left\lfloor#1\right\rfloor}}

\newcommand{\event}{{\cal{E}}}

\newcommand{\opt}{\mathsf{OPT}}

\newcommand{\poefull}{Path-of-Expanders }
\newcommand{\doefull}{Duo-of-Expanders }

\newcommand{\posfull}{Strong Path-of-Sets }
\newcommand{\posexp}{Expanding Path-of-Sets }

\newcommand{\set}[1]{\left\{ #1 \right\}}

\newcommand{\sse}{\subseteq}

\newcommand{\hG}{\hat G}

\newcommand{\tset}{{\mathcal T}}

\newcommand{\iset}{{\mathcal{I}}}
\newcommand{\isetodd}{{\mathcal{I}}_{\text{odd}}}
\newcommand{\iseteven}{{\mathcal{I}}_{\text{even}}}

\newcommand{\hqset}{\hat{\mathcal{Q}}}
\newcommand{\hpset}{\hat{\mathcal{P}}}
\newcommand{\pset}{{\mathcal{P}}}
\newcommand{\qset}{{\mathcal{Q}}}

\newcommand{\dset}{{\mathcal{D}}}

\newcommand{\aset}{{\mathcal{A}}}
\newcommand{\cset}{{\mathcal{C}}}
\newcommand{\fset}{{\mathcal{F}}}
\newcommand{\mset}{{\mathcal M}}

\newcommand{\hmset}{\hat{\mathcal M}}
\newcommand{\tmset}{\tilde{\mathcal M}}

\newcommand{\gset}{{\mathcal G}}

\newcommand{\wset}{{\mathcal{W}}}

\newcommand{\uset}{{\mathcal U}}

\newcommand{\hset}{{\mathcal{H}}}
\newcommand{\sset}{{\mathcal{S}}}

\newcommand{\be}{\begin{enumerate}}
\newcommand{\ee}{\end{enumerate}}
\newcommand{\bd}{\begin{description}}
\newcommand{\ed}{\end{description}}
\newcommand{\bi}{\begin{itemize}}
\newcommand{\ei}{\end{itemize}}

\newtheorem{theorem}{Theorem}[section]
\newtheorem{lemma}[theorem]{Lemma}

\newtheorem{observation}[theorem]{Observation}
\newtheorem{corollary}[theorem]{Corollary}
\newtheorem{claim}[theorem]{Claim}

\newtheorem{definition}{Definition}

\newenvironment{proof}{\par \smallskip{\bf Proof:}}{\hfill\stopproof}
\def\stopproof{\square}
\def\square{\vbox{\hrule height.2pt\hbox{\vrule width.2pt height5pt \kern5pt
\vrule width.2pt} \hrule height.2pt}}

\renewcommand{\phi}{\varphi}
\newcommand{\eps}{\epsilon}

\newcommand{\poly}{\operatorname{poly}}

\newcommand{\reals}{{\mathbb R}}

\newcommand{\expect}[2][]{\text{\bf E}_{#1}\left [#2\right]}
\newcommand{\prob}[2][]{\text{\bf Pr}_{#1}\left [#2\right]}

\newcommand{\bestbound}[1]{\frac{n}{{#1}\log n}\cdot \left(\frac{\alpha}{d}\right )^{#1}}
\newcommand{\simplebound}[1]{\frac{\alpha^3 n}{{#1} d^5\log^2n}}

\begin{document}
\title{Large Minors in Expanders}

\author{
Julia Chuzhoy\vspace{0.5em}\thanks{Toyota Technological Institute at Chicago. Email: {\tt cjulia@ttic.edu}. Part of the work was done while the author was a Weston visiting professor in the Department of Computer Science and Applied Mathematics, Weizmann Institute. Supported in part by NSF grant CCF-1616584.}
\and
Rachit Nimavat\vspace{0.5em}\thanks{Toyota Technological Institute at Chicago. Email: {\tt nimavat@ttic.edu}. Supported in part by NSF grant CCF-1616584.}
}

\begin{titlepage}
\thispagestyle{empty}
\maketitle

\begin{abstract}
In this paper we study expander graphs and their minors. Specifically, we attempt to answer the following question: what is the largest function $f(n,\alpha,d)$, such that every $n$-vertex $\alpha$-expander with maximum vertex degree at most $d$ contains {\bf every} graph $H$ with at most $f(n,\alpha,d)$ edges and vertices as a minor? Our main result is that there is some universal constant $c$, such that $f(n,\alpha,d)\geq \bestbound{c}$.
This bound achieves a tight dependence on $n$: it is well known that there are bounded-degree $n$-vertex expanders, that do not contain any grid with $\Omega(n/\log n)$ vertices and edges as a minor. The best previous result showed that $f(n,\alpha,d) \geq \Omega(n/\log^{\kappa}n)$, where $\kappa$ depends on both $\alpha$ and $d$. Additionally, we provide a randomized algorithm, that, given an $n$-vertex $\alpha$-expander with maximum vertex degree at most $d$, and another graph $H$ containing at most $\bestbound{c}$ vertices and edges, with high probability finds a model of $H$ in $G$, in time $\poly(n)\cdot (d/\alpha)^{O\left( \log(d/\alpha) \right)}$. We also show a simple randomized algorithm with running time $\poly(n, d/\alpha)$, that obtains a similar result with slightly weaker dependence on $n$ but a better dependence on $d$ and $\alpha$, namely: if $G$ is an $n$-vertex $\alpha$-expander with maximum vertex degree at most $d$, and $H$ contains at most $\simplebound{c'}$ edges and vertices, where $c'$ is an absolute constant, then our algorithm with high probability finds a model of $H$ in $G$.

We note that similar but stronger results were independently obtained by Krivelevich and Nenadov: they show that $f(n,\alpha,d)=\Omega(\frac{n\alpha^2}{d^2\log n})$, and provide an efficient algorithm, that, given an $n$-vertex $\alpha$-expander of maximum vertex degree at most $d$, and a graph $H$ with $O(\frac{n\alpha^2}{d^2\log n})$ vertices and edges, finds a model of $H$ in $G$.

Finally, we observe that expanders are the `most minor-rich' family of graphs in the following sense: for every $n$-vertex and $m$-edge graph $G$, there exists a graph $H$ with $O(\frac{n+m}{\log n})$ vertices and edges, such that $H$ is not a minor of $G$.
\end{abstract}

\end{titlepage}

\input{intro}
\input{prelims}
\input{new_main_proof}
\input{routing_in_toe}
\input{routing_in_doe}
\input{poe-construction}
\input{constructive_proof}

\appendix
\input{cor-proof}
\input{appendix_lower_bounds}

\input{short-paths-routing}

\input{appendix-sec7}

\bibliographystyle{alpha}
\bibliography{main}
\end{document}

%% file: intro.tex
\section{Introduction} \label{sec: intro}
In this paper we study large minors in expander graphs. A graph $G$ is an \emph{$\alpha$-expander}, if, for every partition $(A,B)$ of its vertices into non-empty subsets, the number of edges connecting vertices of $A$ to vertices of $B$ is at least $\alpha\cdot \min\set{|A|,|B|}$. We say that $G$ is an \emph{expander}, if it is an $\alpha$-expander for some constant $0<\alpha<1$, that is independent of the graph size. A graph $H$ is a \emph{minor} of a given graph $G$, if one can obtain a graph isomorphic to $H$ from $G$, via a sequence of edge- and vertex-deletions and edge-contractions.

Bounded-degree expanders are graphs that are simultaneously extremely well connected, while being sparse. Expanders are ubiquitous in discrete mathematics, theoretical computer science and beyond, arising in a wide variety of fields ranging from computational complexity to designing robust computer networks (see \cite{avi_survey} for a survey on expanders and their applications).
In this paper we study an extremal problem about expanders: what if the largest function $f(n,\alpha,d)$, such that every $n$-vertex $\alpha$-expander with maximum vertex degree at most $d$ contains \emph{every} graph with at most $f(n,\alpha,d)$ vertices and edges as a minor?

Our main result is that there is a constant $c$, such that
$f(n,\alpha,d)\geq \bestbound{c}$.
As we discuss below, this result achieves an optimal dependence on $n$.
We also provide a randomized algorithm that, given an $n$-vertex $\alpha$-expander with maximum vertex degree at most $d$, and 
another graph $H$ containing at most $\bestbound{c}$ edges and vertices, with high probability finds a model of $H$ in $G$, in time $\poly(n)\cdot (d/\alpha)^{\log(d/\alpha)}$.
Additionally, we show a simple randomized algorithm with running time $\poly(n,d/\alpha)$, that achieves a bound that has a slightly worse dependence on $n$ but a better dependence on $d$ and $\alpha$:
if $G$ is an $n$-vertex $\alpha$-expander with maximum vertex degree at most $d$, and $H$ is any graph  with at most $\simplebound{{c'}}$ edges and vertices, for some universal constant $c'$, the algorithm finds a model of $H$ in $G$ with high probability.

Independently from our work, Krivelevich and Nenadov (see Theorem 8.1 in~\cite{expander-minor}) provide an elegant proof of a similar but stronger result: namely, they show that $f(n,\alpha,d)=\Omega(\frac{n\alpha^2}{d^2\log n})$, and provide an efficient algorithm, that, given an $n$-vertex $\alpha$-expander of maximum vertex degree at most $d$, and a graph $H$ with $O(\frac{n\alpha^2}{d^2\log n})$ vertices and edges, finds a model of $H$ in $G$.

One of our main motivations for studying this question is the Excluded Grid Theorem of Robertson and Seymour. This is a fundamental result in graph theory,  that was proved by Robertson and Seymour~\cite{gmt_5} as part of their Graph Minors series. The theorem states that there is a function $t: \mathbb{Z}^+ \to \mathbb{Z}^+$, such that for every integer $g>0$, every graph of treewidth at least $t(g)$ contains the $(g\times g)$-grid as a minor. The theorem has found many applications in graph theory and algorithms, including routing problems~\cite{robertson1995graph}, fixed-parameter tractability~\cite{bidimensionality,DemaineH07}, and Erd\H{o}s-P\'{o}sa-type results \cite{gmt_5,thomassen1988presence, Reed-chapter, FominST11}. 
For an integer $g>0$, let $t(g)$ be the smallest value, such that  every graph of treewidth at least $t(g)$ contains the $(g \times g)$-grid as a minor. An important open question is establishing tight bounds on the function $t$. Besides being a fundamental graph-theoretic question in its own right, improved upper bounds on $t$ directly affect the running times of numerous algorithms that rely on the theorem, as well as parameters in various graph-theoretic results, such as, for example, Erd\H{o}s-P\'{o}sa-type results.

In a series of works \cite{gmt_5, RST_exclude_planar, KK_gmt, leaf_gmt, CC_gmt, C_gmt, gmt_julia_arxiv,CT18}, it was shown that $t(g) = \tilde O(g^{9})$ holds. The best currently known negative result, due to Robertson et al. \cite{RST_exclude_planar} is that $t(g) = \Omega(g^2 \log g)$. This is shown by employing a family bounded-degree expander graphs of large girth. Specifically, consider an $n$-vertex expander $G$ whose maximum vertex degree is bounded by a constant independent of $n$, and whose girth is $\Omega(\log n)$. It is not hard to show that the treewidth of $G$ is $\Omega(n)$. Assume now that $G$ contains the $(g\times g)$-grid as a minor, for some value $g$. Such a grid contains $\Omega(g^2)$ disjoint cycles, each of which must consume $\Omega(\log n)$ vertices of $G$, and so $g\leq O(\sqrt{n/\log n})$. This simple argument is the best negative result that is currently known for the Excluded Grid Theorem. In fact, Robertson and Seymour conjecture that this bound is tight, that is, $t(g)=\Theta(g^2\log g)$ must hold. A natural question therefore is whether this analysis is tight, and in particular, whether every $n$-vertex bounded-degree expander must contain a $(g\times g)$-grid as a minor, for $g=O(\sqrt {n/\log n})$. In this paper we answer this question in the affirmative, and moreover, we show that \emph{every} graph with at most $O(n/\log n)$ vertices and edges is a minor of such an expander.

The problem of finding large minors in bounded-degree expanders was first considered by Kleinberg and Rubinfield \cite{KR}.
Building on the random walk-based techniques of Broder et al. \cite{BFU}, they showed that every expander $G$ on $n$ vertices contains every graph with $O(n/\log^{\kappa} n)$ vertices and edges as a minor. The exponent $\kappa$ depends on the expansion $\alpha$ and the maximum degree $d$ of the expander; we estimate it to be at least $\Theta(\log^2d/\log^2(1/\alpha))$.
They also show an efficient algorithm for finding a model of  such a graph in $G$. 

Another related direction of research is the existence of large clique minors in graphs.
The study of the size of the largest clique minor in a graph is motivated by Hadwiger's conjecture, which states that, if the chromatic number of a graph is at least $k$, then it contains a clique with $k$ vertices as a minor.
One well-known result in this area, due to Kawarbayashi and Reed \cite{clique_or_sep_opt}, shows that every $\alpha$-expander $G$ with $n$ vertices and maximum vertex-degree bounded by $d$ contains a clique with $\Omega(\alpha\sqrt{n}/d)$ vertices as a minor.
Recently, Krivelevich and Nenadov \cite{KN2018} improved the dependence on the expansion $\alpha$ and the maximum vertex degree $d$ under a somewhat stronger definition of expansion.
We note that both these bounds have tight dependence on $n$, since $G$ contains only $O(n)$ edges.
Our results imply a weaker bound of $\Omega \left( \left(\frac{\alpha}{d} \right)^{c'} \sqrt{n/\log n} \right)$ on the size of the clique minor, for some absolute constant $c'$.

The existence of large clique minors was also studied in the context of random graphs.
Recall that $G \sim \gset(n,p)$ is a random graph on $n$ vertices, whose edges are added independently with probability $p$ each.
Bollob\'{a}s, Catlin and Erd\H{o}s \cite{hadwiger_erdos} showed that Hadwiger's conjecture is true for almost all graphs $\gset(n,p)$ for every constant $p > 0$.
Fountoulakis et al. \cite{ccl_random} later showed that for every $\epsilon > 0$, there is a constant $\hat c_\epsilon$ such that the following is true:
if $q(n, \epsilon)$ is the probability that the graph $G \sim \gset(n, \frac{1+\epsilon}{n})$ does not contain a clique minor on $\ceil{\hat c_{\epsilon} \sqrt{n}}$ vertices,
then $\lim_{n \rightarrow \infty} q(n,\epsilon) = 0$.
Using a theorem from \cite{exp_random}, our results imply a slightly weaker bound of $\Omega(\sqrt{n/\log n})$ on the clique minor size.

\paragraph{Our Results and Techniques.}
All graphs that we consider are finite; they do not have loops or parallel edges.
Given a graph $H$, we define its \emph{size} to be $|V(H)|+|E(H)|$.
Our main result is summarized in the following theorem:
\begin{theorem}\label{thm: general main}
	There is a constant $c^*$, such that for all $0<\alpha<1$ and $d\geq 1$, if $G$ is an $n$-vertex $\alpha$-expander with maximum vertex degree at most $d$, and $H$ is any graph of size at most $\frac{n}{c^*\log n}\cdot \left(\frac{\alpha}{d}\right )^{c^*}$, then $H$ is a minor of $G$.
	Moreover, there is a randomized algorithm, whose running time is $\poly(n)\cdot (d/\alpha)^{O(\log(d/\alpha))}$, that, given $G$ and $H$ as above, with high probability, finds a model of $H$ in $G$.
\end{theorem}

As discussed above, the theorem implies that we cannot get stronger negative results for the Excluded Grid Theorem using bounded-degree $\alpha$-expanders, where $\alpha$ is independent of the graph size. But this leaves open the possibility of obtaining stronger negative results when $\alpha$ is a function of $n$, such as, for example, $\alpha=1/\poly\log n$, or $\alpha=1/n^{\eps}$ for some small constant $\eps$. Our next result provides a simpler algorithm, with better running time and a better dependence on $d$ and $\alpha$, at the cost of slightly weaker dependence on $n$ in the minor size.

\begin{theorem}\label{thm: constructive main}
	There is a constant $\tilde c^*$ and a randomized algorithm, that,
	given an $n$-vertex $\alpha$-expander with maximum vertex degree at most $d$, where $0<\alpha<1$, and
	another graph $H$ of size at most $\simplebound{\tilde c^*}$,
	with high probability computes a model of $H$ in $G$, in time 
	$\poly(n, d/\alpha)$.
\end{theorem}

The following corollary easily follows from Theorem~\ref{thm: general main} and a result of \cite{exp_random}.

\begin{corollary}\label{cor: random}
	For every $\epsilon > 0$, there is a constant $c_{\epsilon}$ depending only on $\epsilon$, such that a random graph $G \sim \gset(n, \frac{1+\epsilon}{n})$  with high probability contains every graph of size at most $c_{\epsilon} n/\log n$ as a minor.
\end{corollary}

As mentioned earlier, similar but somewhat stronger results were obtained independently by Krivelevich and Nenadov (see Theorem 8.1 in~\cite{expander-minor}).

As a final comment, we show in Appendix \ref{sec: lower bound} that expanders are the `most minor-rich' family of graphs:

\begin{observation}	\label{obs: lower bound}
	For every graph $G$ of size $s \geq 2$, there is a graph $H_G$ of size at most $20 s/ \log s$ such that $G$ does not contain $H_G$ as a minor.
\end{observation}

We now turn to describe our techniques, starting with the simpler result: Theorem~\ref{thm: constructive main}. Given an $n$-vertex $\alpha$-expander $G$ with maximum vertex degree at most $d$, we compute a partition of $G$ into two disjoint subgraphs, $G_1$ and $G_2$, such that $G_1$ is a connected graph; $G_2$ is an $\alpha'$-expander for a  somewhat weaker parameter $\alpha'$, and a large matching $\mset$ connecting vertices of $G_1$ to vertices of $G_2$. We refer to the edges of $\mset$, and to their endpoints, as \emph{terminals}. 
Assume now that we are given a graph $H$, containing at most $\simplebound{\tilde c^*}$ vertices and edges.
We can assume w.l.o.g. that the maximum vertex degree in $H$ is at most $3$, as we can compute a graph $H'$ of size at most twice the size of $H$, such that the maximum vertex degree of $H'$ is at most $3$, and $H$ is a minor of $H'$. Using the transitivity of the minor relation, it is now sufficient to show that $H'$ is a minor of $G$. Therefore, we assume that the maximum vertex degree in $H$ is at most $3$, and we denote $|V(H)|=n'$.
Using the standard grouping technique, we partition the graph $G_1$ into connected subgraphs $S_1,\ldots,S_{n'}$, each of which contains at least $\Theta(d^2\log^2n/\alpha^2)$ terminals. Assume that $H=\set{v_1,\ldots,v_{n'}}$. We map the vertex $v_i$ of $H$ to the graph $S_i$. Let $E_i\subseteq \mset$ be the set of edges of $\mset$ incident to the vertices of $S_i$. Every edge $(v_i,v_j)\in E(H)$ is embedded into a path in the expander $G_2$, that connects some edge of $E_i$ to some edge of $E_j$. The paths are found using standard techniques: we use the classical result of Leighton and Rao~\cite{LeightonRao} to show that for every edge $e=(v_i,v_j)$ of $H$, there is a large set $\pset_e$ of paths in $G_2$, connecting edges of $E_i$ to edges of $E_j$, such that all resulting paths in $\pset=\bigcup_{e\in E(H)}\pset_e$ are short, and cause a small vertex-congestion in $G_2$. We then use the constructive proof of the Lovasz Local Lemma by Moser and Tardos~\cite{Moser-Tardos} to select a single path $P_e$ from each such set $\pset_e$, so that the resulting paths are disjoint in their vertices.

\begin{figure}[h]
    \center
    \includegraphics[width=0.6\linewidth]{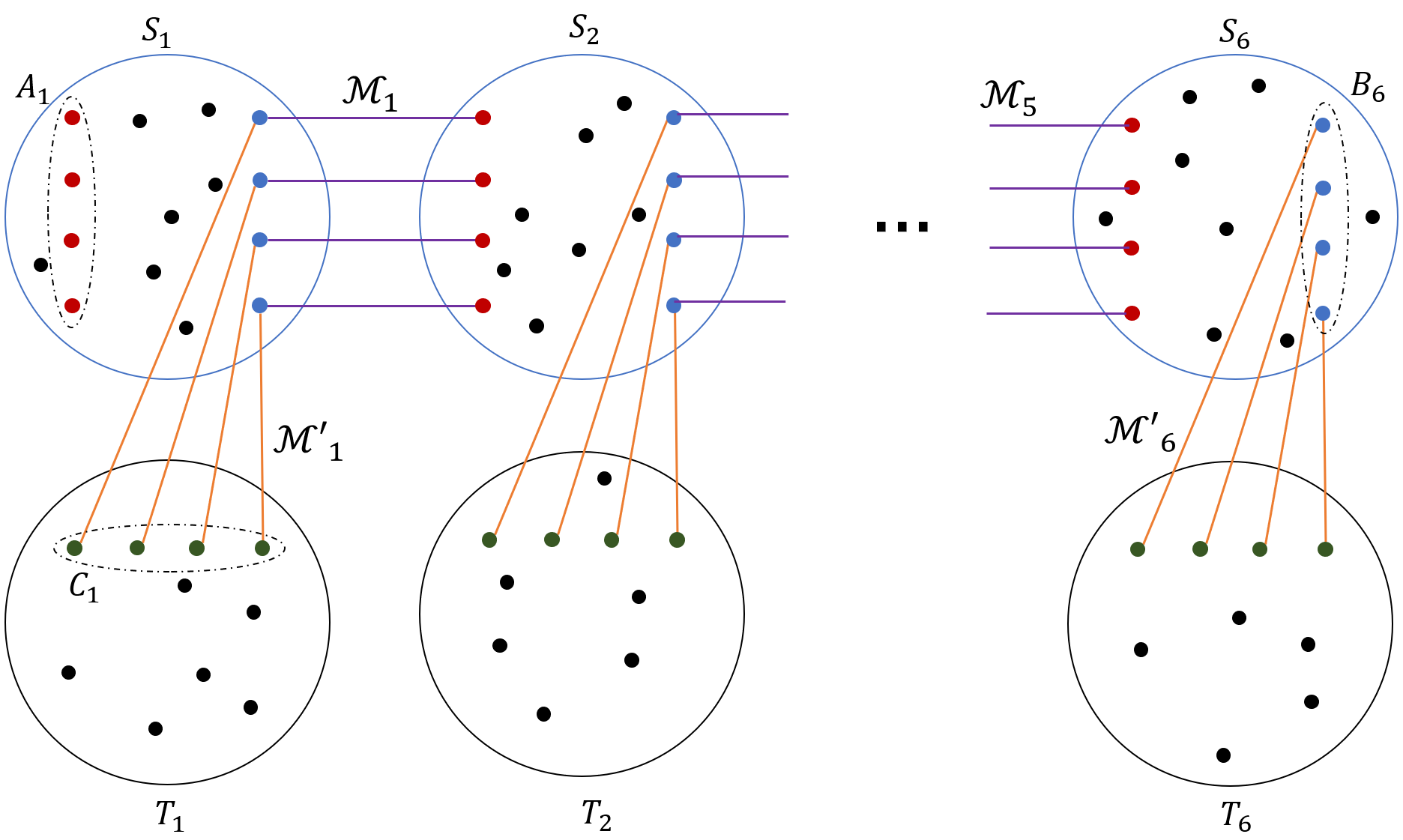}
	\caption{An illustration of the \poefull System $\Pi = (\sset, \mset, A_1, B_{6},\tset, \mset')$.
	For each $1\leq i\leq 6$, the vertices of $A_i$, $B_i$ and $C_i$ are shown in red, blue and green, respectively.
	}
    \label{fig: poe-intro}
\end{figure}

The proof of Theorem~\ref{thm: general main} is somewhat more complex. As before, we assume w.l.o.g. that maximum vertex degree in the graph $H$ is at most $3$. We define a new combinatorial object called a \poefull System (see Figure~\ref{fig: poe-intro}). At a high level, a \poefull System of width $w$ and expansion $\alpha'$ consists of 12 graphs: graphs $T_1,\ldots,T_6$ that are $\alpha'$-expanders, and graphs $S_1,\ldots,S_6$ that are connected graphs. For each $1\leq i\leq 6$, we are also given a matching $\mset'_i$ of cardinality $w$ connecting vertices of $S_i$ to vertices of $T_i$; the endpoints of the edges of $\mset'_i$ in $S_i$ and $T_i$ are denoted by $B_i$ and $C_i$, respectively. For each $1\leq i< 6$, we are given a matching $\mset_i$ connecting every vertex of $B_i$ to some vertex of $S_{i+1}$; the endpoints of the edges of $\mset_i$ that lie in $S_{i+1}$ are denoted by $A_{i+1}$. We show that an $n$-vertex $\alpha$-expander with maximum vertex degree at most $d$ must contain a \poefull System of width $w\geq n(\alpha/d)^c$ and expansion $\alpha'=(\alpha/d)^{c'}$ for some constants $c$ and $c'$, and provide an algorithm with running time $\poly(n)\cdot (d/\alpha)^{O(\log(d/\alpha))}$ to compute it. Next, we split the \poefull System into three parts. The first part is the union of the graphs $S_2,T_2$ and the matching $\mset'_2$. We view the vertices of $B_2$ as terminals, and we use the graph $T_2$ and the matching $\mset'_2$ in order to partition them into large enough groups, and to define a connected sub-graph of $T_2\cup \mset'_2$ spanning each such group, like in the proof of Theorem~\ref{thm: constructive main}. We ensure that the number of groups is equal to the number of vertices in the graph $H$ that we are trying to embed into $G$. Every vertex of $H$ is then embedded into a separate group, together with the corresponding connected sub-graph of $T_2\cup \mset'_2$ spanning the group.

We use the graphs $S_3,\ldots,S_6,T_3\ldots,T_6$ in order to route all but a small fraction of the edges of $H$. The algorithm in this part is inspired by the algorithm of Frieze~\cite{Frieze} for routing a large set of demand pairs in an expander graph via edge-disjoint paths. Lastly, the remaining edges of $H$ are routed in graph $S_1\cup T_1\cup \mset'_1$, using essentially the same algorithm as the one in the proof of Theorem~\ref{thm: constructive main}.

\paragraph{Organization.}
We start with Preliminaries in Section~\ref{sec: prelims}. The proof of Theorem~\ref{thm: general main} is provided in Section~\ref{sec: new main}, with some of the technical details deferred to Sections~\ref{sec:routing in toe} and~\ref{sec: routing in doe}. Section~\ref{sec: pos to poe} contains an algorithm for constructing a \poefull System. The proof of Theorem~\ref{thm: constructive main} appears in Section~\ref{sec: constructive proof}, and the proofs of Corollary~\ref{cor: random} and Observation~\ref{obs: lower bound} appear in Sections~\ref{sec: cor proof} and~\ref{sec: lower bound} of the Appendix, respectively.

%% file: prelims.tex
\section{Preliminaries} \label{sec: prelims}
Throughout the paper, for an integer $\ell\geq 1$, we denote $[\ell]=\set{1,\ldots,\ell}$. All logarithms in the paper are to the base of $2$.

All graphs that we consider are finite; they do not have loops or parallel edges.

We will use the following simple observation, whose proof is deferred to the Appendix.

\begin{observation}\label{obs: simple partition} 
	There is an efficient algorithm, that, given a set
	$\set{x_1,\ldots,x_r}$ of non-negative integers, with
	$\sum_{i}x_i=N$, and $x_i\leq 3N/4$ for all $i$, computes a
	partition $(A,B)$ of $\set{1,\ldots,r}$, such that $\sum_{i\in A}x_i
	\geq N/4$ and $\sum_{i\in B}x_i\geq N/4$.
\end{observation}

Given a graph $G=(V,E)$ and a subset $V' \subseteq V$ of its vertices, we denote by $\delta_G(V')$ the set of all edges that have exactly one  endpoint in $V'$, and by $E_G[V']$ the set of all edges with both endpoints in $V'$.
For readability, we write $\delta_G(v)$ instead of $\delta_G(\set{v})$. Given a pair $V', V'' \subseteq V$  of disjoint subsets of vertices, we denote by $E_G(V', V'')$ the set of all the edges with one endpoint in $V'$ and another in $V''$.
We will omit the subscript $G$ when the underlying graph is clear from context.
For a subset $V'\subseteq V$ of vertices of $G$, 
we denote by $G[V']$ the subgraph of $G$ induced by $V'$.

Given a path $P$ in a graph $G$, we denote by $V_P$ and $E_P$ the sets of all its vertices and edges, respectively. 
Given a path $P$ and a set $V'$ of vertices of $G$, we say that $P$ is \emph{disjoint} from $V'$ iff $V_P\cap V'=\emptyset$. We say that $P$ is \emph{internally disjoint} from $V'$ iff every vertex of $V'\cap V_P$ is an endpoint of $P$.

Similarly, suppose we are given two paths $P,P'$ in a graph $G$. We say that the two paths are  \emph{disjoint} iff $V_P \cap V_{P'} = \emptyset$, and we say that they are  \emph{internally disjoint} iff all vertices in $V_P \cap V_{P'}$ serve as endpoints of both these paths.

Let $\pset$ be any set of paths in a graph $G$. We say that $\pset$ is a set of \emph{disjoint} paths iff every pair  $P,P' \in \pset$ of distinct paths are disjoint.
We say that $\pset$ is a set of \emph{internally disjoint} paths iff every pair  $P,P' \in \pset$ of distinct paths are internally disjoint.
We denote by $V(\pset) = \bigcup_{P \in \pset} V_P$ the set of all vertices participating in the paths of $\pset$.
Given a pair $V',V''$ of subsets of vertices of $V$ (that are not necessarily disjoint), we say that a path $P\in \pset$  \emph{connects} $V'$ to $V''$ iff one of its endpoints is in $V'$ and the other endpoint is in $V''$. We use a shorthand $\pset : V' \rightsquigarrow V''$ to indicate that $\pset$ is a collection of disjoint paths, where each path $P \in \pset$ connects $V'$ to $V''$.
Notice that each path in $\pset$ must originate at a distinct vertex of $V'$ and terminate at a distinct vertex of $V''$.

Finally, assume that we are given a (partial) matching $\mset$ over the vertices of $G$, and a set $\pset$ of $|\mset|$ paths. We say that $\pset$ \emph{routes $\mset$} iff for every pair of vertices $(v',v'')\in \mset$, there is a path $P \in \pset$, whose endpoints are $v'$ and $v''$. 

\paragraph{Sparsest Cut and Expansion.}
A \emph{cut}  in $G$ is a bipartition $(S, S')$ of its vertices, that is, $S \cup S' = V$, $S \cap S' = \emptyset$ and $S, S' \neq \emptyset$.
The \emph{sparsity} of the cut $(S, S')$ is $|E(S,S')|/\min{ \set{|S|, |S'|}}$. The \emph{expansion} of a graph $G$, denoted by $\phi(G)$, is the minimum sparsity of any cut in $G$. 

\begin{definition}
Given a parameter $\alpha>0$, we say that a graph $G$ is an $\alpha$-expander iff $\phi(G)\geq \alpha$. Equivalently, for every subset $S$ of at most $|V(G)|/2$ vertices of $G$, $|\delta_G(S)|\geq \alpha |S|$.
\end{definition}

The following theorem follows from the standard Cheeger's inequality, that shows that for any graph $G$, whose maximum vertex degree is bounded by $d$, $\frac{ \lambda(G)}{2} \leq \varphi(G) \leq  \sqrt{2d \lambda(G)}$, where $\lambda(G)$ is the second smallest eigenvalue of the Laplacian of $G$, and from the algorithm of~\cite{cheeger-alg} (see also~\cite{Cheeger1,Cheeger2,Alon}).

\begin{theorem}\label{thm: spectral}
There is an efficient algorithm, that, given an $n$-vertex graph $G$ with maximum vertex degree at most $d$, computes a cut $(A,B)$ in $G$ of sparsity $O(\sqrt{d\phi(G)})$.
\end{theorem}

Finally, we use the following simple claim several times; the claim allows one to ``fix'' an expander, after a small number of edges were deleted from it. The proof appears in Appendix.

\begin{claim}\label{claim: large expanding subgraph-edges}
     Let $T$ be an $\alpha$-expander, and let $E'$ be any subset of edges of $T$. Then there is an $\alpha/4$-expander $T'\subseteq T\setminus E'$, with $|V(T')|\geq |V(T)|-\frac{4|E'|}{\alpha}$.
\end{claim}

\paragraph{Graph Minors.}

\begin{definition}[Graph Minors] \label{def: minor}
	We say that a graph $H = (U, F)$ is a \emph{minor} of a graph $G = (V, E)$ iff there is a map $f$, called a \emph{model of $H$ in $G$}, mapping every vertex $u \in U$ to a subset $X_u\subseteq V$ of vertices, and mapping every edge $e \in F$ to a path $P_e$ in $G$, such that:
	\begin{itemize}
		\item For every vertex $u \in U$, $G[X_u]$ is connected;
		\item For every edge $e = (u,v) \in F$, the path $P_e$ connects $X_u$ to $X_v$;
		\item For every pair $u,v \in U$ of distinct vertices, $X_u\cap X_v=\emptyset$; and
		\item Paths $\set{P_e \mid e \in F}$ are internally disjoint from each other and they are internally disjoint from the set $\bigcup_{u \in U}X_u$ of vertices. 
	\end{itemize}
For a vertex $u\in U$ we sometimes call $G[X_u]$ the \emph{embedding of $u$ into $G$}, and for an edge $e\in F$, we sometimes refer to $P_e$ as the \emph{embedding of $e$ into $G$}.
\end{definition}

\paragraph{Well-Linkedness and Path-of-Sets System.}

We use a slight variation of the standard definition of (node)-well-linkedness.

\begin{definition}[Well-Linkedness]
We say that a set $A$ of vertices in a graph $G$ is \emph{well-linked} iff for every pair $A',A''$ of disjoint equal-cardinality subsets of $A$, there is a set  $\pset : A' \rightsquigarrow A''$ of $|A'|$ paths in $G$, that are internally disjoint from $A$. (Note that the paths in $\pset$ must be disjoint).
\end{definition}

Next, we define a Path-of-Sets system, that was first introduced in~\cite{CC_gmt} (a somewhat similar object called \emph{grill} was introduced by \cite{leaf_gmt}), and was used since then in a number of graph theoretic results.

\begin{definition}[Path-of-Sets System] \label{def: pos}
	Given integers $w,\ell>0$ a Path-of-Sets System of width $w$ and length $\ell$  (see Figure \ref{fig: pos}) consists of:
	\begin{itemize}
		\item a sequence $\sset = (S_1, \ldots, S_{\ell})$ of $\ell$ disjoint connected graphs, that we refer to as \emph{clusters};

		\item for each $1\leq i \leq \ell$, two disjoint subsets, $A_i, B_i \subseteq V(S_i)$ of $w$ vertices each; and

		\item For each $1 \leq i < \ell$, a collection $\mset_i$ of edges, connecting every vertex of $B_i$ to a distinct vertex of $A_{i+1}$.
	\end{itemize}
	We denote the Path-of-Sets  System by $\Sigma = (\sset, \mset, A_1, B_{\ell})$, where $\mset = \bigcup_i \mset_i$. We also denote by $G_{\Sigma}
	$ the graph defined by the Path-of-Sets System, that is, $G_{\Sigma}=\left(\bigcup_{i=1}^{\ell}S_i\right )\cup \mset$.
	
	We say that a given Path-of-Sets System is \emph{a \posfull System} iff all $1\leq i\leq \ell$, the vertices of $A_i\cup B_i$ are well-linked in $S_i$. We say that it is $\alpha$-expanding, iff for all $1\leq i\leq \ell$, graph $S_i$ is an $\alpha$-expander. Note that a \posfull System is not necessarily $\alpha$-expanding and vice versa.
\end{definition}

\begin{minipage}{\linewidth}
    \centering
    \begin{minipage}{0.6\linewidth}
		\begin{figure}[H]
			\includegraphics[width=\linewidth]{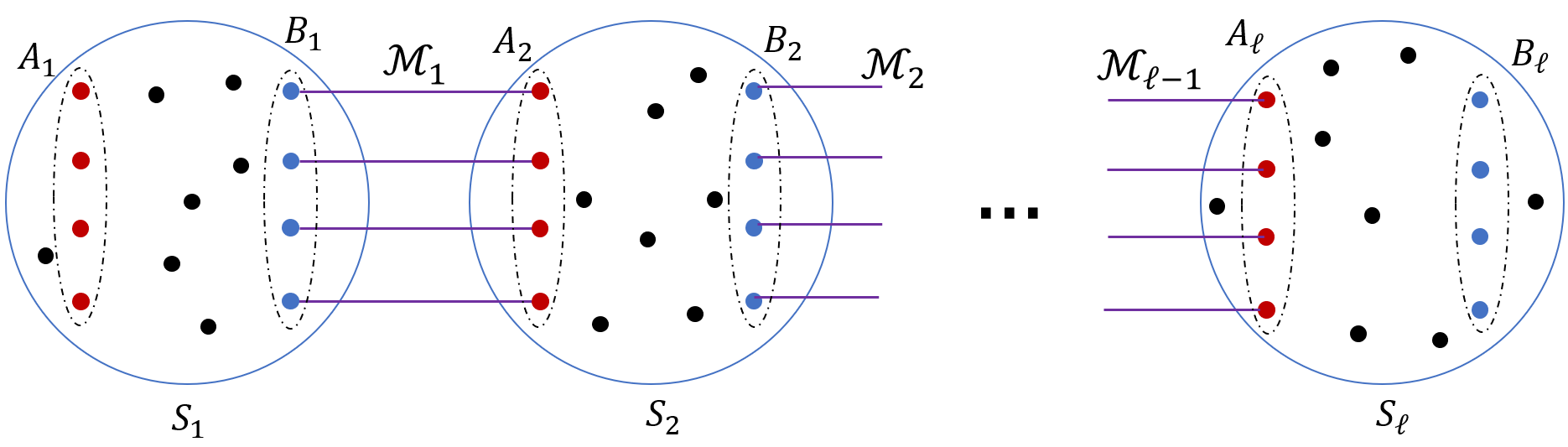}
			\caption{An illustration of a Path-of-Sets System $(\sset, \mset, A_1, B_{\ell})$.
				For each $i \in [\ell]$, the vertices of $A_i$ and $B_i$ are shown in red and blue respectively.
			}
    		\label{fig: pos}
        \end{figure}
    \end{minipage}
    \hspace{0.04\linewidth}
    \begin{minipage}{0.5\linewidth}
        \begin{figure}[H]
            \includegraphics[width=\linewidth]{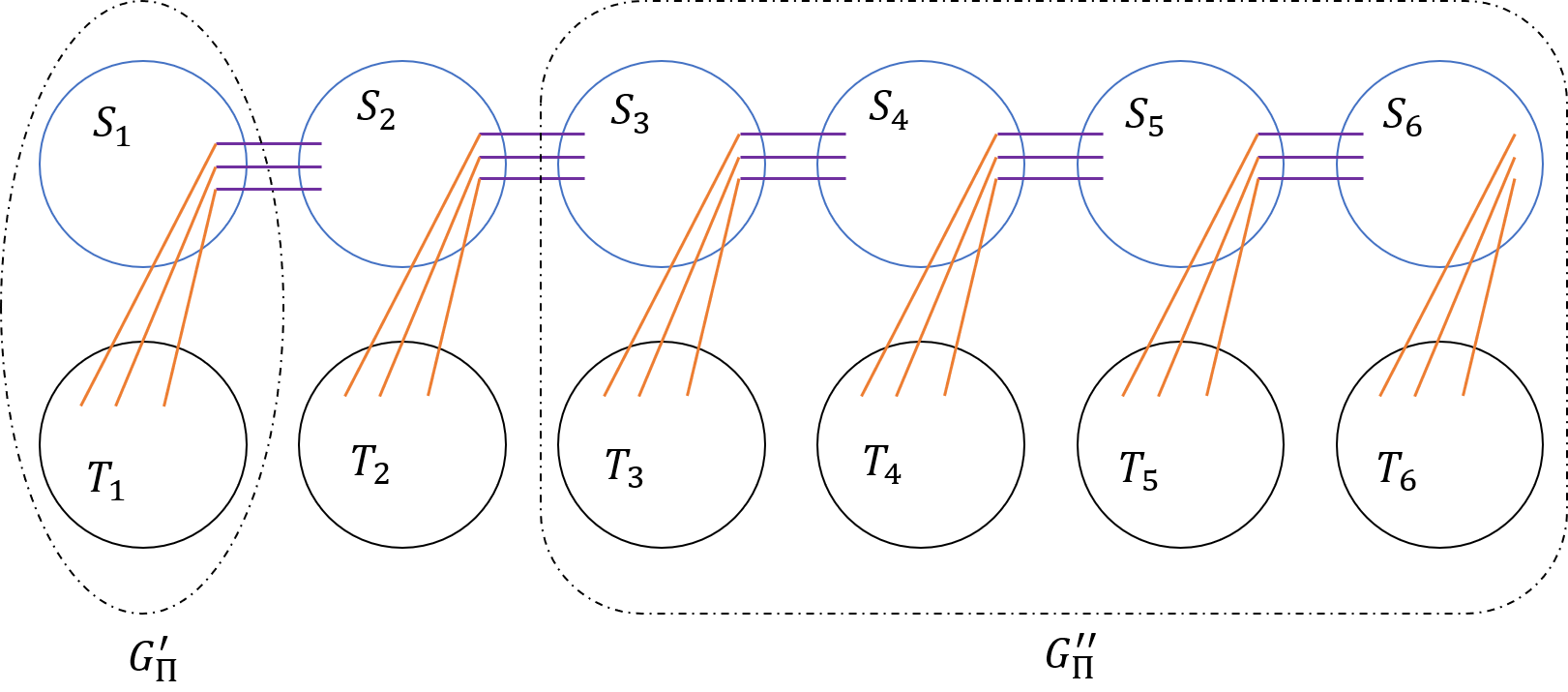}
			\caption{An illustration of the subgraphs $G'_{\Pi}$ and $G''_{\Pi}$ of $G_{\Pi}$.}
    		\label{fig: g_prime_and_double_prime}
        \end{figure}
    \end{minipage}
\end{minipage}

\subsection{\poefull System}\label{subsec: poe}
\poefull System is the main new structural object that we use.

\begin{definition}[\poefull System]\label{def: poe}
	Given an integer $w>0$ and a parameter $0<\alpha<1$, a  \poefull System of width $w$ and expansion $\alpha$  (see Figure \ref{fig: poe-intro}) consists of:
	\begin{itemize}
		\item a \posfull System $\Sigma = (\sset, \mset, A_1, B_6)$ of width $w$ and length $6$;

		\item a sequence $\tset = (T_1, \ldots, T_{6})$ of $6$ disjoint connected graphs, such that for each $1\leq i\leq 6$, $T_i$ is disjoint from $S_1,\ldots,S_6$, and it is an $\alpha$-expander; and

		\item for each $1\leq i\leq 6$, a perfect matching $\mset'_i$ between $B_i$ and some subset $C_i$ of $w$ vertices of $T_i$.
	\end{itemize}
	We denote the \poefull System by $\Pi = (\sset, \mset,  A_1,B_6,\tset, \mset')$, where $\mset' = \bigcup_i \mset'_i$.
	For convenience, for each $1\leq i \leq 6$, we denote by $W_i$ be the graph obtained from the union of the graphs $S_i$ and $T_i$, and the matching $\mset'_i$.
\end{definition}

Similarly to the Path-of-Sets System, we associate with the \poefull System $\Pi$ a graph $G_{\Pi}$, obtained by taking the union of the graphs $S_1,\ldots,S_6$, $T_1,\ldots,T_6$ and the sets $\mset,\mset'$ of edges.

We will be interested in three subgraphs of $G_\Pi$ (see Figure \ref{fig: g_prime_and_double_prime}): 
(i) Graph $W_1$, that we denote by $G'_{\Pi}$;
(ii) Graph $W_2$; and
(iii) Graph $G''_{\Pi}$, obtained by taking the union of $W_3 \cup W_4 \cup W_5 \cup W_6$ and the edges of $\mset_3\cup\mset_4\cup \mset_5$.

\begin{definition}
We say that a graph $G$ contains a Path-of-Sets System of width $w$ and length $\ell$ as a minor iff there is a Path-of-Sets System  $\Sigma$ of width $w$ and length $\ell$, such that its corresponding graph $G_{\Sigma}$ is a minor of $G$. Similarly, 
we say that  a graph $G$ contains a \poefull  System of width $w$ and expansion $\alpha$ as a minor iff there is a \poefull  System $\Pi$ of width $w$ and expansion $\alpha$, such that its corresponding graph $G_{\Pi}$ is a minor of $G$.
\end{definition}

The following theorem, that we prove in Section \ref{sec: pos to poe}, shows that an expander must contain a \poefull System with suitably chosen parameters, and provides an algorithm to compute its model in the expander. 

\begin{theorem} \label{thm: poe}
There are constants $\hat c_1,\hat c_2$,  and an algorithm, that, given an $\alpha$-expander $G$ with $|V(G)|=n$, whose maximum vertex degree is at most $d$, and $0<\alpha<1$, constructs a \poefull System $\Pi$ of expansion $\tilde \alpha \geq \left(\frac{\alpha}{d}\right )^{\hat c_1}$ and width $w \geq n \cdot \left(\frac{\alpha}{d}\right )^{\hat c_2} $, such that the corresponding graph $G_{\Pi}$ has maximum vertex degree at most $d+1$ and is a minor of $G$. Moreover, the algorithm computes a model of $G_{\Pi}$ in $G$. The running time of the algorithm is $\poly(n)\cdot (d/\alpha)^{O(\log(d/\alpha))}$.
\end{theorem}

%% file: new_main_proof.tex
\section{Proof of Theorem \ref{thm: general main}} \label{sec: new main}
The goal of this section is to prove Theorem \ref{thm: general main}.
We prove it by using the following theorem.

\begin{theorem} \label{thm: embedding algo}
    There is constant $c_0$ and a randomized algorithm, that, 
    given a \poefull System $\Pi$ with expansion $\alpha$ and width $w$,
    such that the maximum vertex degree in $G_{\Pi}$ is at most $d$ and $|V(G_{\Pi})|\leq n$ for some $n>c_0$,
    together with a graph $H$ of maximum vertex degree at most $3$ and $|V(H)|\leq \frac{w^2\alpha^2}{2^{19}d^4n \log n}$,
    with high probability,
    in time $\poly(n)$,
    finds a model of $H$ in $G_{\Pi}$. 
\end{theorem}

Before proving Theorem \ref{thm: embedding algo}, we complete the proof of Theorem \ref{thm: general main} using it.
Let $G$ be the given $\alpha$-expander with $|V(G)|=n$, and maximum vertex degree at most $d$. 
Recall that $0<\alpha < 1$.
By letting $c^*$ be a sufficiently large constant, we can assume that $n$ is sufficiently large,
so that, for example, $n>c_0$, where $c_0$ is the constant from Theorem~\ref{thm: embedding algo}.
Indeed, otherwise, it is enough to show that the graph with $1$ vertex is a minor of $G$, which is trivially true.
Therefore, we assume from now on that $n$ is sufficiently large. 
From Theorem~\ref{thm: poe}, $G$ contains as a minor a \poefull System $\Pi$
of width $w \geq n \cdot \left(\frac{\alpha}{d}\right )^{\hat c_2} $ and
expansion  $\tilde \alpha \geq \left(\frac{\alpha}{d}\right )^{\hat c_1}$,
such that the maximum vertex degree in $G_{\Pi}$ is at most $d+1$. 
Using these bounds, we get that:

\[\begin{split}
\frac{w^2 \tilde \alpha^2}{2^{19}(d+1)^4n\log n}&\geq  n^2 \cdot \left(\frac{\alpha}{d}\right )^{2\hat c_2} \cdot  \left(\frac{\alpha}{d}\right )^{2\hat c_1}\cdot \frac{1}{2^{23}d^4 n\log n}\\
&=\frac{n \alpha^{2(\hat c_1+\hat c_2)}}{2^{23}d^{4+2(\hat c_1+\hat c_2)}\log n}\\
&\geq \frac{3n}{c^*\log n}\cdot \left(\frac{\alpha}{d} \right)^{c^*},
\end{split}\]
for $c^* \geq \max{\set{4 + 2 (\hat c_1 + \hat c_2), c_0, 2^{25}}}$.
Therefore, if $H'$ is a graph with maximum vertex degree at most $3$, and
$|V(H')|\leq \frac{3n}{c^*\log n}\cdot \left(\frac{\alpha}{d} \right)^{c^*}$,
then, from Theorem~\ref{thm: embedding algo},
$G$ contains $H'$ as a minor, and from Theorems~\ref{thm: poe} and~\ref{thm: embedding algo}, its model in $G$ can be computed with high probability by a randomized algorithm, in time $\poly(n)\cdot (d/\alpha)^{O(\log(d/\alpha))}$.

Consider now any graph $H=(U,F)$ of size at most
$\frac{n}{c^*\log n}\cdot \left(\frac{\alpha}{\Delta} \right)^{c^*}$.
Let $n'=|U|$ and $m'=|F|$, so
$n'+m'\leq \frac{n}{c^*\log n}\cdot \left(\frac{\alpha}{d} \right)^{c^*}$.
We construct another graph $H'$, whose maximum vertex degree is at most $3$ and $|V(H')|\leq n'+2m'$, such that $H$ is a minor of $H'$. Since $H'$ must be a minor of $G$, it follows that $H$ is a minor of $G$. In order to construct graph $H'$ from $H$, we consider every vertex $u\in U$ of degree $d_u>3$ in turn, and replace it with a cycle $C_u$ on $d_u$ vertices, such that every edge incident to $u$ in $H$ is incident to a distinct vertex of $C_u$. It is easy to verify that the resulting graph $H'$ has maximum vertex degree at most $3$, that $H$ is a minor of $H'$, and that $|V(H')|\leq 2m'+n'$, completing the proof of Theorem~\ref{thm: general main}. Notice that this proof is constructive, that is, there is a randomized algorithm that constructs a model of $H$ in $G$ in time $\poly(n)\cdot (d/\alpha)^{O(\log(d/\alpha))}$. The remainder of this section is dedicated to proving Theorem~\ref{thm: embedding algo}, with some details deferred to subsequent sections.

\subsection{Large Minors in \poefull System} \label{subsec: proofof embedding algo}

This subsection is devoted to the proof Theorem \ref{thm: embedding algo}.
We assume that we are given a  \poefull System $\Pi = (\sset, \mset, A_1, B_{6},\tset, \mset')$ of width $w$ and expansion $\alpha$, whose corresponding graph $G_{\Pi}$ contains at most $n$ vertices, where $n>c_0$ for some large enough constant $c_0$, and its maximum vertex degree is bounded by $d$. 
In order to simplify the notation, we denote $G_{\Pi}$ by $G$. 
We also use the following parameter: $\rho = 2^{16}\floor{\frac{d^3n\log n}{\alpha^2w}}$. 

We are also given a graph $H = (U,F)$ of maximum degree $3$, with
$ |U| \leq \frac{w^2\alpha^2}{2^{19}d^4n\log n} \leq \frac{w}{8d \rho}$.

Our goal is to find a model of $H$ in $G$.
Our algorithm consists of three steps. In the first step, we associate with each vertex $u\in U$, a subset $X_u$ of vertices of $W_2$, such that $W_2[X_u]$ is a connected graph. This defines the embeddings of the vertices of $H$ into $G$ for the model of $H$ that we are computing.
In the second step, we embed all but a small fraction of the edges of $H$ into $G''_{\Pi}$, and in the last step, we embed the remaining edges of $H$ into $G'_{\Pi}$. 
We now describe each step in detail.

\paragraph{Step 1: Embedding the Vertices of $H$.}
In this step we compute an embedding of every vertex of $H$ into a connected subgraph of $W_2$. Recall that graph $W_2$ is the union of the graphs $S_2$ and $T_2$, and the matching $\mset'_2$, connecting the vertices of $B_2\subseteq V(S_2)$ to the vertices of $C_2\subseteq V(T_2)$, where $|B_2|=|C_2|=w$.
We use the following simple observation, that was used extensively in the literature (often under the name of ``grouping technique'') (see e.g. ~\cite{CKS, RaoZhou, Andrews,Chuzhoy11}). The proof is deferred to Section \ref{subsec: proof of decompose by spanning tree} of the Appendix.

\begin{observation} \label{obs: decompose by spanning tree}
    There is an efficient algorithm that, given a connected graph $\hG$ with maximum vertex degree at most $d$, an integer $r\geq 1$, and a subset  $R \subseteq V(\hG)$ of vertices of $\hG$ with $|R|\geq r$, computes a collection $\set{V_1, \ldots, V_r}$ of $r$ mutually disjoint subsets of $V(\hat G)$, such that:
    \begin{itemize}
        \item For each $i \in [r]$, the induced graph $\hG[V_i]$ is connected; and
        \item For each $i \in [r]$, $|V_i \cap R| \geq \floor{|R|/(d r)}$.
    \end{itemize}
\end{observation}

We apply the above observation to the graph $T_2$, together with vertex set $R=C_2$ and parameter $r=\floor{\frac{w}{8d\rho}}$. Let $\uset$ be the resulting collection of $r$ subsets of vertices of $T_2$. Recall that for each set $V_i\in \uset$, $|V_i\cap C_2|\geq \floor{\frac{|C_2|}{dr}}\geq \floor{\frac{w}{d\floor{w/8d\rho}}} \geq 3\rho$. Since $|U|\leq \frac{w}{8d\rho}$, we can choose $|U|$ distinct sets $V_1,\ldots,V_{|U|}\in \uset$. We also denote $U=\set{u_1,\ldots,u_{|U|}}$. Finally, for each $1\leq i\leq |U|$, we let $E^i\subseteq \mset'_2$ be the subset of edges that have an endpoint in $V_i$, and we let $B^i_2$ be the subset of vertices of $B_2$ that serve as endpoints of the edges in $E^i$. Since  $|V_i\cap C_2|\geq 3\rho$, $|B^i_2|\geq 3\rho$ for all $i$. We are now ready to define the embeddings of the vertices of $H$ into $G$. For each $1\leq i\leq |U|$, we let $f(u_i)=G[B_2^i\cup V_i]$. Notice that for all $1\leq i\leq |U|$, $f(u_i)$ is a connected graph, and for all $1\leq i<j\leq | U|$, $f(u_i)\cap f(u_j)=\emptyset$. 
In the remaining steps, we focus on embedding  the edges of $H$ into $G$, such that the resulting paths are internally disjoint from $B_2 \cup T_2$.

\begin{figure}[h]
    \center
    \includegraphics[width=0.45\linewidth]{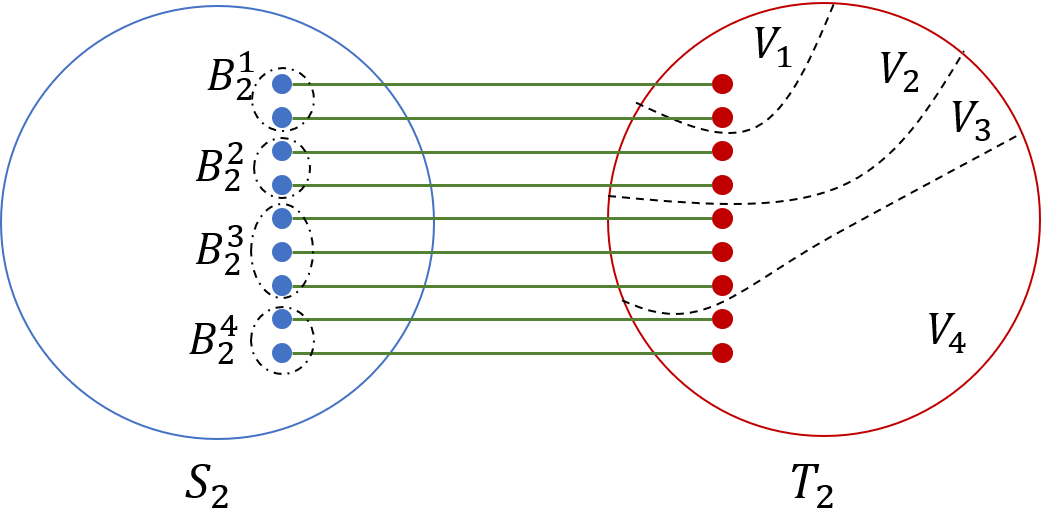}
    \caption{A sketch of the partition of $T_2$ and $B_2$. Vertices of $B_2$ and $C_2$ are shown in blue and red respectively. }
    \label{fig: embedding in w2}
\end{figure}

\paragraph{Step 2: Routing in $G''_{\Pi}$.}
Consider some vertex $u_i \in U$, its corresponding graph $f(u_i)$, and the set $B_2^i\subseteq B_2$ of vertices that lie in $f(u_i)$; recall that $|B_2^i|\geq 3\rho$. Recall that the maximum vertex degree in $H$ is at most $3$. For every edge $e\in \delta_H(u_i)$, we select an arbitrary subset $B_2^i(e)\subseteq B_2^i$ of $\rho$ vertices, so that all resulting sets $\set{B_2^i(e)}_{e\in \delta_H(u_i)}$ are mutually disjoint.

Recall that graph $G_{\Pi}$ contains a perfect matching $\mset_2$ between the vertices of $B_2$ and the vertices of $A_3$. We let $\hat E^i\subseteq \mset_2$ be the subset of edges whose endpoints lie in $B_2^i$, and denote by $A_3^i\subseteq A_3$ the set of endpoints of the edges of $E^i$ that lie in $A_3$.
For every edge $e\in \delta(v_i)$, we let $A_3^i(e)\subseteq A_3^i$ be the set of $\rho$ vertices that are connected to the vertices of $B_2^i(e)$ with an edge of $\mset_2$. Clearly, all resulting vertex sets $\set{A_3^i(e)}_{e\in \delta_H(u_i)}$ are mutually disjoint.
Let $A_3' = \bigcup_{u_i \in U} \bigcup_{e \in \delta_H(u_i)} A_3^i(e)$, and notice that
\[ |A_3'| \leq 3 \rho \cdot |U| \leq 3 \rho \cdot \frac{w}{8d\rho}  = \frac{3w}{8d} \leq \frac{w}{2}.\]

The following lemma, whose proof is deferred to Section \ref{sec:routing in toe}, allows us to embed a large number of edges of $H$ in $G''_{\Pi}$.

\begin{lemma}\label{lem: routing in toe}
    There is an efficient algorithm, that, given a \poefull System $\Pi = (\sset, \mset,  A_1, B_{6}, \tset, \mset')$ of expansion $\alpha$ and width $w$, where $0<\alpha<1$ and $w$ is an integral multiple of $4$, whose corresponding graph $G_{\Pi}$ contains at most $n$ vertices and has  maximum vertex degree at most $d$, together with a subset $A'_3 \subseteq A_3$ of at most $w/2$ vertices, and a collection $\set{A_3^1, \ldots, A_3^{2 r}}$ of mutually disjoint subsets of $A'_3$ of cardinality $\rho= 2^{16}\floor{\frac{d^3n\log n}{\alpha^2w}}$ each, where $r> \frac{w \alpha^2 (\log \log n)^2}{d^3 \log^3 n}$, returns 
 a partition $\iset',\iset''$ of $\set{1,\ldots,r}$, and a set $\pset^* = \set{P^*_j \mid j \in \iset'}$ of disjoint paths in $G''_{\Pi}$, such that for each $j \in \iset'$, path $P^*_j$ connects $A_3^j$ to $A_3^{j+r}$, and $|\iset''|\leq  \frac{w \alpha^2 (\log \log n)^2}{d^3 \log^3 n}$.
 \end{lemma}

We obtain the following immediate corollary of the lemma.

\begin{corollary}\label{cor: step 2}
There is an efficient algorithm to compute a partition $(F_1,F_2)$ of the set $F$ of edges of $H$, and for each edge $e=(u_i,u_j)\in F_1$, a path $P^*_e$ in graph $G''_{\Pi}$, connecting a vertex of $A_3^i(e)$ to a vertex of $A_3^j(e)$, such that all paths in set $\pset^*_1=\set{P^*(e)\mid e\in F_1}$ are disjoint, and $|F_2|\leq \frac{w \alpha^2 (\log \log n)^2}{d^3 \log^3 n}$.
\end{corollary}
\begin{proof}
By appropriately ordering the collection $\set{A_3^i(e)\mid u_i\in U,e\in \delta_H(e)}$ of vertex subsets, and applying Lemma~\ref{lem: routing in toe} to the resulting sequence of subsets of $A_3'$, we obtain a set $F_1\subseteq F$ of edges of $H$, and for each edge $e=(u_i,u_j)\in F_1$, a path $P^*_e$, connecting a vertex of $A_3^i(e)$ to a vertex of $A_3^j(e)$ in graph $G''_{\Pi}$, such that all paths in set $\pset^*_1=\set{P^*(e)\mid e\in F_1}$ are disjoint. Let $F_2=F\setminus F_1$. From Lemma~\ref{lem: routing in toe}, $|F_2| \leq  \frac{w \alpha^2 (\log \log n)^2}{d^3 \log^3 n}$. 
\end{proof}

For each edge $e=(u_i,u_j)\in F_1$, we extend the path $P^*_e$ to include the two edges of $\mset_2$ incident to its endpoints, so that $P^*_e$ now connects a vertex of $B_2^i$ to a vertex of $B_2^j$. Path $P^*_e$ becomes the embedding $f(e)$ of $e$ in the model $f$ of $H$ that we are constructing. For convenience, the resulting set of paths $\set{P^*_e\mid e\in F_1}$ is still denoted by $\pset^*_1$. The paths in $\pset^*_1$ remain disjoint from each other; they are internally disjoint from $W_2$, and completely disjoint from  $W_1$ (see Figure \ref{fig: pset 1}).

\paragraph{Step 3: Routing in $G'_{\Pi}$.}
In this step we complete the construction of a minor of $H$ in $G$, by embedding the edges of $F_2$.
The main tool that we use is the following lemma, whose proof is deferred to Section~\ref{sec: routing in doe}.

\begin{lemma}\label{lem: routable b}
    There is a universal constant $c$, and an efficient algorithm that, given a \poefull System $\Pi = (\sset, \mset,A_1, B_{6}, \tset, \mset')$ of expansion $\alpha$ and width $w$, such that the corresponding graph $G_{\Pi}$ contains at most $n$ vertices and has  maximum vertex degree at most $d$, computes a subset $B'_1 \subseteq B_1$ of at least $ \frac{cw \alpha^2}{d^3 \log^2 n}$ vertices, such that the following holds.
    There is an efficient randomized algorithm, that given any matching $\mset^*$ over the vertices of $B'_1$, with high probability returns a set $\pset$ of disjoint paths in $W_1$, routing $\mset^*$.
\end{lemma}

We now conclude the last step using the above lemma.
Let $B'_1 \subseteq B_1$ be the subset of at least $\frac{c w \alpha^2}{d^3 \log^2 n}$ vertices, computed by algorithm from Lemma~\ref{lem: routable b}.
Let $A'_2 \subseteq A_2$ be the set of all the vertices connected to the vertices of $B'_1$ by the edges of the matching $\mset_1$.
Observe that $|A'_2|\geq 2|F_2|$, since:

\[ 2|F_2|\leq \frac{2 w \alpha^2 (\log \log n)^2}{d^3 \log^3 n} \leq \frac{c w \alpha^2}{d^3 \log^2 n} = |B'_1| = |A'_2|,\]

since we have assumed that $n$ is sufficiently large. We let $A''_2$ be an arbitrary subset of $2|F_2|$ vertices of $A'_2$.

Recall that every vertex $u_i \in U$, and edge $e\in \delta_H(u_i)$, we have defined a subset $B_2^i(e)\subseteq B_2^i$ of vertices. 
We select an arbitrary representative vertex $b_2^i(e)\in B_2^i(e)$, and we let $B'_2=\set{b_2^i(e)\mid u_i\in U, e\in \delta(u_i)\cap F_2}$ be the resulting set of representative vertices, so that $|B'_2|=2|F|$. 

Since $(A_2\cup B_2)$ are well-linked in $S_2$, there is a set $\qset_2$ of $2|F_2|$ disjoint paths in $S_2$, connecting every vertex of $B'_2$ to some vertex of $A''_2$, such that the paths in $\qset_2$ are internally disjoint from $A_2 \cup B_2$.
For each vertex $b_2^i(e) \in B'_2$, let $a_2^i(e) \in A''_2$ be the corresponding endpoint of the path of $\qset_2$ that originates at $b_2^i(e)$ (see Figure \ref{fig: pset 2}).
Let $b_1^i(e) \in B'_1$ be the vertex of $B_1$ that is connected to $a_2^i(e)$ with an edge from $\mset_1$. We can now naturally define a matching $\mset^*$ over the vertices of $B'_1$, where for every edge $e=(u_i,u_j)\in F_2$, we add the pair $(b_1^i(e),b_1^j(e))$ of vertices to the matching.
From Lemma \ref{lem: routable b}, with high probability we obtain a collection $\pset^*_2 = \set{P^*_e \mid e  \in F_2}$ of disjoint paths in $W_1$, such that, for every edge $e = (u_i, u_j) \in F_2$, the  corresponding path $P^*_e$ connects $b_1^i(e)$ to $b_1^j(e)$. We extend this path to connect the vertex $b_2^i(e)$ to the vertex $b_2^j(e)$, by using the edges of $\mset_1$ that are incident to $b_1^i(e)$ to $b_1^j(e)$, and the paths of $\qset_2$ that are incident to $a_2^i(e)$ to $a_2^j(e)$. 
Notice that  the resulting extended paths are internally disjoint from $B_2$, and are completely disjoint from $T_2\cup G''_{\Pi}$. We now embed each edge $e\in F_2$ into the path $P^*_e$, that is, we set $f(e)=P^*_e$. This completes the construction of the model of $H$ in $G_{\Pi}$, except for the proofs of Lemmas~\ref{lem: routing in toe} and \ref{lem: routable b}, that are provided in Sections~\ref{sec:routing in toe} and~\ref{sec: routing in doe}, respectively.

\begin{minipage}{\linewidth}
    \centering
    \begin{minipage}{0.55\linewidth}
        \begin{figure}[H]
            \includegraphics[width=\linewidth]{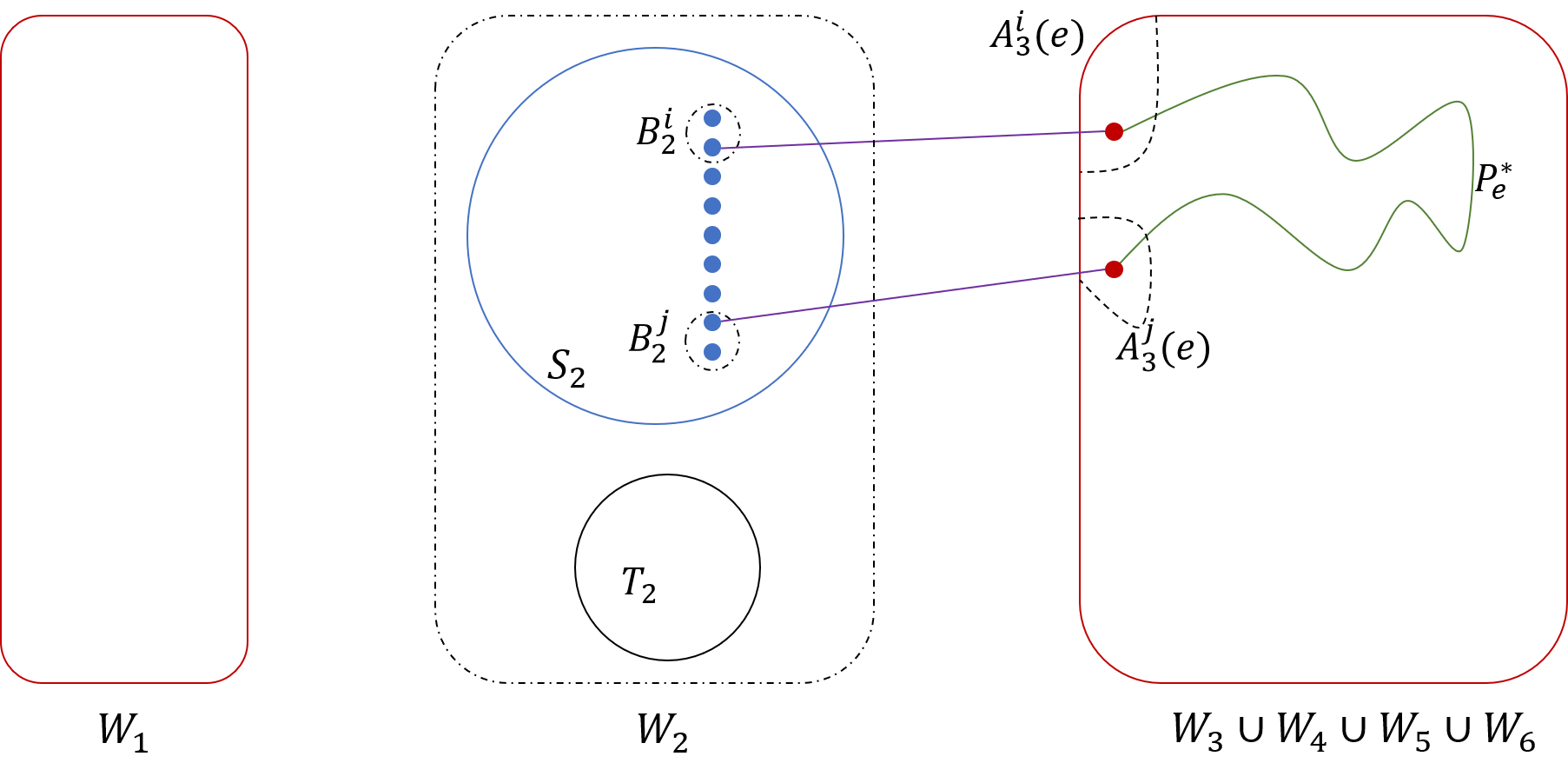}
            \caption{An illustration of a path $P^*_e \in \pset^*_1$ routing an edge $e=(u_i, u_j) \in F_1$. Dashed boundaries represent the labeled subsets.}
            \label{fig: pset 1}
        \end{figure}
    \end{minipage}
    \hspace{0.04\linewidth}
    \begin{minipage}{0.55\linewidth}
        \begin{figure}[H]
            \includegraphics[width=\linewidth]{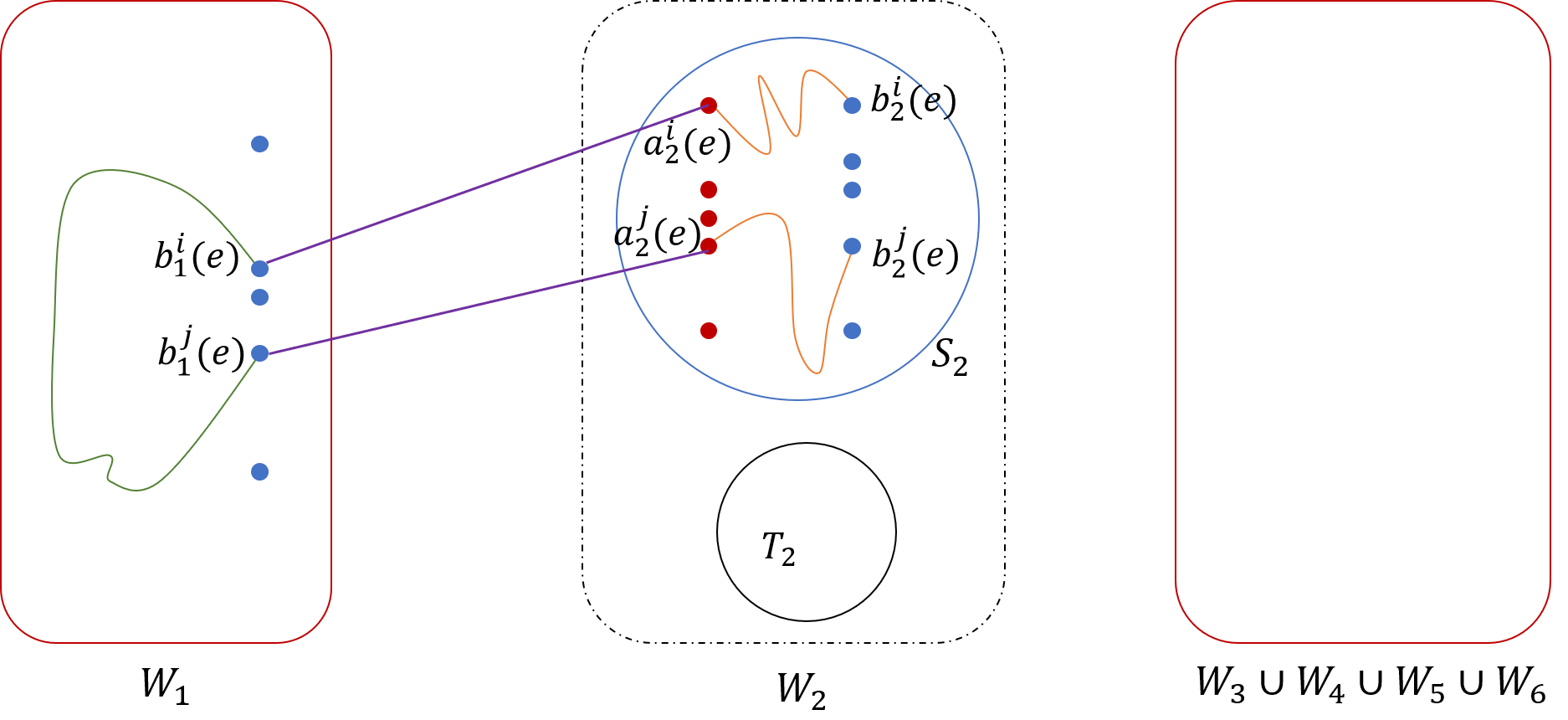}
            \caption{An illustration of the path $P^*_e \in \pset^*_2$ connecting $e=(u_i, u_j) \in F_2$.}
            \label{fig: pset 2}
        \end{figure}
    \end{minipage}
\end{minipage}

%% file: routing_in_toe.tex
\section{Routing in $G''_{\Pi}$} \label{sec:routing in toe}
This section is dedicated to the proof of Lemma \ref{lem: routing in toe}.
We define a new combinatorial object, called a \doefull System.

\begin{figure}[h]
    \center
    \includegraphics[width=0.45\linewidth]{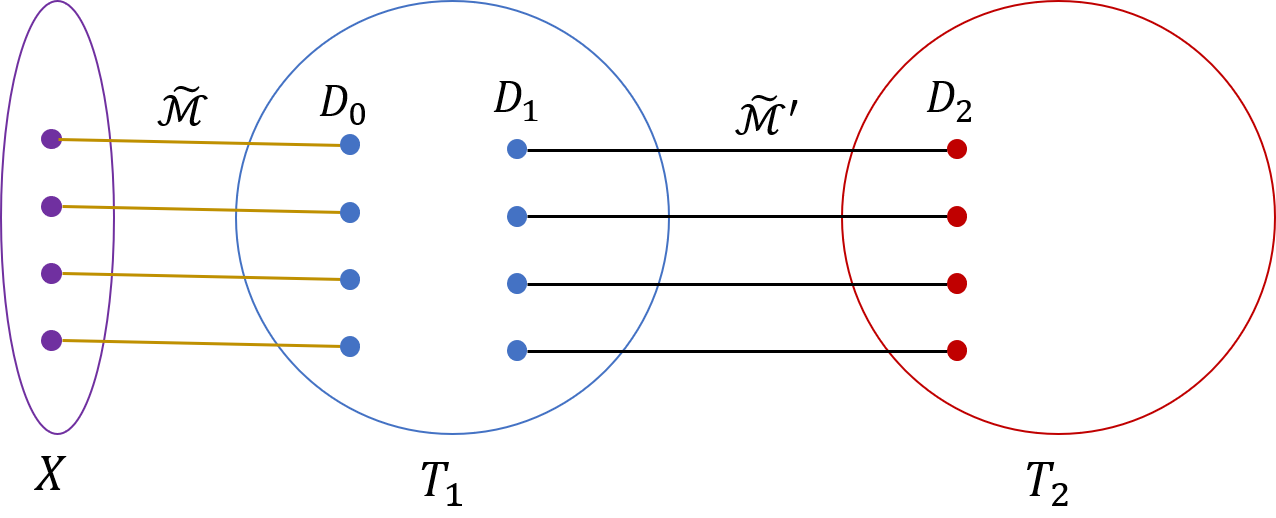}
    \caption{An illustration of the \doefull System.}
    \label{fig: new doe}
\end{figure}

\begin{definition}
    A \doefull System of width $w$, expansion $\alpha$ (see Figure \ref{fig: new doe}) consists of:
    \begin{itemize}
        \item two disjoint graphs $T_1,T_2$, each of which is an $\alpha$-expander;
        \item a set $X$ of $w$ vertices that are disjoint from $T_1\cup T_2$, and three subsets $D_0,D_1\subseteq V(T_1)$ and $D_2\subseteq V(T_2)$ of $w$ vertices each, where all three subsets are disjoint; and
        \item a complete matching $\tmset$ between the vertices of $X$ and the vertices of $D_0$, and a complete matching $\tmset'$ between the vertices of $D_1$ and the vertices of $D_2$, so $|\tmset|=|\tmset'|=w$.
    \end{itemize}

    We denote the \doefull System by $\dset = (T_1, T_2, X, \tmset, \tmset')$. The set $X$ of vertices is called the \emph{backbone} of $\dset$.
Let $G_{\dset}$ be the graph corresponding to the \doefull System $\dset$, so $G_{\dset}$ is the union of graphs $T_1,T_2$, the set $X$ of vertices, and the set $\tmset\cup \tmset'$ of edges.
\end{definition}

Similarly to \poefull System, given a graph $G$, we say that it contains a \doefull System $\dset$ as a minor iff $G_{\dset}$ is a minor of $G$.

The following lemma is central to the proof of Lemma~ \ref{lem: routing in toe}.

\begin{lemma}\label{lem: routing in new doe}
    There is an efficient algorithm that, given a \doefull System $\dset$ of width $w/4$ and expansion $\alpha$, for some $0<\alpha<1$, such that the corresponding graph $G_{\dset}$ contains at most $n$ vertices and has maximum vertex degree at most $d$, together with a collection $\set{X_1, \ldots, X_{2 r}}$ of  mutually disjoint subsets of the backbone $X$ of cardinality $\sigma = 2^{15}\floor{\frac{d^3n\log n}{\alpha^2w}}$ each, where $r > \frac{w \alpha^2 (\log \log n)^2}{d^3 \log^3 n}$, returns a partition $\iset',\iset''$ of $\set{1,\ldots,r}$, and for each $j\in \iset'$, a path $P_j$ connecting a vertex of $X_j$ to a vertex of $X_{j+r}$ in $G_{\dset}$, such that the paths in set $\pset = \set{P_j \mid j \in \iset'}$ are disjoint, and $|\iset''|\leq r \cdot \frac{\log \log n}{\log n}$.
    \end{lemma}

We defer the proof of Lemma~\ref{lem: routing in new doe} to Section~\ref{subsec: routing in new doe}, after we complete the proof of Lemma~ \ref{lem: routing in toe} using it. 
Recall that we are given a \poefull System $\Pi = (\sset, \mset, A_1, B_6,\tset, \mset')$, together with its corresponding graph $G_{\Pi}$. 
Recall  that we are also given a subset $A'_3\subseteq A_3$ of at most $w/2$ vertices, and a partition of $A'_3$ into $2r$ disjoint subsets $A^1_3,\ldots,A^{2r}_3$, of cardinality $\rho=2^{16}\floor{\frac{d^3n\log n}{\alpha^2w}}$ each, where $r\geq \frac{w\alpha^2(\log\log n)^2}{d^3\log^3n}$. For each $1\leq i\leq 2r$, we arbitrarily partition $A_3^i$ into two subsets, $W^i_1,W^i_2$, of cardinality $\rho/2$ each (note that $\rho$ is an even integer). Let $W_1=\bigcup_{i=1}^{2r}W_1^i$ and let $W_2=\bigcup_{i=1}^{2r}W^i_2$. Note that $|W_1|,|W_2|\leq  |A'_3|/2\leq w/4$. We add arbitrary vertices of $A_3\setminus A_3'$ to $W_1$ and $W_2$, until each of them contains $w/4$ vertices (recall that $w/4$ is an integer), while keeping them disjoint. The vertices of $A_3\setminus (W_1\cup W_2)$ are then arbitrarily partitioned into two subsets, $Y_1$ and $Y_2$, of cardinality $w/4$ each.

Next, we show that graph $G''_{\Pi}$ contains two disjoint \doefull Systems as minors. We will then use Lemma~\ref{lem: routing in new doe} in each of the two \doefull Systems in turn in order to obtain the desired routing.

\begin{claim}\label{claim: two does}
There is an efficient algorithm to compute two disjoint subgraphs, $G^{(1)}$ and $G^{(2)}$ of $G''_{\Pi}$, and for each $z\in \set{1,2}$, to compute a model $f^{(z)}$ of a \doefull System $\dset^{(z)}=(T_1^{(z)},T_2^{(z)},X^{(z)},\tmset^{(z)},(\tmset')^{(z)})$ of width $w/4$ and expansion $\alpha$ in $G^{(z)}$,  such that the corresponding graph $G_{\dset^{(z)}}$ has maximum vertex degree at most $d$, and for every vertex $w\in W_z$, there is a distinct vertex $v(w)$ in the backbone $X^{(z)}$, such that $w\in f^{(z)}(v(w))$.
\end{claim}

\begin{proofof}{Claim~\ref{claim: two does}}
From the definition of the \poefull System, for $3\leq j\leq 6$, the set $A_j\cup B_j$ of vertices is well-linked in $S_j$. Therefore, there is a set $\pset_j$ of $w$ node-disjoint paths in $S_j$, connecting $A_j$ to $B_j$. By concatenating the path sets $\pset_3,\pset_4,\pset_5,\pset_6$, and the edge sets $\mset_3,\mset_4,\mset_5$, we obtain a collection $\pset$ of $w$ node-disjoint paths in $G''_{\Pi}$, connecting $A_3$ to $B_6$. We partition $\pset$ into two subsets: set $\pset^{(1)}$ contains all paths originating at the vertices of $W_1\cup Y_1$, and set $\pset^{(2)}$ contains all paths originating at the vertices of $W_2\cup Y_2$. 

We are now ready to define the two graphs $G^{(1)}$ and $G^{(2)}$. Graph $G^{(1)}$ is obtained from the union of the expanders $T_3$ and $T_4$, the paths of $\pset^{(1)}$, and the edges of $\mset'_3\cup \mset'_4$ that have an endpoint lying on the paths of $\pset^{(1)}$. Graph $G^{(2)}$ is defined similarly by using $T_5,T_6$, the paths of $\pset^{(2)}$, and the edges of $\mset'_5\cup\mset'_6$ that have an endpoint lying on the paths of $\pset^{(2)}$. It is immediate to verify that the graphs $G^{(1)}$ and $G^{(2)}$ are disjoint.

It now remains to show that each of the resulting graphs contains a \doefull System as a minor, with the required properties. We show this for $G^{(1)}$; the proof for $G^{(2)}$ is symmetric. Our first step is to contract every path of $\pset^{(1)}$ into a single vertex. For each such path $P\in \pset^{(1)}$, let $w\in W_1\cup Y_1$ be the first vertex of $P$. We denote the new vertex obtained by contracting $P$ by $v(w)$.
We let the backbone $X^{(1)}$ of the new \doefull System $\dset^{(1)}$ be $X^{(1)}=\set{v(w)\mid w\in W_1}$, so $|X^{(1)}|=w/4$. We map every vertex $w\in W_1$ to the corresponding vertex $v(w)$ in the model of $G_{\dset^{(1)}}$ that we are constructing in $G^{(1)}$; that is, we set $f^{(1)}(w)=v(w)$.
We also map the two expanders $T^{(1)}_1,T^{(1)}_2$ of $\dset^{(1)}$ to $T_3$ and $T_4$, respectively, by setting $T^{(1)}_1=T_3$ and $T^{(1)}_2=T_4$.

Consider some vertex $w\in W_1\cup Y_1$ and the path $P\in \pset^{(1)}$ originating from $w$. Let $w'$ be the unique vertex of $P$ that belongs to $B_3$, and let $w''$ be the unique vertex of $P$ that belongs to $B_4$, in the original \poefull System $\Pi$. Recall that there is an edge of $\mset'_3$, connecting $w'$ to some vertex $u_w\in C_3$, and there is an edge of $\mset'_4$, connecting $w''$ to some vertex $u'_{w}\in C_4$. Therefore, there are edges $(v(w),u_w)$ and $(v(w),u'(w))$ in the new contracted graph.

We set $D_0^{(1)}=\set{u_{w}\mid w\in  W_1}$, and we let $\tmset^{(1)}=\set{(v(w),u_w)\mid w\in W_1}$. 
We also set $D_1^{(1)}=\set{u_{y}\mid y\in \hat Y_1}$, and $D_1^{(2)}=\set{u'_{y}\mid y\in \hat Y_1}$. Observe that all three sets $D^{(1)}_0,D^{(1)}_1,D^{(1)}_2$ of vertices are disjoint, and they contain $w/4$ vertices each. It now remains to define the set $(\tmset')^{(1)}$ of edges, that connect vertices of $D_1^{(1)}$ and $D_2^{(1)}$. In order to do so, for every vertex $y\in Y_1$, we merge the two edges $(v(y),u_{y})$ and $(v(y),u'_{y})$ into a single edge, by contracting one of these two edges. The resulting edge is added to $(\tmset')^{(1)}$. It is easy to see that we have obtained a \doefull System $\dset^{(1)}$, whose width is $w/4$ and expansion $\alpha$. It is easy to verify that the maximum vertex degree in the corresponding graph $G_{\dset^{(1)}}$ is bounded by $d$.  Notice that for every vertex $w\in W_1$, there is a distinct vertex $v(w)\in X^{(1)}$, such that $w\in f^{(1)}(v(w))$.
\end{proofof}

We apply Lemma~\ref{lem: routing in new doe} to $\dset^{(1)}$, together with vertex sets $W_1^1,\ldots,W_1^{2r}$, each of which now contains $\rho/2=2^{15}\floor{\frac{d^3n\log n}{\alpha^2w}}$ vertices obtaining a partition $(\iset',\iset'')$ of $\set{1,\ldots,r}$, together with a set $\pset_1= \set{P_j \mid j \in \iset'}$ of disjoint paths in $G_{\dset^{(1)}}$, such that for all $j \in \iset'$ path $P_j $ connects a vertex of $W_1^j$ to a vertex of $W_1^{r+j}$, and $|\iset''|\leq r \cdot \frac{\log \log n}{\log n}$. Since $G_{\dset^{(1)}}$ is a minor of $G^{(1)}$, it is immediate to obtain a collection $ 
\pset_1'= \set{P'_j \mid j \in \iset'}$ of disjoint paths in $G^{(1)}$, such that for all $j \in \iset'$ path $P'_j $ connects a vertex of $A_3^j$ to a vertex of $A_3^{j+r}$.

 If $|\iset''|\leq \frac{w \alpha^2 (\log \log n)^2}{d^3 \log^3 n}$, then we terminate the algorithm, and return the set $\pset'$ of paths, together with the partition $(\iset',\iset'')$ of $\iset$. Next, we denote $|\iset''|=r'$, and we assume that $r'>\frac{w \alpha^2 (\log \log n)^2}{d^3 \log^3 n}$.

 We apply Lemma~\ref{lem: routing in new doe} to $\dset^{(2)}$, together with vertex sets $\set{W^j_2,W^{j+r}_2 \mid j\in \iset''}$, that are appropriately ordered.
 We then obtain a partition $\iset_1,\iset_2$ of $\iset''$,  and a set $\pset_2= \set{P_j \mid j \in \iset_1}$ of disjoint paths in $G_{\dset^{(2)}}$, such that for each $j \in \iset_2$ path $P_j $ connects a vertex of $W^j_2$ to a vertex of $W_2^{j+r}$, and  $|\iset_2|\leq r' \cdot \frac{\log \log n}{\log n}\leq r \cdot \frac{(\log \log n)^2}{\log^2 n}$. As before, since $G_{\dset^{(2)}}$ is a minor of $G^{(2)}$, it is immediate to obtain a collection $ 
\pset_2'= \set{P'_j \mid j \in \iset_1}$ of disjoint paths in $G^{(2)}$, such that for all $j \in \iset_1$ path $P'_j $ connects a vertex of $A_3^j$ to a vertex of $A_3^{j+r}$.

We return the partition $(\iset'\cup \iset_1,\iset_2)$ of $\set{1,\ldots,r}$, together with the set $\pset_1'\cup \pset_2'$ of paths. Since the graphs $G^{(1)}$ and $G^{(2)}$ are disjoint, all paths in $\pset_1'\cup \pset_2'$ are disjoint. It now only remains to show that $|\iset_2|\leq \frac{w \alpha^2 (\log \log n)^2}{d^3 \log^3 n}$.

Recall that the set $A_3'$ of at most $w/2$ vertices is partitioned into $2r$ subsets of cardinality $\rho=2^{16}\floor{\frac{d^3n\log n}{\alpha^2w}}$ each. Therefore:

\[\begin{split}
r&\leq \frac{w}{4\rho}\\
&=\frac{w}{2^{18}\floor{d^3n\log n/(\alpha^2w)}}\\
&\leq \frac{w^2\alpha^2}{2^{17}d^3n\log n}\\
&\leq \frac{w\alpha^2}{2^{17}d^3\log n}.
\end{split}
\]

Therefore, $|\iset_2|\leq r \cdot \frac{(\log \log n)^2}{\log^2 n}\leq \frac{w\alpha^2(\log \log n)^2}{d^3\log^3 n}$, as required.

\subsection{Routing in \doefull | Proof of Lemma \ref{lem: routing in new doe}} \label{subsec: routing in new doe}

The goal of this section is to prove Lemma \ref{lem: routing in new doe}. The proof is inspired by the algorithm of Frieze~\cite{Frieze} for routing a large set of demand pairs in an expander graph via edge-disjoint paths.
Recall that we are given a \doefull System $\dset$ of width $w/4$ and expansion $\alpha$, for some $0<\alpha<1$, such that the maximum vertex degree in the corresponding graph $G_{\dset}=(V,E)$ is at most $d$, and $|V|\leq n$. We are also given   mutually disjoint subsets $\set{X_1, \ldots, X_{2 r}}$ of the backbone $X$, of cardinality $\sigma=2^{15}\floor{\frac{d^3n\log n}{w\alpha^2}}$ each, where $r>\frac{w \alpha^2 (\log \log n)^2}{d^3 \log^3 n}$. In particular, since $|X|=w/4$, we get that $2r\sigma\leq w/4$, and so $r\leq \frac{w}{8\sigma}\leq  \frac{w}{8\cdot 2^{15}\floor{d^3n\log n/(w\alpha^2)}}\leq \frac{w^2\alpha^2}{2^{17}d^3n\log n}$.
 Therefore, we obtain the following bounds on $r$ that we will use throughout the proof:
 
 \begin{equation}\label{eq: bounds on r}
\frac{w \alpha^2}{d^3}\cdot\frac{ (\log \log n)^2}{\log^3 n}<r\leq \frac{w^2\alpha^2}{2^{17}d^3n\log n}.
\end{equation}

For convenience, we will denote $G_{\dset}$ by $G$ for the rest of this subsection.

We will iteratively construct the set $\pset$ of disjoint paths in $G$, where for each path $P\in \pset$, there is some index $j\in \set{1,\ldots,r}$, such that $P$ connects $X_j$ to $X_{j+r}$. Whenever a path $P$ is added to $\pset$, we delete all vertices of $P$ from $G$.
Throughout the algorithm, we say that an index $j \in [r]$ is \emph{settled} iff there is a path $P_j \in \pset$ connecting $X_j$ to $X_{j+r}$, and otherwise we say that it is \emph{not settled}.
We use a parameter $\gamma = 512 n d^2/w\alpha$.
We say that a path $P$ in $G$ is \emph{permissible} iff $P$ contains at most $\gamma \log \log n$ nodes of $T_1$ and at most $\gamma \log n$ nodes of $T_2$.

\begin{framed}
\vspace{-1em}
\paragraph{The Algorithm.}
Start with $\pset = \emptyset$.
While there is an index $j \in [r]$ and a permissible path $P^*_j$ in the current graph $G$ such that:
\begin{itemize}
\item $j$ is not settled;
\item $P^*_j$ connects $X_j$ to $X_{j+r}$; and
\item $P^*_j$ is internally disjoint from $X$:
\end{itemize}
add $P^*_j$ to $\pset$ and delete all vertices of $P^*_j$ from $G$.
\end{framed}

In order to complete the proof of Lemma~\ref{lem: routing in new doe}, it is enough to show that, when the algorithm terminates, at most $\frac{r \log \log n}{\log n}$ indices $j \in [r]$ are not settled. Assume for contradiction that this is not true.
Let $\pset$ be the path set obtained at the end of the algorithm, and let $\tilde V = V(\pset)$ be the set of vertices participating in the paths of $\pset$. We further partition $\tilde V$ into three subsets: $\tilde V_1=\tilde V\cap V(T_1)$; $\tilde V_2=\tilde V\cap V(T_2)$; and $\tilde X=\tilde V\cap X$. Note that, since $|\pset|\leq r$, we are guaranteed that $|\tilde V_1|\leq \gamma  r\log\log n$; $|\tilde V_2|\leq \gamma r\log n$, and, since we have assumed that $|\pset|\leq r(1-\log \log n/\log n)$, and all paths in $\pset$ are internally disjoint from $X$, we get that $|\tilde X|\leq 2r-\frac{2r\log \log n}{\log n}$.

We now proceed as follows. First, we show that $T_1\setminus \tilde V_1$ and $T_2\setminus \tilde V_2$ both contain very large $\alpha/4$-expanders. We also show that there is a large number of edges in $\tmset'$ that connect these two expanders. This will be used to show that there must still be a permissible path $P^*_j$, connecting two sets $X_j$ and $X_{j+r}$ for some index $j$ that is not settled yet, leading to a contradiction. We start with the following claim that allows us to find large expanders in $T_1\setminus \tilde V_1$ and $T_2\setminus \tilde V_2$. 

\begin{claim}\label{claim: large expanding subgraph}
     Let $T$ be an $\alpha$-expander with maximum vertex degree at most $d$, and let $Z$ be any subset of vertices of $T$. Then there is an $\alpha/4$-expander $T'\subseteq T\setminus Z$, with $|V(T')|\geq |V(T)|-\frac{4d|Z|}{\alpha}$.
\end{claim}

The proof of Claim~\ref{claim: large expanding subgraph} follows immediately from Claim~\ref{claim: large expanding subgraph-edges}, by letting $E'$ be the set of all edges incident to the vertices of $Z$.
The following corollary follows immediately from Claim~\ref{claim: large expanding subgraph}

\begin{corollary}\label{clm: surviving step 1}
There is a subgraph $T'_1\subseteq T_1\setminus \tilde V_1$ that is an $\alpha/4$-expander, and $|V(T_1)\setminus V(T'_1)|\leq 4dr\gamma\log\log n/\alpha$. Similarly, there is a subgraph $T'_2\subseteq T_2\setminus \tilde V_2$ that is an $\alpha/4$-expander, and $|V(T_2)\setminus V(T'_2)|\leq 4dr\gamma\log n/\alpha$.
\end{corollary}

Let $R_1=V(T_1)\setminus V(T_1')$ and let $R_2=V(T_2)\setminus V(T_2')$. We refer to the vertices of $R_1$ and $R_2$ as the vertices that were \emph{discarded} from $T_1$ and $T_2$, respectively. The vertices that belong to $T_1'$ and $T_2'$ are called \emph{surviving} vertices. It is easy to verify that $|R_1|,|R_2|\leq w/64$. Indeed, observe that $|R_1|,|R_2|\leq 4dr\gamma\log n/\alpha$. Since, from Equation~(\ref{eq: bounds on r}), $r\leq \frac{w^2\alpha^2}{2^{17}d^3n\log n}$, we get that altogether:

\[|R_1|,|R_2|\leq \frac{4dr\gamma\log n}{\alpha}\leq \frac{\gamma w^2\alpha}{2^{15} d^2 n}\leq \frac{w}{64},\]

since $\gamma=512nd^2/w\alpha$.

Recall that the \doefull $\dset$ contains a matching $\tmset'$ between the set $D_1\subseteq V(T_1)$ of $w/4$ vertices and the set $D_2\subseteq V(T_2)$ of $w/4$ vertices. Next, we show that there are large subsets $D'_1\subseteq D_1$ and $D'_2\subseteq D_2$ of surviving vertices, such that a subset of $\tmset'$ defines a complete matching between them.

\begin{observation}
There are two sets $D'_1\subseteq D_1$ and $D_2'\subseteq D_2$ containing at least $w/16$ vertices each, and a subset $\hmset\subseteq \tmset'$ of edges, such that $\hmset$ is a complete matching between $D_1'$ and $D_2'$.
\end{observation}
\begin{proof}
Let $\hat D_1=D_1\setminus R_1$. Since $|R_1|\leq w/64$, $|\hat D_1|\geq w/8$. Let $\hmset'\subseteq \mset'$ be the set of edges whose endpoints lie in $\hat D_1$, and let $\hat D_2\subseteq D_2$ be the set of vertices that serve as endpoints for the edges in $\hmset'$, so $|\hat D_2|\geq w/8$. Finally, let $D_2'=D_2\setminus R_2$, so $|D_2'|\geq w/8-|R_2|\geq w/16$. We let $\hmset\subseteq \hmset'$ be the set of all edges incident to the vertices of $D_2'$, and we let $D_1'$ be the set of endpoints of these edges.
\end{proof}

Our second main tool is the following claim, that shows that for any pair of large enough sets of vertices in an expander, there is a short path connecting them. The proof uses standard methods and is deferred to Appendix.

\begin{claim}\label{claim: short paths in expanders}
Let $T$ be an $\alpha'$-expander for some $0<\alpha'<1$, such that $|V(T)|\leq n$, and the maximum vertex degree in $T$ is at most $d$. Let $Z,Z'\subseteq V(T)$ be two vertex subsets, with $|Z|=z$ and $|Z'|=z'$. Then there is a path in $T$, connecting a vertex of $Z$ to a vertex of $Z'$, whose length is at most $\frac{8d}{\alpha'}(\log (n/z)+\log(n/z'))$. In particular, for every pair $v,v'$ of vertices in $T$, there is a path of length at most $16d\log n/\alpha'$ connecting $v$ to $v'$ in $T$.
\end{claim}

Let $J\subseteq \set{1,\ldots,r}$ be the set of indices that are not settled yet. From our assumption, $|J|\geq \frac{r\log\log n}{\log n}$. For every index $j\in J$, consider the corresponding sets $X_j,X_{j+r}$ of vertices of $X$, and let $Y_j,Y_{j+r}$ be the sets of vertices of $D_0$, that are connected to $X_j$ and $X_{j+r}$ via the matching $\tmset$. Let $Y_j'=Y_j\setminus R_1$ and let $Y_{j+r}'=Y_{j+r}\setminus R_1$ be the subsets  of surviving vertices in $Y_j$ and $Y_{j+r}$ respectively. We say that index $j$ is \emph{bad} iff $|Y'_j|<\sigma/2$ or $|Y'_{j+r}|<\sigma/2$; otherwise we say that it is a \emph{good index}. Recall that $|R_1|\leq 4dr\gamma\log\log n/\alpha$. Therefore, the total number of bad indices is at most:

\[\begin{split}
\frac{2|R_1|}{\sigma}&\leq \frac{8dr\gamma\log\log n}{\alpha\cdot 2^{15}\floor{d^3n\log n/(w\alpha^2)}}\\
&\leq \frac{w\alpha r\gamma\log\log n}{2^{11}d^2 n\log n}\\
&\leq \frac{r\log\log n}{4\log n}\cdot \frac{w\alpha \gamma}{512 d^2n}\\
&\leq \frac{r\log\log n}{4\log n},\end{split}\]

since $\gamma=512 n d^2/w\alpha$.

Let $J'\subseteq J$ be the set of all good indices, so $|J'|\geq \frac{r\log\log n}{2\log n}$.
We say that an index $j\in J'$ is \emph{happy} iff there is a path $P_1(j)$ in $T'_1$, of length at most $(\gamma\log\log n)/4$, connecting a vertex of $Y'_j$ to a vertex of $D'_1$, and there is a path $P_2(j)$ in $T'_1$,  of length at most $(\gamma\log\log n)/4$, connecting a vertex of $Y'_{j+r}$ to a vertex of $D'_1$. The following claim will finish the proof of Lemma~\ref{lem: routing in toe}.

\begin{claim}\label{claim: happy index}
At least one index of $J'$ is happy.
\end{claim}

Assume first that the claim is correct. Consider the paths $P_1(j)$ and $P_2(j)$ in $T'$, given by Claim~\ref{claim: happy index}, and assume that path $P_1(j)$ connects a vertex $v\in Y'_j$ to a vertex $v'\in D'_1$. Let $v''\in D'_2$ be the vertex connected to $v'$ by an edge of $\hat \mset$, that we denote by $e_v$. Similarly, assume that path $P_2(j)$  connects a vertex $u\in Y'_j$ to a vertex $u'\in D'_1$. Let $u''\in D'_2$ be the vertex connected to $u'$ by an edge of $\hat \mset$, that we denote by $e_u$.  From Claim~\ref{claim: short paths in expanders}, there is a path $P$ in $T_2'$, of length at most $64d\log n/\alpha<\gamma\log n$, connecting $v''$ to $u''$. By combining $P_1(j),e_v,P',e_u,P_2(j)$, together with the edges of $\tmset$ incident to $u$ and $v$, we obtain an admissible path, connecting a vertex of $X_j$ to a vertex of $X_{j+r}$, a contradiction.
It now remains to prove Claim~\ref{claim: happy index}.

\begin{proofof}{\ref{claim: happy index}}
We say that a vertex $v$ of $D_0\cap V(T_1')$ is \emph{happy} iff there is a path in $T_1'$, of length at most $(\gamma\log\log n)/4$,
connecting $v$ to a vertex of $D_1'$. Assume for contradiction that the claim is false. Then for each good index $j$, either all vertices of $Y_{j}'$ are unhappy, or all vertices of $Y_{j+r}'$ are unhappy. Let $Z\subseteq D_0\cap V(T_1')$ be the set of all unhappy vertices. Since $|Y_{j}'|,|Y_{j+1}'|\geq \sigma/2$, and $|J'|\geq  \frac{r\log\log n}{2\log n}$, we get that:

\[\begin{split}
|Z|&\geq \frac{r\log\log n}{2\log n}\cdot \frac{\sigma}{2}\\
&\geq \frac{w\alpha^2(\log\log n)^3}{2d^3\log^4n}\cdot 2^{14}\cdot\floor{\frac{d^3n\log n}{w\alpha^2}}\\
&\geq \frac{2^{12} n (\log\log n)^3}{\log^3n}.
\end{split}
\]

Let $Z'=D_1'$, so $|Z'|\geq w/16$. From Claim~\ref{claim: short paths in expanders}, there is a path in $T_1'$, connecting a vertex of $Z$ to a vertex of $Z'$, of length at most: $\frac{32d}{\alpha}(\log (n/|Z|)+\log(n/|Z'|))\leq \frac{32d}{\alpha} \left (\log\left (\frac{\log^3n}{2^{13}(\log\log n)^3}   \right )+\log(\frac{16n}{w})\right )\leq \frac{32d}{\alpha}(3\log\log n+\log(\frac{16n}{w}))\leq (\gamma\log\log n)/4$, since $\gamma=512 n d^2/(w\alpha)$.
\end{proofof}

%% file: routing_in_doe.tex
\section{Routing in $G'_{\Pi}$}\label{sec: routing in doe}
The goal of this section is to prove Lemma \ref{lem: routable b}.
We use the following lemma, whose proof uses  standard techniques and is deferred to Section \ref{sec: routing along short paths} of Appendix.

\begin{restatable}{lemma}{routeInExpanders}
    \label{lem: can route large disjoint subsets in expanders}
    There is a universal constant $c$, and an efficient randomized algorithm, that,
    given graph $G = (V,E)$ with $|V|\leq n$,
    such that the maximum vertex degree in $G$ is at most $d$ and
    a parameter $0<\alpha<1$,
    together with a collection $\set{C_1, \ldots, C_{2 r}}$ of mutually disjoint subsets of $V$ of cardinality $q =\ceil{ cd^2 \log^2n/\alpha^2}$ each,
    computes one of the following:
    \begin{itemize}
        \item either a collection $\qset = \set{Q_1, \ldots, Q_{r}}$ of paths in $G$,
        where for each $1\leq j\leq r$, path $Q_j$ connects a vertex of $C_{j}$ to a vertex of $C_{r + j}$, and
        with high probability the paths in $\qset$ are disjoint; or
        \item a cut $(S,S')$ in $G$ of sparsity less than $\alpha$.
    \end{itemize}
\end{restatable}

Consider the subgraph $W'$ of $G_{\Pi}$; recall that it consists of two graphs, $S_1$ and $T_1$, where $S_1$ is a connected graph and $T_1$ is an $\alpha$-expander. Recall that $S_1$ contains a set $B_1$ of $w$ vertices; $T_1$ contains a set $C_1$ of $w$ vertices, and $\mset'_1$ is a perfect matching between these two sets.

We let $q=\ceil{cd \log^2n/\alpha^2}$, where $c$ is the constant from Lemma~\ref{lem: can route large disjoint subsets in expanders}, and we let $r=\floor{w/dq}=\Omega(w\alpha^2/d^3\log^2n)$.
Observe that $q\leq \floor{w/dr}$.
We use Observation \ref{obs: decompose by spanning tree} to compute $r$ connected subgraphs $S^1,\ldots,S^{r}$ of $S_1$, each of which contains at least $\floor{w/dr}\geq q$ vertices of $B_1$.
For $1\leq i\leq r$, we denote $B^i=B_1 \cap V(S^i)$.
We also let $\mset^i\subseteq \mset'_1$ be the set of edges incident to the vertices of $B^i$ in $\mset'_1$, and
we let $C^i\subseteq C_1$ be the set of the endpoints of the edges of 
$\mset^i$
that lie in $C_1$.
Observe that for all $1\leq i\leq 2r$, $|C^i|\geq q$.
For each $1\leq i\leq 2r$, we select an arbitrary vertex $b_i\in B^i$, and we let $B'=\set{b_i\mid 1\leq i\leq 2r}$,
so that $|B'|=2r=\Omega(w\alpha^2/d^3\log^2n)$, as required.

Assume now that we are given an arbitrary matching $\mset^*$ over the vertices of $B'$. By appropriately re-indexing the sets $B^i$, we can assume w.l.o.g. that $\mset^*=\set{(b_i, b_{r+i})}_{i=1}^r$.
Since $T_1$ is an $\alpha$-expander, the algorithm of Lemma~\ref{lem: can route large disjoint subsets in expanders} computes a collection $\qset = \set{Q_1, \ldots, Q_{r}}$ of paths in
$T_1$,
where for each $1\leq j\leq r$, path $Q_j$ connects some vertex $c^*_j\in C^j$ to some vertex $c^*_{j+r}\in C^{j+r}$, and with high probability the paths in $\qset$ are disjoint. 

Consider now some index $1\leq j\leq 2r$.
We let $e_j$ be the unique edge of the matching $\mset'_1$ incident to $c^*_j$, and we let $b^*_j\in B^j$ be the other endpoint of this edge.
Since graph $S^j$ is connected, and it contains both $b_j$ and $b^*_j$, we can find a path $P_j$ in $S^j$, connecting $b_j$ to $b^*_j$. 
For each $1\leq j\leq r$, let $P^*_j$ be the path obtained by concatenating $P_j,e_j,Q_j,e_{j+r},P_{j+r}$, and let $\pset^*=\set{P^*_j\mid 1\leq j\leq r}$.
It is immediate to verify that, if the paths in $\qset$ are disjoint from each other, then so are the paths in $\pset^*$, since all graphs in $\set{S^j\mid 1\leq j\leq 2r}$ are disjoint from each other and from $T_1$.
Moreover, for each $1\leq j\leq r$, path $P^*_j$ connects $b_j$ to $b_{j+r}$, as required.

%% file: poe-construction.tex
\section{Constructing a \poefull System}\label{sec: pos to poe}
The goal of this section is to prove Theorem \ref{thm: poe}. The proof consists of three parts. In the first part, we construct an $\alpha'$-expanding Path-of-Sets System of length $24$ in $G$, for some $\alpha'$. In the second part, we transform it into a \posfull System of the same length. In the third and the final part, we turn the \posfull System into a \poefull System.

\input{exp-to-pos}
\input{expanding-to-strong}
\input{pos-to-poe}

%% file: exp-to-pos.tex
\subsection{Part 1: Constructing an Expanding Path-of-Sets System} \label{subsec: exp to pos}

The main technical result of this section is the following theorem.

\begin{theorem}\label{thm: split an expander}
There is a constant $c_x>3$, and a deterministic algorithm, that,  given an $n$-vertex $\alpha$-expander $G$ with maximum vertex degree at most $d$, where $0<\alpha<1$, computes, in time $\poly(n)\cdot \left ( \frac{d}{\alpha}\right)^{O(\log(d/\alpha))}$ a partition $(V',V'')$ of $V(G)$, such that $|V'|,|V''|\geq  \frac{\alpha |V(G)|}{256d}$, and each graph $G[V'],G[V'']$ is an $\alpha^*$-expander, for $\alpha^*\geq \left(\frac{\alpha}{d}\right )^{c_x}$. 
\end{theorem}

The main tool that we use in the proof of the theorem is the following lemma.
\begin{lemma}\label{lem: first cluster}
There is a constant $c'_x$, and deterministic algorithm, that,  given an $n$-vertex $\alpha$-expander $G$ with maximum vertex degree at most $d$, where $0<\alpha<1$, computes, in time $\poly(n)\cdot \left ( \frac{d}{\alpha}\right)^{O(\log(d/\alpha))}$,  a subset $V'\subseteq V(G)$ of vertices, such that $\frac{\alpha |V(G)|}{256d}\leq |V'|\leq \frac{\alpha |V(G)|}{8d}$, and $G[V']$ is an $\hat \alpha^*$-expander, for $\hat \alpha^*\geq \left(\frac{\alpha}{d}\right )^{c'_x}$.
\end{lemma}

\begin{proof}
Given a graph $G$, we say that a partition $(U',U'')$ of $V(G)$ is a \emph{balanced cut} iff $|U'|,|U''|\geq |V(G)|/4$.

Our starting point is the following claim.

\begin{claim} \label{clm: cheeger toy}
	There is an efficient algorithm that, given an $n$-vertex graph $G=(V,E)$, and a parameter $\beta$, returns one of the following:
	\begin{itemize}
		\item either a subset $V'\subseteq V$ of vertices, such that $n/2\leq |V'|\leq 3n/4$ and $G[V']$ is an $\Omega(\frac{\beta^2}{d})$-expander; 
		\item or a partition $(S, T)$ of $V$ with $|E_G(S,T)| < \beta \cdot \min\set{|S|,|T|}$.
	\end{itemize} 
\end{claim}
\begin{proof}
We start with an arbitrary balanced cut $(U',U'')$ in $G$ with $|U'|\geq |U''|$, and perform a number of iterations. In every iteration, we will either establish that $G[U']$
is an $\Omega(\frac{\beta^2}{d})$-expander, or compute the desired partition $(S,T)$ of $V$, or find a new balanced cut $(J',J'')$ in $G$ with $|E(J',J'')|<|E(U',U'')|$. In the first two cases, we terminate the algorithm and return either $V'=U'$ (in the first case), or the cut $(S,T)$ (in the second case). In the last case, we replace $(U',U'')$ with $(J',J'')$, and continue to the next iteration.

We now describe the execution of an iteration. Recall that we are given a balanced cut $(U',U'')$ of $G$ with  $|U'|\geq |U''|$. If $|E(U',U'')|<\beta\cdot \min\set{|U'|,|U''|}$, then we return the cut $(S,T)=(U',U'')$ and terminate the algorithm. Therefore, we assume that $|E(U',U'')|\geq \beta\cdot \min\set{|U'|,|U''|}$. We apply the algorithm from Theorem~\ref{thm: spectral} to graph $G[U']$, and consider the cut $(S,T)$ of $G[U']$ computed by the algorithm. We then consider two cases. First, if $|E(S,T)|\geq \frac{\beta}{4}\min\set{|S|,|T|}$, then from Theorem~\ref{thm: spectral}, we are guaranteed that $G[U']$ is an $\Omega(\frac{\beta^2}{d})$-expander. We terminate the algorithm and return $V'=U'$. 

We assume that $|E(S,T)|<\frac{\beta}{4}\min\set{|S|,|T|}$ from now on, and we assume w.l.o.g. that $|T|\leq |S|$. We consider again two cases. First, if $|E(T,U'')|\leq \frac{\beta}{2} |T|$, we define a new cut $(S',T)$ in $G$, where $S'=S\cup U''$. We then get that $|T|\leq |S'|$, and moreover, $|E_G(S',T)|=|E_G(S,T)|+|E_G(U'',T)|<\beta |T|$. We return the cut $(S',T)$ and terminate the algorithm.

The final case is when $|E(T,U'')|>\frac{\beta}{2}|T|$. In this case, we are guaranteed that $|E(T,U'')|>|E(S,T)|$. Therefore, if we consider the cut $(J',J'')$, where $J'=S$ and $J''=T\cup U''$, then $(J',J'')$ is a balanced cut in $G$, and moreover:

\[|E(J',J'')|=|E(S,U'')|+|E(S,T)|<|E(S,U'')|+|E(T,U'')|=|E(U',U'')|.\]

We then replace $(U',U'')$ with the new cut $(J',J'')$, and continue to the next iteration.
It is easy to verify that every iteration can be executed in time $\poly(n)$. Since the number of the edges in the set $E(U',U'')$ decreases in every iteration, the number of iterations is also bounded by $\poly(n)$.
\end{proof}

By combining Claim~\ref{clm: cheeger toy} with Observation~\ref{obs: simple partition}, we obtain the following simple corollary.

\begin{corollary}\label{cor: expander or balanced}
	There is an efficient algorithm that, given an $n$-vertex graph $G=(V,E)$ with maximum vertex degree at most $d$, and a parameter $\beta$, returns one of the following:
	\begin{itemize}
		\item either a subset $V'\subseteq V$ of vertices, such that $n/4\leq |V'|\leq 3n/4$ and $G[V']$ is an $\Omega(\frac{\beta^2}{d})$-expander; 
		\item or a {\bf balanced} partition $(S, T)$ of $V$ with $|E_G(S,T)| < \beta \cdot \min\set{|S|,|T|}$.
	\end{itemize} 
\end{corollary}

\begin{proof}
Throughout the algorithm, we maintain a set $E'$ of edges of $G$ that we remove from the graph, starting with $E'=\emptyset$, and a collection $\gset$ of disjoint induced subgraphs of $G\setminus E'$, starting with $\gset=\set{G}$. The algorithm continues as long as there is some graph $H\in \gset$, with $|V(H)|>3|V(G)|/4$. In every iteration, we select the unique graph $H\in \gset$ with  $|V(H)|>3|V(G)|/4$, and apply Claim~\ref{clm: cheeger toy} to it, with the parameter $\beta/4$. If the outcome is a subset $V'\subseteq V(H)$ of vertices, such that $|V(H)|/2\leq |V'|\leq 3|V(H)|/4$, and $H[V']$ is an $\Omega(\frac{\beta^2}{d})$-expander, then we return $V'$: it is easy to verify that $n/4\leq |V'|\leq 3n/4$, so $V'$ is a valid output. Otherwise, we obtain a partition $(S',T')$ of $V(H)$ with $|E(S',T')| < \frac{\beta}{4} \cdot \min\set{|S'|,|T'|}$. We add the edges of $E(S',T')$ to $E'$,  remove $H$ from $\gset$, and add $H[S']$ and $H[T']$ to $\gset$ instead. If $|S'|<|T'|$, then our algorithm will never attempt to process the graph $H[S']$ again, so we \emph{charge} the edges of $E(S',T')$ to the vertices of $S'$, where every vertex of $S'$ is charged fewer than $\beta/4$ units. The algorithm terminates when every graph $H\in \gset$ has $|V(H)|\leq 3n/4$ (unless it terminates earlier with an expander). Notice that from our charging scheme, at the end of the algorithm, $|E'|< n\beta/4$. Moreover, using Observation~\ref{obs: simple partition}, we can partition the final collection $\hset$ of graphs into two subsets, $\hset',\hset''$, such that $\sum_{H\in \hset'}|V(H)|,\sum_{H\in \hset''}|V(H)|\geq n/4$. Letting $S=\bigcup_{H\in \hset'}V(H)$ and $T=\bigcup_{H\in \hset''}V(H)$, we obtain a balanced partition $(S,T)$ of $V(G)$. Since $E(S,T)\subseteq E'$, we get that $|E(S,T)|< \frac{\beta n}{4}\leq \beta \cdot \min\set{|S|,|T|}$.
\end{proof}

We now turn to complete the proof of Lemma~\ref{lem: first cluster}.
We denote $|V(G)|=n$, and we let $n^*=\alpha |V(G)|/(8d)$. 
Our goal now is to compute a subset $V'\subseteq V(G)$ of vertices, with $n^*/32\leq |V'|\leq n^*$, such that $G[V']$ is an $\hat \alpha^*$-expander, where $\hat\alpha^*\geq \left(\frac{\alpha}{d}\right )^{c'_x}$ for some constant $c'_x$.
Our algorithm is recursive. Over the course of the algorithm, we will consider smaller and smaller sub-graphs of $G$, containing at least $n^*/4$ vertices each. For each such subgraph $G'\subseteq G$, we define its \emph{level} $L(G')$ as follows. Let $n'=|V(G')|$. If $n'\leq 4n^*/3$, then $L(G')=0$; otherwise, $L(G')=\ceil{\log_{4/3}(n'/n^*)}$. Intuitively, $L(G')$ is the number of recursive levels that we will use for processing $G'$.
Notice that, from the definition of $n^*$, $L(G)\leq O(\log (d/\alpha))$.
 We use the following claim.

\begin{claim}\label{claim: recursion}
There is a deterministic algorithm, that, given a subgraph
 $G'\subseteq G$, such that $|V(G')|\geq n^*/4$, and a parameter $0<\beta<1$, returns one of the following:
\begin{itemize}
\item Either a balanced cut $(S,T)$ in $G'$ with $|E_{G'}(S,T)|<\beta\cdot\min\set{|S|,|T|}$; or
\item A subset $V'\subseteq V(G')$ of vertices of $G'$, such that $n^*/32\leq |V'|\leq n^*$, and $G'[V']$ is an $\hat \beta$-expander, for $\hat \beta\geq \Omega\left(\frac{\beta^2}{d\cdot 2^{10L(G')}}\right )$.
\end{itemize}
The running time of the algorithm is $\poly(n)\cdot \left (\frac{256 d}{\hat \beta}\right)^{L(G')}$.
\end{claim}

We prove the claim below, after we complete the proof of Lemma~\ref{lem: first cluster} using it.
We apply Claim~\ref{claim: recursion} to the input graph $G$ and the parameter $\alpha$. Since $G$ is an $\alpha$-expander, we cannot obtain a cut $(S,T)$ in $G$ with $|E(S,T)|<\alpha\min\set{|S|,|T|}$. Therefore, the outcome of the algorithm is a subset $V'\subseteq V$ of vertices of $G$, with $n^*/32\leq V'\leq n^*$, such that $G[V']$ is a $\hat\alpha$-expander, for $\hat \alpha=\Omega\left(\frac{\alpha^2}{d\cdot 2^{10 L(G)}}\right )$, in time $\poly(n)\cdot \left (\frac{256 d}{\hat \alpha}\right)^{L(G)}$. Recall that $L(G)\leq O(\log (d/\alpha))$. Therefore, we get that $\hat \alpha=\Omega\left(\frac{\alpha^2}{d\cdot 2^{O(\log(d/\alpha))}}\right )\geq (\alpha/d)^{c'_x}$ for some constant $c'_x$, and the running time of the algorithm is $\poly(n)\cdot \left ( \frac{d}{\alpha}\right)^{O(\log(d/\alpha))}$. It now remains to prove Claim~\ref{claim: recursion}.

\begin{proofof}{Claim~\ref{claim: recursion}}
We denote $|V(G')|=n'$.
We let $c$ be a large enough constant. We prove by induction on $L(G')$ that the claim is true, with the running time of the algorithm bounded by $n^c\cdot  \left (256d/\beta \right)^{L(G')}$. The base of the recursion is when $L(G')=0$, and so $n^*/4\leq n'\leq 4n^*/3$. We apply Corollary~\ref{cor: expander or balanced} to graph $G'$ with the parameter $\beta$. If the outcome of the corollary is a subset $V'\subseteq V(G')$ of vertices with $n'/4\leq |V'|\leq 3n'/4$, such that $G'[V']$ is an $\Omega(\beta^2/d)$-expander, then we terminate the algorithm and return $V'$. Notice that in this case, we are
 guaranteed that $n^*/16\leq |V'|\leq n^*$. Otherwise, the algorithm returns a balanced cut $(S,T)$ in $G'$, with $|E_{G'}(S,T)|<\beta\cdot\min\set{|S|,|T|}$. We then return this cut. The running time of the algorithm is $\poly(n)$.
 
 We now assume that the theorem holds for all graphs $G'$ with $L(G')<i$, for some integer $i>0$, and prove it for a given graph $G'$ with $L(G')=i$. Let $n'=|V(G')|$. The proof is somewhat similar to the proof of Corollary~\ref{cor: expander or balanced}. Throughout the algorithm, we maintain a balanced cut $(U',U'')$ of $G'$, with $|U'|\geq |U''|$. Initially, we start with an arbitrary such balanced cut. Notice that $|E(U',U'')|\leq |E(G')|\leq n'd$. While $|E(U',U'')|\geq \beta n'/4$, we perform iterations (that we call phases for convenience, since each of them consists of a number of iterations). At the end of every phase, we either compute a subset $V'\subseteq V(G')$ of vertices of $G'$, such that $n^*/32 \leq |V'|\leq n^*$, and $G'[V']$ is an $\hat \beta$-expander, in which case we terminate the algorithm and return $V'$; or we compute a new balanced cut $(J',J'')$ in $G'$, such that $|E(J',J'')|\leq |E(U',U'')|-\frac{\beta n'}{32}$.  If $|E(J',J'')|<\beta n'/4$, then we return this cut; it is easy to verify that $|E(J',J'')|<\beta\cdot\min\set{|J'|,|J''|}$. Otherwise, we replace $(U',U'')$ with the new cut $(J',J'')$, and continue to the next iteration. Since initially $|E(U',U'')|\leq n'd$, and since $|E(U',U'')|$ decreases by at least $\frac{\beta n'}{32}$ in every phase, the number of phases is bounded by $\frac{32d}{\beta}$. We now proceed to describe a single phase.

 \paragraph{An execution of a phase.}
 We assume that we are given a balanced cut $(U',U'')$ in $G'$, with $|U'|\geq |U''|$, and $|E(U',U'')|\geq \beta n'/4$. Our goal is to either compute a subset $V'$ of vertices of $G'$ such that $n^*/32\leq |V'|\leq n^*$ and $G'[V']$ is an $\hat \beta$-expander, or return another balanced cut $(J',J'')$ in $G'$, with $|E(J',J'')|\leq |E(U',U'')|-\frac{\beta n'}{32}$. Let $\beta'=\beta/32$. Over the course of the algorithm, we will maintain a set $E'$ of edges that we remove from the graph, starting with $E'=\emptyset$, and a collection $\gset$ of subgraphs of $G[U']$ (that will contain at most $4$ such subgraphs). As each graph $H\in \gset$ is a subgraph of $G[U']$, we are guaranteed that $|V(H)|\leq 3n'/4$, and so $L(H)\leq L(G')-1$. We start with $\hset$ containing a single graph, the graph $G'[U']$. We then iterate, while there is a graph $H\in \hset$ with $|V(H)|>|U'|/2$.

 In every iteration, we let $H\in \hset$ be the unique graph with $|V(H)|> |U'|/2$. Notice that $|V(H)|\geq n'/4 \geq n^*/3$, since we have assumed that $L(G')>0$ and so $n'\geq 4n^*/3$.
We apply the algorithm from the induction hypothesis to $H$, with the parameter $\beta'=\beta/32$. If the outcome is a subset $V'\subseteq V(H)$ of vertices of $G'$, such that  $n^*/32 \leq |V'|\leq n^*$ and $H[V']$ is a $\hat \beta'$-expander, for $\hat \beta'\geq \Omega\left(\frac{(\beta')^2}{d\cdot 2^{10L(H)}}\right )$ then we terminate the algorithm and return $V'$. Notice that, since $L(H)\leq L(G)-1$, and $\beta'=\beta/32$, we get that $\frac{(\beta')^2}{d\cdot 2^{10L(H)}}\geq \frac{\beta^2}{d\cdot 2^{10L(G')}}$, so $G'[V']$ is a $\hat \beta$-expander. Otherwise, the algorithm returns a balanced cut $(S,T)$ of $V(H)$, such that $|E(S,T)|<\beta'\cdot\min\set{|S|,|T|}$. We add the edges of $E(S,T)$ to $E'$, remove $H$ from $\hset$, and add $H[S]$ and $H[T]$ to $\hset$. The algorithm terminates once for every graph $H\in \hset$, $|V(H)|\leq |U'|/2$. Let $r=|\hset|$ at the end of the algorithm. Since the cuts $(S,T)$ that we compute in every iteration are balanced, it is easy to verify that we run the algorithm from the induction hypothesis at most $3$ times, and that $r\leq 4$, since in every iteration the size of the largest graph in $\hset$ decreases by at least factor $3/4$, and $(3/4)^3<1/2$. Denote $\hset=\set{H_1,\ldots,H_r}$, and for each $1\leq j\leq r$, let $V_j=V(H_j)$, and let $m_j=|E(V_j,U'')|$. Since $|E(U',U'')|\geq \beta n'/4$, there is some index $1\leq j\leq r$, such that $|E(V_j,U'')|\geq \beta n'/16$. We define a new balanced cut $(J',J'')$, by setting $J'=U'\setminus V_j$ and $J''=U''\cup V_j$. Since $|V_j|\leq |U'|/2$, it is immediate to verify that it is a balanced cut. Moreover, it is immediate to verify that $|E'|\leq \beta'|U'|\leq 3\beta'n'/4\leq \beta n'/32$, and so:

\[|E(J',J'')|\leq |E(U',U'')|-|E(V_j,U'')|+|E'|\leq |E(U',U'')|-\frac{\beta n'}{16}+\frac{\beta n'}{32}\leq |E(U',U'')|-\frac{\beta n'}{32}.\]

Finally, we bound the running time of the algorithm. The running time is at most $\poly(n)$ plus the time required for the recursive calls to the same procedure.
Recall that the number of phases in the algorithm is at most $32d/\beta$, and every phase requires up to $3$ recursive calls. Therefore, the total number of recursive calls is bounded by $100d/\beta$. Each recursive call is to a graph $H$ that has $L(H)<L(G)$. From the induction hypothesis, the running time of each recursive call is bounded by $n^c\cdot  \left (256d/\hat \beta' \right)^{L(G)-1}\leq n^c\cdot  \left (256d/\hat \beta \right)^{L(G)-1}$, and so the total running time of the algorithm is bounded by:

\[n^c+\frac{100 d}{\beta}\cdot n^c\cdot  \left (\frac{256d}{\hat \beta} \right)^{L(G)-1}\leq n^c\cdot  \left (\frac{256d}{\hat \beta }\right)^{L(G)},\]

since $\beta>\hat \beta$.
\end{proofof}
\end{proof}

We are now ready to complete the proof of Theorem~\ref{thm: split an expander}.

\begin{proofof}{Theorem~\ref{thm: split an expander}}
We start with the input $n$-vertex $\alpha$-expander $G$ and apply Lemma~\ref{lem: first cluster} to it, obtaining a subset $V_1\subseteq V(G)$ of vertices, such that $G[V_1]$ is a $\hat \alpha^*$-expander and $\frac{\alpha n}{256d}\leq |V_1|\leq \frac{\alpha n}{8d}$. Let $E'=\delta_G(V_1)$. Since the maximum vertex degree in $G$ is at most $d$, $|E'|\leq \frac{\alpha n}{8}$. 

We use the following claim, which is similar to Claim~\ref{claim: large expanding subgraph-edges}, except that it provides an efficient algorithm instead of the existential result of Claim~\ref{claim: large expanding subgraph-edges}, at the expense of obtaining  somewhat weaker parameters. The proof appears in the Appendix.

\begin{claim} \label{claim: large expanding subgraph cheeger}
There is an efficient algorithm, that given an $\alpha$-expander $G = (V,E)$ with maximum vertex degree at most $d$ and a subset $E' \subseteq E$ of its edges, computes a subgraph $H\subseteq G\setminus E'$ that is an $\Omega\left(\frac{\alpha^2}{d}\right)$-expander, and $|V(H)|\geq |V| - \frac{4|E'|}{\alpha}$.
\end{claim}

We apply Claim~\ref{claim: large expanding subgraph cheeger} to graph $G$ and the set $E'$ of edges computed above. Let $H\subseteq G\setminus E'$ be the resulting graph, and let $V_2=V(H)$. From Claim~\ref{claim: large expanding subgraph cheeger},  $|V_2|\geq n-\frac{4|E'|}{\alpha}\geq n/2$. Since $|V_1|<n/2$ and the set $E'$ of edges disconnects the vertices of $V_1$ from the rest of the graph, while $H$ is an $\Omega\left(\frac{\alpha^2}{d}\right)$-expander and therefore a connected graph, $V_1\cap V_2=\emptyset$. 

We are now ready to define the final partition $(V',V'')$ of $V(G)$, by letting it be the minimum cut separating the vertices of $V_1$ from the vertices of $V_2$ in $G$: that is, we require that $V_1\subseteq V'$, $V_2\subseteq V''$, and among all such partitions $(V',V'')$ of $V(G)$, we select the one minimizing $|E(V',V'')|$. The partition $(V',V'')$ can be computed efficiently using standard techniques: we construct a new graph $\hat G$ by starting with $G$, contracting all vertices of $V_1$ into a source $s$, contracting all vertices of $V_2$ into a destination $t$, and computing a minimum $s$-$t$ cut in the resulting graph. The resulting cut naturally defines the partition $(V',V'')$ of $V(G)$. Let $E''=E(V',V'')$, and denote $|E''|=z$. From Menger's theorem, there is a set $\pset$ of $z$ edge-disjoint paths in $G$, connecting $V_1$ to $V_2$. Therefore, there is a set $\pset_1$ of $z$ edge-disjoint paths in $G[V']\cup E''$, where each path in $\pset_1$ connects a distinct edge of $E''$ to a vertex of $V_1$, and similarly, there is a set $\pset_2$ of $z$ edge-disjoint paths in $G[V'']\cup E''$, where each path in $\pset_2$ connects a distinct edge of $E''$ to a vertex of $V_2$.

We claim that each of the graphs $G[V'],G[V'']$ is an $\alpha^*$-expander, for $\alpha^*=\frac{\alpha\hat \alpha^*}{512d}$. We prove this for $G[V']$; the proof for $G[V'']$ is similar. Assume for contradiction that $G[V']$ is not an $\alpha^*$-expander. Then there is a cut $(X,Y)$ in $G[V']$, such that $|E(X,Y)|<\alpha^*\cdot \min\set{|X|,|Y|}$.
Assume w.l.o.g. that $|X\cap V_1|\leq |Y\cap V_1|$. We now consider two cases.

The first case happens when $|X\cap V_1|\geq \frac{\alpha |X|}{512d}$. In that case, since $G[V_1]$ is an $\hat \alpha^*$-expander, there are at least $\hat \alpha^*\cdot |X\cap V_1|\geq \frac{\hat \alpha^*\cdot \alpha |X|}{512d}\geq \alpha^*|X|$ edges connecting $X\cap V_1$ to $Y\cap V_1$, and so $|E(X,Y)|>\alpha^*\cdot \min\set{|X|,|Y|}$, a contradiction. Therefore, we assume now that $|X\cap V_1|<\frac{\alpha |X|}{512d}$.

\begin{figure}[h]
        \center
        \scalebox{0.30}{\includegraphics{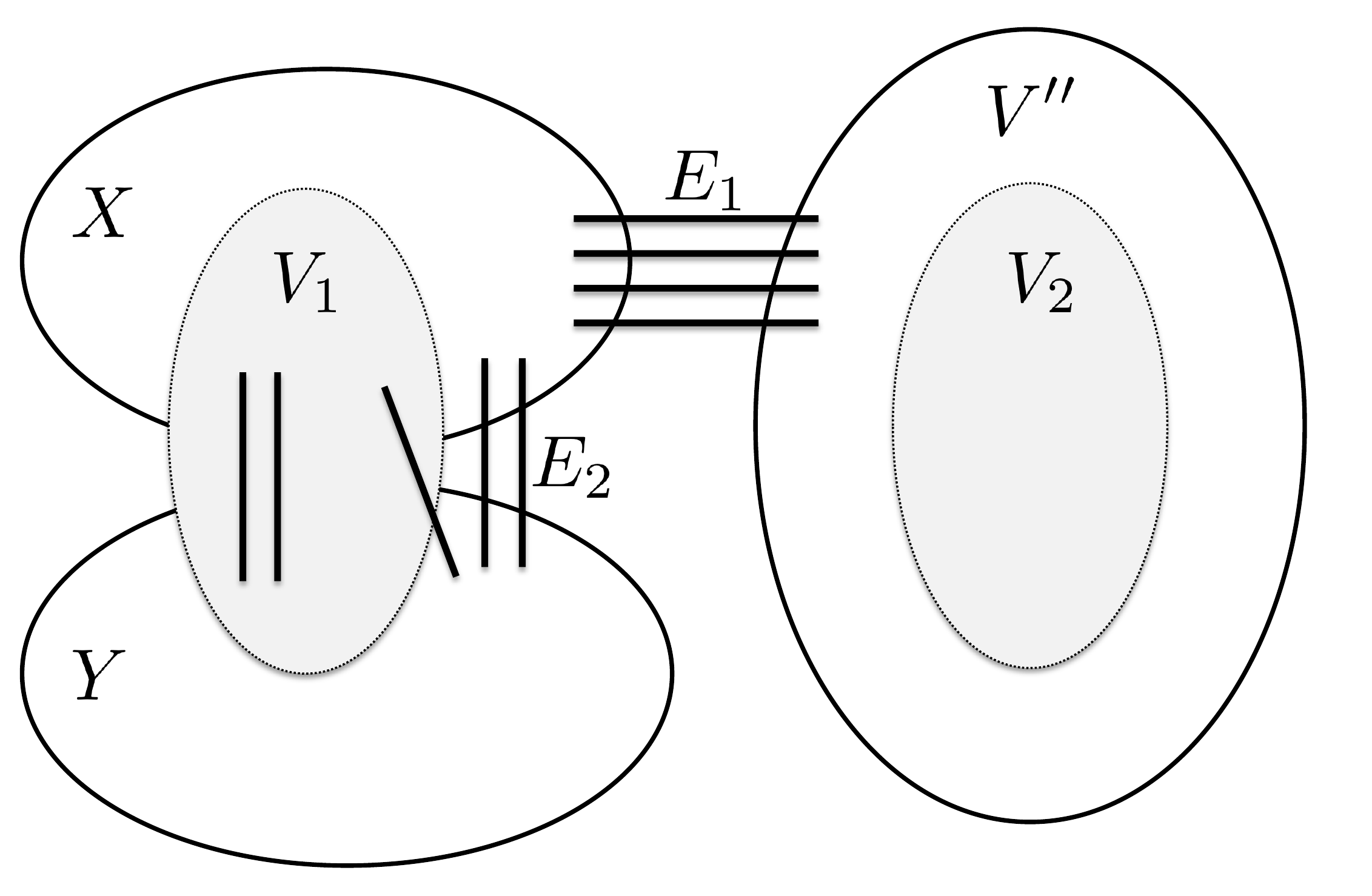}}
        \caption{An illustration for the proof of Theorem~\ref{thm: split an expander}}
        \label{fig: expansion proof}
    \end{figure}

 We partition the edges of $\delta_G(X)$ into two subsets: set $E_1$ contains all edges that lie in $E(V',V'')$, and set $E_2$ contains all remaining edges, so $E_2=E(X,Y)$ (see Figure~\ref{fig: expansion proof}). Note that from the definition of the cut $(X,Y)$, $|E_2|<\alpha^*|X|$.
 Recall that for every edge $e\in E(V',V'')$, there is a path $P_e\in \pset_1$ contained in $G[V']\cup E(V',V'')$, connecting $e$ to a vertex of $V_1$, such that all paths in $\pset_1$ are edge-disjoint. Let $\tilde \pset\subseteq \pset_1$ be the set of paths originating at the edges of $E_1$. We further partition $\tilde \pset$ into two subsets: set $\tilde \pset'$ contains all paths $P_e$ that contain an edge of $E_2$, and $\tilde \pset''$ contains all remaining paths. Notice that $|\tilde \pset'|\leq |E_2|< \alpha^*|X|$. On the other hand, every path $P_e\in \tilde \pset''$ is contained in $G[X]\cup E_1$, and contains a vertex of $V_1\cap X$ -- the endpoint of $P_e$. Since we have assumed that $|V_1\cap X|<\frac{\alpha |X|}{512d}$, and since the maximum vertex degree in $G$ is at most $d$, while the paths in $\tilde \pset''$ are edge-disjoint, we get that $|\tilde \pset''|<\frac{\alpha |X|}{512}$. Altogether, we get that $|E_1|=|\tilde \pset|\leq \alpha^*|X|+\frac{\alpha |X|}{512}$, and $|\delta_G(X)|=|E_1|+|E_2|\leq 2\alpha^*|X|+\frac{\alpha |X|}{512}\leq \frac{\alpha |X|}{256}<\alpha\cdot \min\set{|X|,n/256}\leq \alpha\cdot\min\set{|X|,|V(G)\setminus X|}$, since $|V(G)\setminus X|\geq n/256$, as $V_2\cap X=\emptyset$. This contradicts the fact that $G$ is an $\alpha$-expander.
\end{proofof}

\begin{corollary} \label{corr: exp to pos}
    There is an algorithm, that, given, an $n$-vertex $\alpha$-expander $G$ with maximum vertex degree at most $d$ and an integer $\ell\geq 1$, where $0<\alpha<1/3$, 
    computes an $\alpha_{\ell}$-expanding Path-of-Sets system $\Sigma$ of length $\ell$ and width $w_{\ell}=\ceil{ \alpha_{\ell}n}$, together with a subgraph $G_{\Sigma}$ of $G$, where $\alpha_{\ell}=\alpha^{{c_x}^{\ell-1}}/d^{c_x^{2\ell-2}}$, and $c_x\geq 3$ is the constant from Theorem~\ref{thm: split an expander}. The running time of the algorithm is $\poly(n)\cdot\left(\frac{d}{\alpha_{\ell}}\right )^{O(\log(d/\alpha_{\ell}))}$.
\end{corollary}

We note that we will use the corollary for with $\ell=48$, and so the resulting Path-of-Sets System will have expansion $(\alpha/d)^{O(1)}$, and the running time of the algorithm from Corollary~\ref{corr: exp to pos} is $\poly(n)\cdot \left(\frac{d}{\alpha}\right)^{O(\log(d/\alpha))}$.

\begin{proof}
The proof is by induction on $\ell$. The base case is when $\ell=1$. We choose two arbitrary disjoint subsets $A_1,B_1$ of $\ceil{w_1}<n/2$ of vertices, and we let $S_1=G$. This defines an $\alpha$-expanding Path-of-Sets System of length $1$ and width $w_1$.

We now assume that we are given an integer $\ell>1$, and an $\alpha_{\ell-1}$-expanding Path-of-Sets System $\Sigma=(\sset,\mset,A_1,B_{\ell-1})$ of length $\ell-1$ and width $w_{\ell-1}$, where $G_{\Sigma}\subseteq G$. We assume that $\sset=(S_1,\ldots,S_{\ell-1})$.  We compute an $\alpha_{\ell}$-expanding Path-of-Sets System $\Sigma'=(\sset',\mset',A'_1,B'_{\ell})$ of length $\ell$ and width $w_{\ell}$. 
We will denote $\sset'=(S_1',\ldots,S_{\ell}')$, and for each $1\leq i\leq \ell'$, the corresponding vertex sets $A_i$ and $B_i$ in $S'_i$ are denoted by $A'_i$ and $B'_i$, respectively.

For all $1\leq i<\ell-1$, we set $S'_i=S_i$. We also let $A'_1\subseteq A_1$ be any subset of $w_{\ell}$ vertices, and for $1\leq i<\ell-2$, we let $\mset'_i\subseteq \mset_i$ be any subset of $w_{\ell}$ edges; the endpoints of these edges lying in $B_i$ and $A_{i+1}$ are denoted by $B'_i$ and $A'_{i+1}$ respectively. It remains to define $S'_{\ell-1},S'_{\ell}$, the matchings $\mset'_{\ell-2}$ and $\mset'_{\ell-1}$ (that implicitly define the sets $B_{\ell-2}',A_{\ell-1}',B_{\ell-1}',A_{\ell}'$ of vertices), and the set $B'_{\ell}$ of vertices.

We apply Theorem~\ref{thm: split an expander} to graph $S_{\ell-1}$, and compute, in time $\poly(n)\cdot \left ( \frac{d}{\alpha_{\ell-1}}\right)^{O(\log(d/\alpha_{\ell-1}))}$ a partition $(V',V'')$ of $V(S_{\ell-1})$, such that $|V'|,|V''|\geq  \frac{\alpha_{\ell-1} |V(S_{\ell-1})|}{256d}$, and each graph $G[V'],G[V'']$ is an $\alpha^*$-expander, for $\alpha^*\geq \left(\frac{\alpha_{\ell-1}}{d}\right )^{c_x}$. 

One of the two subsets, say $V'$, must contain at least half of the vertices of $A_{\ell-1}$. We set $S'_{\ell-1}=S_{\ell-1}[V']$ and $S'_{\ell}=S_{\ell-1}[V'']$. Recall that:
$|V'|,|V''|\geq \frac{\alpha_{\ell-1} |V(S_{\ell-1})|}{256d}\geq   \frac{\alpha_{\ell-1} w_{\ell-1}}{128d}$. Since graph $S_{\ell-1}$ is an $\alpha_{\ell-1}$-expander, there are at least $\frac{\alpha_{\ell-1}^2 w_{\ell-1}}{128d}$ edges connecting $V'$ to $V''$. Since maximum vertex degree in $G$ is at most $d$, there is a matching $\mset$, between vertices of $V'$ and vertices of $V''$, with $|\mset|\geq \frac{\alpha_{\ell-1}^2 w_{\ell-1}}{128d^2}$. We claim that $|\mset|\geq w_{\ell}$. In order to see this, it is enough to prove that $w_{\ell}\leq \frac{\alpha_{\ell-1}^2 w_{\ell-1}}{128d^2}$. Since $w_{\ell}=\ceil{\alpha_{\ell}n}$, this is equivalent to proving that:

\[\alpha_{\ell}\leq \frac{\alpha_{\ell-1}^3}{256 d^2}.\] 

This is easy to verify from the definition of $\alpha_{\ell}$ and the fact that $c_x\geq 3$.
We let $\mset'_{\ell-1}$ be any subset of $\mset$ containing $w_{\ell}$ edges. The endpoints of the edges of $\mset'_{\ell-1}$ lying in $V'$ and $V''$ are denoted by $B'_{\ell-1}$ and $A'_{\ell}$ respectively. We let $B_{\ell}'$ be any subset of $w_{\ell}$ vertices of $V''\setminus A_{\ell}'$. Finally, we let $A_{\ell-1}'$ any subset of $w_{\ell}$ vertices of $(V'\cap A_{\ell-1})\setminus B'_{\ell-1}$; $\mset'_{\ell-2}\subseteq \mset_{\ell-2}$ the subset of edges whose endpoints lie in $A'_{\ell-1}$; and $B'_{\ell-2}$ the set of endpoints of the edges of $\mset'_{\ell-2}$ lying in $B_{\ell-2}$. This completes the construction of the Path-of-Sets System $\Sigma'$. It is immediate to verify that it has length $\ell$, width $w_{\ell}$, and that $G_{\Sigma'}\subseteq G$. It remains to prove that it is $\alpha_{\ell}$-expanding, or equivalently, that $S'_{\ell-1}$ and $S'_{\ell}$ are $\alpha_{\ell}$-expanders. Recall that Theorem~\ref{thm: split an expander} guarantees that both these graphs are $\alpha^*$-expanders, where $\alpha^*\geq \left(\frac{\alpha_{\ell-1}}{d}\right )^{c_x}$. It is now enough to verify that $\alpha^*\geq \alpha_{\ell}$, which is immediate to do from the definition of $\alpha_{\ell}$:

\[
\alpha^*\geq \left(\frac{\alpha_{\ell-1}}{d}\right )^{c_x}=\frac{\left(\alpha^{{c_x}^{\ell-2}}/d^{c_x^{2\ell-4}}\right)^{c_x}}{d^{c_x}}
=\frac{\alpha^{{c_x}^{\ell-1}}}{d^{c_x^{2\ell-3}}\cdot d^{c_x}}\geq \frac{\alpha^{{c_x}^{\ell-1}}}{d^{c_x^{2\ell-2}}}=\alpha_{\ell}
\]

Lastly, the running time of the algorithm is dominated by partitioning $S_{\ell-1}$, and is bounded by $\poly(n)\cdot \left ( \frac{d}{\alpha_{\ell-1}}\right)^{O(\log(d/\alpha_{\ell-1}))}\leq \poly(n)\cdot \left ( \frac{d}{\alpha_{\ell}}\right)^{O(\log(d/\alpha_{\ell}))}$, as required.
\end{proof}

We apply Corollary~\ref{corr: exp to pos} to the input graph $G$, with the parameter $\ell=48$, obtaining a sub-graph $G_{\Sigma}\subseteq G$, and an $\alpha'$-expanding Path-of-Sets System $\Sigma$ of length $48$ and width $w'=\ceil{\alpha' n}$, where $\alpha'=(\alpha/d)^{O(1)}$. The running time of the algorithm is $\poly(n)\cdot \left(\frac{d}{\alpha}\right)^{O(\log(d/\alpha))}$.

%% file: expanding-to-strong.tex
\subsection{Part 2: From Expanding to Strong Path-of-Sets System} \label{subsec: exp to strong}

The goal of this subsection is to prove the following theorem:

\begin{theorem}\label{thm: exp pos to strong pos}
 There is an efficient algorithm, that, given a parameter $\ell>0$, and an $\alpha$-\posexp System $\Sigma$ of
    width $w$ and length $4 \ell$,
    where $0<\alpha <1$,
    such that the corresponding graph $G_{\Sigma}$ has
     maximum vertex-degree at most $d$,
    computes a \posfull System $\Sigma'$, of
    width $w' =\Omega(\alpha^3w/d^4)$ and length $\ell$, such that the maximum vertex degree in the corresponding graph $G_{\Sigma'}$ is at most $d$, and $G_{\Sigma'}$ is a minor of $G_{\Sigma}$. Moreover, the algorithm computes a model of $G_{\Sigma'}$ in $G_{\Sigma}$.
\end{theorem}

We use the following simple claim, whose proof is deferred to the Appendix.

 \begin{claim} \label{claim: flow in expander}
There is an efficient algorithm, that, given an $\alpha$-expander $G$, whose maximum vertex degree is at most $d$, where $0<\alpha<1$, together with two disjoint subsets $A,B$ of its vertices of cardinality $z$ each, computes a collection $\pset$ of $\ceil{\alpha z/d}$ disjoint paths, connecting vertices of $A$ to vertices of $B$ in $G$.
\end{claim}

We will also use the following theorem, whose proof is similar to some arguments that appeared in~\cite{CC_gmt}, and is deferred to the Appendix.

\begin{theorem} \label{thm: exp to wl}
        There is an efficient algorithm, that, given an $\alpha$-\posexp System $\Sigma = (\sset, \mset, A_1, B_3)$ of width $w$ and length $3$, where $0<\alpha <1$, and the corresponding graph $G_{\Sigma}$ has maximum vertex degree at most $d$, computes subsets $\hat A_1\subseteq A_1,\hat B_3\subseteq B_3$ of $\Omega(\alpha^2 w/d^3)$ vertices each, such that $\hat A_1\cup \hat B_3$ is well-linked in $G_{\Sigma}$.
\end{theorem}

We are now ready to complete the proof of Theorem \ref{thm: exp pos to strong pos}.

\begin{proofof}{Theorem \ref{thm: exp pos to strong pos}}
We construct a \posfull System $\Sigma'=(\sset',\mset',A_1',B_{\ell}')$ of length $\ell$ and width $w'$, denoting $\sset'=(S_1',\ldots,S_{\ell}')$. For all $1\leq i\leq\ell$, the corresponding vertex sets $A_i$ and $B_i$ are denoted by $A'_i$ and $B'_i$, respectively. 

For all $1\leq i\leq \ell$, we let $\Sigma_i$ be the $\alpha$-expanding Path-of-Sets System of width $w$ and length $3$ obtained by using the clusters $S_{4i-3},S_{4i-2}$, $S_{4i-1}$, and the matchings $\mset_{4i-3}$ and $\mset_{4i-2}$. In order to define the new Path-of-Sets System, for each $1\leq i\leq \ell$, we set $S'_i=G_{\Sigma_i}$. We apply Theorem~\ref{thm: exp to wl} to $\Sigma_i$, to obtain subsets $\hat A_i\subseteq A_{4i-3}$, $\hat B_i\subseteq B_{4i-1}$ of $\Omega(\alpha^2w/d^3)$ vertices each, such that $\hat A_i\cup \hat B_i$ are well-linked in $S'_i$.

In order to complete the construction of the Path-of-Sets System $\Sigma'$, we let $A_1'\subseteq \hat A_1$ be any subset of $w'$ vertices, and we define $B'_{\ell}\subseteq \hat B_{\ell}$ similarly. It remains to define, for each $1\leq i<\ell$, the matching $\mset'_i$. We will ensure that the endpoints of the resulting matching are contained in $\hat B_i$ and $\hat A_{i+1}$, respectively, ensuring that the resulting Path-of-Sets System is strong.

Consider some index $1\leq i<\ell$. Recall that we have computed the sets $\hat B_i\subseteq B_{4i-1}$, $\hat A_{i+1}\subseteq A_{4i+1}$ of vertices. We let $E'_i\subseteq \mset_{4i-1}$ be the set of edges incident to the vertices of $\hat B_i$, and we denote by $\tilde A_{4i}\subseteq A_{4i}$ the set of vertices in $A_{4i}$ that serve as their endpoints. Similarly, we let $E''_i\subseteq \mset_{4i}$ be the set of edges incident to the vertices of $\hat A_{i+1}$, and we denote by $\tilde B_{4i}\subseteq B_{4i}$ the set of vertices in $B_{4i}$ that serve as their endpoints. From Claim~\ref{claim: flow in expander}, there is a set $\qset_i$ of disjoint paths in $S_{4i}$, connecting vertices of $\tilde A_{4i}$ to vertices of $\tilde B_{4i}$, of cardinality $w'=\Omega(\alpha^3w/d^4)$. By extending the paths in $\qset_i$ to include the edges of $E'_i\cup E''_i$ incident to them, we obtain a collection $\qset'_i$ of $w'$ disjoint paths in $S_{4i}\cup \mset_{4i-1}\cup\mset_{4i}$, connecting vertices of $\hat B_i$ to vertices of $\hat A_{i+1}$. We denote the endpoints of the paths in $\qset'_i$ lying in $\hat B_i$ by $B_i'$, and the endpoints of the paths in $\qset'_i$ lying in $\hat A_{i+1}$ by $A'_{i+1}$. The paths in $\qset'_i$ naturally define the matching $\mset'_i$ between the vertices of $B_i'$ and the vertices of $A_{i+1}'$. This concludes the definition of the Path-of-Sets System $\Sigma'$. It is immediate to verify that it is a strong Path-of-Sets System of length $\ell $ and width $w'$, and to obtain a model of $G_{\Sigma'}$ in $G_{\Sigma}$. Note that graph $G_{\Sigma'}$ has maximum vertex degree at most $d$.
\end{proofof}

Recall that in Part 1 of the algorithm, we have obtained  a sub-graph $G_{\Sigma}\subseteq G$, and an $\alpha'$-expanding Path-of-Sets System $\Sigma$ of length $48$ and width $w'=\ceil{\alpha' n}$, where $\alpha'=(\alpha/d)^{O(1)}$. 
Applying Theorem~\ref{thm: exp pos to strong pos} to $\Sigma$, we obtain a \posfull System $\Sigma'$ of length $12$ and width $w''=\Omega\left(\frac{(\alpha')^3w'}{d^4}\right )=\Omega\left(\frac{(\alpha')^4}{d^4}n\right )=\left(\frac{\alpha}{d}\right )^{O(1)}\cdot n$. We have also computed a model of $G_{\Sigma'}$ in $G$, and established that the maximum vertex degree in $G_{\Sigma'}$ is at most $d$.
For convenience, we let $c'$ be a constant, such that $w''\geq \frac{\alpha^{c'}}{d^{c'}}n$.

%% file: pos-to-poe.tex
\subsection{From \posfull System to \poefull System} \label{subsec: pos to poe}

The goal of this subsection is to prove the following theorem:
\begin{theorem} \label{thm: pos to poe}
    There is an efficient algorithm, that, given a  \posfull System $\Sigma$ of width $w$ and length $12$, such that the corresponding graph $G_{\Sigma}$ has at most $n$ vertices and has maximum vertex degree at most $d$,  computes a \poefull System $\Pi$ of width $\hat w =\Omega\left(\frac{w^4}{d^2n^3}\right )$ and expansion $\hat \alpha \geq \Omega\left (\frac{w^2}{n^2d}\right )$, whose corresponding graph $G_{\Pi}$ has maximum vertex degree at most $d+1$ and is a minor of $G_{\Sigma}$. Moreover, the algorithm computes a model of $G_{\Pi}$ in $G_{\Sigma}$.
\end{theorem}

Before we prove Theorem~\ref{thm: pos to poe}, we complete the proof of Theorem~\ref{thm: poe} using it. Recall that our input is an $\alpha$-expander $G$, for some $0<\alpha<1$, with $|V(G)|=n$, such that the maximum vertex degree in $G$ is at most $d$. Our goal is to provide an algorithm that computes a \poefull System $\Pi$ of expansion $\tilde \alpha \geq \left(\frac{\alpha}{d}\right )^{\hat c_1}$ and width $\tilde w \geq n \cdot \left(\frac{\alpha}{d}\right )^{\hat c_2}$, such that the maximum vertex degree in $G_{\Pi}$ is at most $d+1$, and to compute a minor of $G_{\Pi}$ in $G$.

Recall that in Step 2 we have constructed  a \posfull System $\Sigma'$ of length $12$ and width $w''\geq \frac{\alpha^{c'}}{d^{c'}}n$, for some constant $c'$, such that $G_{\Sigma'}$ has maximum vertex degree at most $d$. We have also computed a model of $G_{\Sigma'}$ in $G$.
Our last step is to apply Theorem~\ref{thm: pos to poe} to $\Sigma'$.
 As a result, we obtain a  \poefull System $\Pi$ of width $\hat w =\Omega\left(\frac{(w'')^4}{d^2n^3}\right )$ and expansion $\hat \alpha \geq \Omega\left (\frac{(w'')^2}{n^2d}\right )$, whose corresponding graph $G_{\Pi}$ has maximum vertex degree at most $d+1$. We also obtain a model of $G_{\Pi}$ in $G_{\Sigma}$. 

Substituting the value $w''\geq \frac{\alpha^{c'}}{d^{c'}}n$, we get that the width of the \poefull System is $\Omega\left(\frac{\alpha^{4c'}}{d^{2+4c'}}\right )\cdot n$, and that its expansion is $\Omega \left(\frac{\alpha^{2c'}}{d^{2c'+1}}\right )$. By appropriately setting the constants $\hat c_1$ and $\hat c_2$, we ensure that the width of the \poefull System is at least $n \cdot \left(\frac{\alpha}{d}\right )^{\hat c_2}$ and its expansion is at least  $\left(\frac{\alpha}{d}\right )^{\hat c_1}$.

In the remainder of this section, we prove Theorem~\ref{thm: pos to poe}.
We can assume w.l.o.g. that  $w^4\geq 2^{14}n^3d^2$, since otherwise it is sufficient to produce a \poefull System of width $1$, which is trivial to do.
 We denote the input \posfull System by $\Sigma = (\sset, \mset, A_1, B_{12})$, where $\sset=(S_1,\ldots,S_{12})$, and we let $G_{\Sigma}$ be its corresponding graph.
    For convenience, we denote by $\iseteven$ and $\isetodd$ the sets of all even and all odd indices in $\set{1,\ldots,12}$, respectively. The algorithm consists of three steps. In the first step, for every index $i\in \iseteven$, we find a large set $\pset_i$ of disjoint paths connecting $A_i$ to $B_i$ in $S_i$, and a subgraph $T_i\subseteq S_i$ that is an $\hat \alpha$-expander, such that the paths in $\pset_i$ are disjoint from $T_i$. In the second step, for each such index $i\in\iseteven$, we compute another set $\qset_i$ of disjoint paths in $S_i$, and a large enough subset $\pset'_i\subseteq \pset_i$ of paths, such that every path in $\qset_i$ connects a vertex on a distinct path of $\pset'_i$ to a distinct vertex of $T_i$. In the third and the final step we compute the \poefull System $\Pi$ and a model of $G_{\Pi}$ in $G_{\Sigma}$.

\paragraph{Step 1.} In this step, we prove the following lemma.

\begin{lemma}\label{lem: step 1}
There is an efficient algorithm, that, given an index $i\in \iseteven$, computes a set $\pset_i$ of $\floor{\frac{w^2}{16nd}}$ paths in $S_i$, and a subgraph $T_i\subseteq S_i$, such that:

\begin{itemize}
\item graph $T_i$ is an $\hat \alpha$-expander, and it contains at least $w/2$ vertices of $A_i$;
\item the paths in $\pset_i$ are disjoint from each other; they are also disjoint from $T_i$ and internally disjoint from $A_i\cup B_i$; 
\item every path in $\pset_i$ connects a vertex of $A_i$ to a vertex of $B_i$; and
\item every path in $\pset_i$  has length at most $2n/w$.
\end{itemize}
\end{lemma}

\begin{proof}
For convenience, we omit the subscript $i$ in this proof. We are given a graph $S$ that contains at most $n$ vertices and has maximum vertex degree at most $d$, and two disjoint subsets $A,B$ of $V(S)$ of cardinality $w$ each, such that each of $A\cup B$ is well-linked in $S$. Therefore, there is a set $\pset$ of $w$ disjoint paths in $S$, connecting vertices of $A$ to vertices of $B$, such that the paths in $\pset$ are internally disjoint from $A\cup B$. We say that a path in $\pset$ is \emph{short} if it contains at most $2n/w$ vertices, and otherwise it is long. Since $|V(S)|\leq n$, at most $w/2$ paths in $\pset$ can be long, and the remaining paths must be short. Let $\pset'\subseteq \pset$ be any subset of $\floor{\frac{w^2}{16nd}}$ paths in $\pset$. It is now sufficient to show an algorithm that computes an $\hat \alpha$-expander $T\subseteq S$, such that $T$ is disjoint from the paths in $\pset'$. In order to do so, we let $E'$ be the set of all edges lying on the paths in $\pset'$, so $|E'|\leq |\pset'|\cdot \frac{2n}{w}\leq \floor{\frac{w^2}{16nd}}\cdot \frac{2n}{w}\leq \frac{w}{8}$. 
 
We start with $T=S\setminus E'$, and then iteratively remove edges from $T$, until we obtain a connected component of the resulting graph that is an $\hat \alpha$-expander, containing at least $w/2$ vertices of $A$. Notice that the original graph $T$ is not necessarily connected. We also maintain a set $E''$ of edges that we remove from $T$, initialized to $E''=\emptyset$. 
Our algorithm is iterative. In every iteration, we apply Theorem~\ref{thm: spectral} to the current graph $T$, to obtain a cut $(Z,Z')$ in $T$. If the sparsity of the cut is at least $\frac{w}{16n}$, that is, $|E_T(Z,Z')|\geq \frac{w}{16n}\min\set{|Z|,|Z'|}$, then we terminate the algorithm. Theorem~\ref{thm: spectral} then guarantees that the expansion of $T$ is $\Omega\left (\frac{w^2}{n^2d}\right )$, that is, $T$ is a $\hat \alpha$-expander. Otherwise, $|E_T(Z,Z')|< \frac{w}{16n}\min\set{|Z|,|Z'|}$. Assume w.l.o.g. that $|Z\cap A|\geq |Z'\cap A|$. We then add the edges of $E_{T}(Z,Z')$ to $E''$, set $T=T[Z]$, and continue to the next iteration. Note that the number of edges added to $E''$ during this iteration is at most $\frac{|Z'|w}{16n}$.

Clearly, the graph $T$ we obtain at the end of the algorithm is an $\hat \alpha$-expander, and it is disjoint from all paths in $\pset'$. It now only remans to show that $T$ contains at least $w/2$ vertices of $A$. Assume for contradiction that this is false.

Assume that the algorithm performs $r$ iterations, and for each $1\leq j\leq r$, let $(Z_j,Z'_j)$ be the cut computed by the algorithm in iteration $j$, where $|Z_j\cap A|\geq |Z'_j\cap A|$. But then for all $1\leq j\leq r$, $|Z'_j\cap A|\leq w/2$ must hold. 
Let $n_j=|Z'_j\cap A|$. Since the vertices of $A$ are well-linked in $S$, $\delta_S(Z'_j)\geq n_j$. Therefore:

\[\sum_{j=1}^r|\delta_S(Z'_j)|\geq \sum_{j=1}^rn_j\geq w/2,\]

since we have assumed that the final graph $T$ has fewer than $w/2$ vertices of $A$. On the other hand, all edges in $\bigcup_{j=1}^r\delta_S(Z'_j)$ are contained in $E'\cup E''$, and so:

\[\sum_{j=1}^r|\delta_S(Z'_j)|\leq 2|E'\cup E''|.\]

Recall that $|E'|\leq \frac{w}{8}$, and it is easy to verify that $|E''|\leq \frac{w}{16n}\cdot n=\frac{w}{16}$. Therefore, $\sum_{j=1}^r|\delta_S(Z'_j)|<\frac w 2$, a contradiction.
\end{proof}

\paragraph{Step 2.} For every index $i\in \iseteven$, let $A'_i\subseteq A_i$ be the subset of vertices that serve as endpoints for the paths in $\pset_i$. The goal of this step is to prove the following lemma.

\begin{lemma}\label{lem: Q-paths}
There is an efficient algorithm, that, given an index $i\in \iseteven$, computes a subset $\pset'_i\subseteq \pset_i$ of $\hat w$ paths, and, for each path $P\in \pset'_i$, a path $Q_P$ in $S_i$, that connects a vertex of $P$ to a vertex of $T_i$, such that the paths in set $\qset_i=\set{Q_P\mid P\in \pset'_i}$ are disjoint from each other, internally disjoint from $T_i$, and internally disjoint from the paths in $\pset'_i$.
\end{lemma}

\begin{proof}
We fix an index $i\in \iseteven$, and for convenience omit the subscript $i$ for the remainder of the proof. Recall that we are given a set $A'\subseteq A$ of $\floor{\frac{w^2}{16nd}}$ vertices, that serve as endpoints of the paths in $\pset$. Recall that $T$ contains at least $w/2$ vertices of $A$. We let $A''\subseteq A$ be any set of $\floor{\frac{w^2}{16nd}}$ vertices of $A$ lying in $T$. Since the set $A$ of vertices is well-linked in $S$, there is a set $\qset$ of $\floor{\frac{w^2}{16nd}}$ node-disjoint paths, connecting the vertices of $A'$ to the vertices of $A''$ in $S$. We say that a path in $\qset$ is \emph{short} if it contains fewer than $\frac{64n^2d}{w^2}$ vertices, and otherwise we say that it is \emph{long}. Since $S$ contains at most $n$ vertices, and the paths in $\qset$ are disjoint, at most $\frac{w^2}{64nd}$ paths of $\qset$ are long. We let $\hqset\subseteq \qset$ be the set of all short paths, so $|\hqset|\geq \frac{w^2}{64nd}$, and we let $\hat A\subseteq A'$ be the set of vertices that serve as endpoints of the paths in $\hqset$. We also let $\hpset\subseteq \pset$ the set of paths originating from the vertices in $\hat A$. We are now ready to compute the set $\pset'$ of paths, and the corresponding paths $Q_P$ for all $P\in \pset'$.

We start with $\pset'=\emptyset$, and then iterate. While $\hpset\neq \emptyset$, let $P$ be any path in $\hpset$, and let $a\in \hat A$ be the vertex from which it originates. Let $Q$ be the path of $\hqset$ originating at $a$. We prune the path $Q$ as needed, so that it connects a vertex of $P$ to a vertex of $T$, but is internally disjoint from $P$ and $T$. Let $Q'$ be the resulting path. We then add $P$ to $\pset'$, and we let $Q_P=Q'$. Next, we delete from $\hpset$ all paths that intersect $Q'$ (since the length of $Q'$ is at most $\frac{64n^2d}{w^2}$, we delete at most $\frac{64n^2d}{w^2}$ paths from $\hpset$), and for every path $P^*$ that we delete from $\hpset$, we delete from $\hqset$ the path sharing an endpoint with $P^*$ (so at most $\frac{64n^2d}{w^2}$ paths are deleted from $\hqset$). Similarly, we delete from $\hqset$ every path that intersects $P$ (since the length of $P$ is at most $2n/w$, we delete at most $\frac{2n}{w}\leq \frac{64n^2d}{w^2}$ paths from $\hqset$), and for every path $Q^*$ that we delete from $\hqset$, we delete from $\hpset$ the path sharing an endpoint with $Q^*$ (again, at most $\frac{64n^2d}{w^2}$ paths are deleted from $\hpset$). Overall, we delete at most $\frac{128n^2d}{w^2}$ paths from $\hpset$, and at most $\frac{128n^2d}{w^2}$ paths from $\hqset$. The paths that remain in both sets form pairs -- that is, for every path $P^*\in \hpset$, there is a path $Q^*\in \hqset$ originating at the same vertex of $A$, and vice versa. Furthermore, and all paths in $\hpset\cup \hqset$ are disjoint from the paths in $\pset'\cup \set{Q_P\mid P\in \pset'}$. 

At the end of the algorithm, we obtain a subset $\pset'\subseteq \pset$ of paths, and for each path $P\in \pset'$, a path $Q_P$ in $S$, connecting a vertex of $P$ to a vertex of $T$, such that  the paths in set $\qset'=\set{Q_P\mid P\in \pset'}$ are disjoint from each other, internally disjoint from $T$, and internally disjoint from the paths in $\pset'$. It now only remains to show that $|\pset'|\geq \hat w$. 

Recall that we start with $|\hpset|\geq \frac{w^2}{64nd}$. In every iteration, we add one path to $\pset'$, and delete at most $\frac{128n^2d}{w^2}$ paths from $\hpset$. Since we have assumed that $w^4\geq 2^{14}n^3d^2$, we get that $\frac{256 n^2d}{w^2}\leq \frac{w^2}{64nd}$. 
It is then easy to verify that at the end of the algorithm, $|\pset'|\geq \floor{\frac{|\hpset|}{256n^2d/w^2}}\geq \Omega\left(\frac{w^4}{n^3d^2}\right )=\hat w$.
\end{proof}

\paragraph{Step 3.} In this step we complete the construction of the \poefull System $\Pi$. We will also define a minor $G'$ of $G_{\Sigma}$ and compute a model of $G_{\Pi}$ in $G'$; it is then easy to obtain a model of $G_\Pi$ in $G_{\Sigma}$.

\begin{figure}[h]
\centering
\subfigure[Paths $P$ (shown in blue) and $Q_P$ (shown in red) before edge contractions]{\scalebox{0.3}{\includegraphics{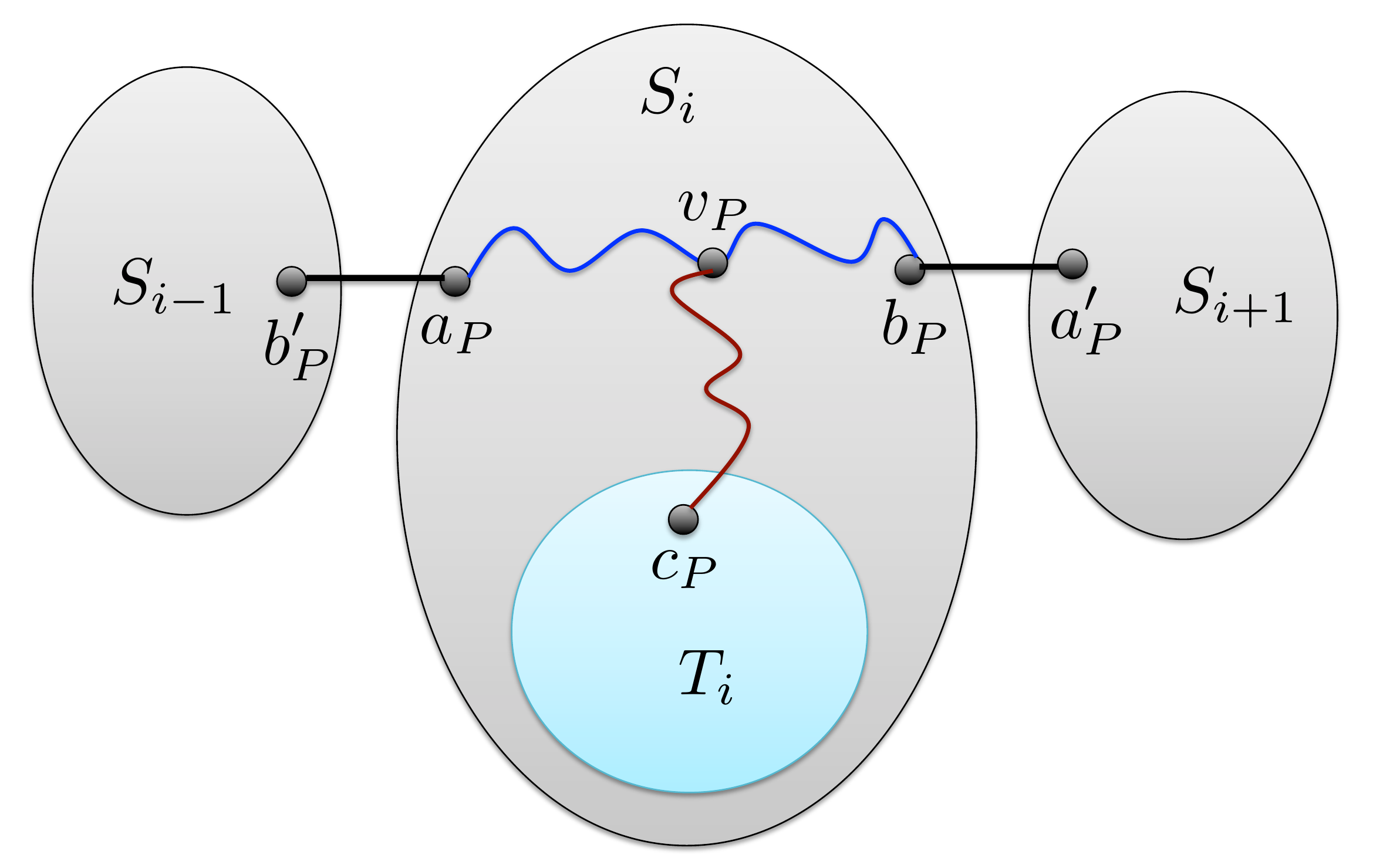}}\label{fig: before contraction}}
\hspace{1cm}
\subfigure[After edge contractions]{
\scalebox{0.3}{\includegraphics{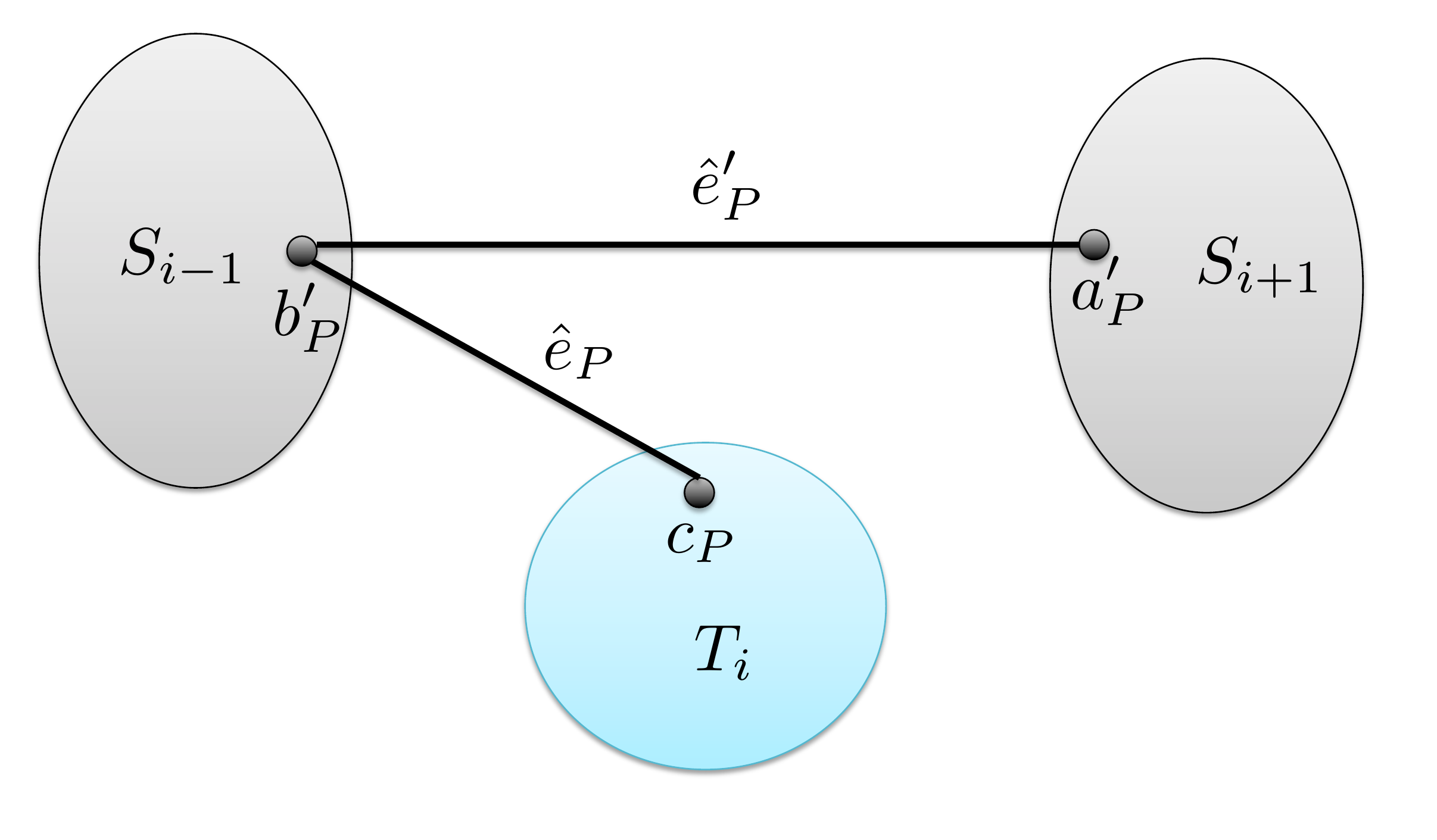}}\label{fig: after contraction}}
\caption{The contractions of the edges on paths $P$ and $Q_P$. \label{fig: edge contractions}}
\end{figure}

Consider some index $i\in \iseteven$, and the sets $\pset'_i,\qset_i$ of paths computed in Step 2. Let $P\in \pset'_i$ be any such path, and assume that it connects a vertex $a_P\in A_i$ to a vertex $b_P\in B_i$. Let $v_P\in P$ be the endpoint of $Q_P$ lying on $P$, and let $c_P$ be its other endpoint. Finally, let $e_P$ be the edge of $\mset_{i-1}$ incident to $a_P$ and let $b'_P\in B_{i-1}$ be its other endpoint. Similarly, if $i\neq 12$, let $e'_P$ be the edge of $\mset_i$ incident to $b_P$, and let $a'_P\in A_{i+1}$ be its other endpoint (see Figure~\ref{fig: before contraction}).

We contract the edge $e_P$ and all edges lying on the sub-path of $P$ between $a_P$ and $v_P$, so that $v_P$ and $b'_P$ merge. The resulting vertex is denoted by $b'_P$. We also suppress all inner vertices on the path $Q_P$, obtaining an edge $\hat e_P$, connecting $b'_P$ to $c_P$. Finally, if $i\neq 12$, then we contract all edges on the sub-path of $P$ between $v_P$ and $b_P$, obtaining an edge $\hat e'_P=(b_P,a'_P)$. We let $\hat E_i=\set{\hat e_P\mid P\in \pset'_i}$ and we let $\hat E_i'=\set{\hat e_P'\mid P\in \pset'_i}$ be the sets of these newly defined edges. 
Notice that the edges of $\hat E_i$ connect a subset of $\hat w$ vertices of $B_{i-1}$ (that we denote by $\hat B_{i-1}$) to a subset of $\hat w$ vertices of $T_{i}$ (that we denote by $\hat C_{i})$, and for $i\neq 12$,  the edges of $\hat E_i'$  connect every vertex of $\hat B_{i-1}$ to some vertex of $A_{i+1}$; we denote the set of endpoints of these edges that lie in $A_{i+1}$ by $\hat A_{i+1}$.

Once we perform this procedure for every path $P\in \pset'_i$, for all $i\in \iseteven$, we delete from the resulting graph all edges and vertices except those lying in graphs $S_i$ for $i\in \isetodd$, graphs $T_i$ for $i\in \iseteven$, and the edges in $\hat E_i\cup \hat E_i'$ for $i\in \iseteven$. The resulting graph, denoted by $G'$, is a minor of $G$, and it is easy to verify that its maximum vertex degree is at most $d+1$. 

We now define a \poefull System
$\Pi=(\tilde \Sigma,\tilde \mset, \tilde A_1,\tilde B_6, \tilde \tset,\tilde \mset')$, where the clusters of $\tilde \Sigma $ are denoted by $\tilde S_1,\ldots,\tilde S_6$; for each $1\leq i\leq 6$ the corresponding sets $A_i,B_i,C_i$ of vertices are denoted by $\tilde A_i$, $\tilde B_i$ and $\tilde C_i$ respectively; the matching $\mset'_i$ is denoted by $\tmset'_i$ and the expander $T_i$ is denoted by $\tilde T_i$. For all $1\leq i<6$, we also denote the matching $\mset_i$ by $\tmset_i$.
    
For each $1\leq i\leq 6$, we let the cluster $\tilde S_i$ of $\tilde \Sigma$ be $S_{2i-1}$, and we let the expander $\tilde T_i$ be $T_{2i}$. We also set $C_i=\hat C_{2i}$, and $\tmset'_i=\hat E_{2i}$.
If $i>1$, then we let $\tilde A_i=\hat A_{2i-1}$, and we let $\tilde A_1$ be any subset of $\hat w$ vertices of $A_1$. Similarly, if  $i<6$, then we let $\tilde B_i=\hat B_{2i-1}$, and we let $\tilde B_6$ be any subset of $\hat w$ vertices of $B_6$. Finally, for $i<6$, we let $\tilde \mset_i=\hat E_{2i}'$.
 It is immediate to verify that we have obtained a \poefull  System of width $\hat w$ and expansion $\hat \alpha$, and a model of $G_{\Pi}$ in $G'$. It is now immediate  to obtain a model of $G_{\Pi}$ in $G_{\Sigma}$.

%% file: constructive_proof.tex
\section{Proof of Theorem~\ref{thm: constructive main}} \label{sec: constructive proof}

The goal of this section is to provide the proof of Theorem~\ref{thm: constructive main}. Notice that Theorem~\ref{thm: constructive main} provides slightly weaker dependence on $n$ in the minor size than Theorem~\ref{thm: general main}, but it has several advantages: its proof is much simpler, the algorithm's running time is polynomial in $n,d$ and $\alpha$, and it provides a better dependence on $\alpha$ and $d$ in the bound on the minor size.
Our algorithm also has an additional useful property:  if it fails to find the required model, then with high probability it certifies that the input graph is not an $\alpha$-expander by exhibiting a cut of sparsity less than $\alpha$.

\newcommand{\consbound}{\floor{\frac{n}{\tilde c^*\log^2 n}\cdot \frac{\alpha^{3}}{d^{5}}}}

Let $G=(V,E)$ be the given $n$-vertex $\alpha$-expander with maximum vertex degree at most $d$.
As in the proof of Theorem~\ref{thm: general main}, given a graph $H$ with $n'$ vertices and $m'$ edges,
we can construct another graph $H'$, whose maximum vertex-degree is at most $3$ and $|V(H')| \leq n' + 2m' \leq 2 \consbound$,
such that $H$ is a minor of $H'$.
It is now enough to provide an efficient algorithm that computes a model of $H'$ in $G$.
For convenience of notation, we denote $H'$ by $H=(U,F)$, and we denote $U= \set{u_1, \ldots, u_{|U|}}$.
We can assume that $n>c_0$ for a large enough constant $c_0$ by appropriately setting the constant $\tilde c^*$, as otherwise it is enough to show that every graph of size $1$ is a minor of $G$, which is trivial.

Our algorithm consists of a number of iterations.
We say that a partition $(V', V'')$ of $V$ is \emph{good} iff $|V'|, |V''| \geq n/(4d)$; and $G[V'], G[V'']$ are both connected graphs.
We start with an arbitrary good partition $(V_1, V_2)$ of $V$, obtained by using the algorithm from \Cref{obs: decompose by spanning tree} with $r=2$.
Assume without loss of generality that $|V_1| \geq |V_2|$.
We now try to compute a model of $H$ in $G$, by first embedding the vertices of $H$ into connected sub-graphs of $G[V_2]$, and then routing the edges of $H$ in $G[V_1]$.
We show an efficient algorithm, that with high probability returns one of the following:
\begin{itemize}
    \item either a good partition $(V'_1, V'_2)$ such that $|E(V'_1, V'_2)| < |E(V_1, V_2)|$
            (in this case, we proceed to the next iteration); or
    \item a model of $H$ in $G$ (in this case, we terminate the algorithm and return the model).
\end{itemize}
Clearly, we terminate after $|E|$ iterations, succeeding with high probability.
We now describe a single iteration in detail.
Recall that we are given a good partition $(V_1, V_2)$ of $V$ with $|V_1|\geq |V_2|$.
Since $G$ is an $\alpha$-expander, we have $|E(V_1, V_2)| \geq \alpha n/(4d)$ (note that, if this is not the case, we have found a cut $(V_1, V_2)$ of sparsity less than $\alpha$).
Since the maximum vertex-degree in $G$ is bounded by $d$, we can efficiently find a matching $\mset\subseteq E(V_1,V_2)$ of cardinality at least $\alpha n/(8d^2)$.
We denote the endpoints of the edges in $\mset$ lying in $V_1$ and $V_2$ by $Z$ and $Z'$, respectively.
Let $\rho := 3 \cdot \ceil{4 c d^2 \log^2 n/\alpha^2}$,
where $c$ is the constant from Lemma \ref{lem: can route large disjoint subsets in expanders}.

    Recall that $U$ is the set of vertices in the graph $H$.
We apply Observation \ref{obs: decompose by spanning tree} to the graph $G[V_2]$, together with $R = Z'$ and parameter $r = |U|$, 
to obtain a collection $\wset=\set{W_1,\ldots,W_{|U|}}$ of disjoint connected subgraphs of $G[V_2]$,
such that for all $1\leq i\leq |U|$,
\[ |V(W_i)\cap Z'|\geq \floor{\frac{|Z'|}{d|U|}}
    \geq \floor{\frac{\alpha n}{8d^3|U|}}
    \geq \floor{\frac{\alpha n}{8d^3} \cdot \frac{\tilde c^* d^5\log^2n}{2n\alpha^3}}
    = \floor{\frac{ \tilde c^* d^2\log^2n}{16\alpha^2}}
\]

Here, we have used the fact that $|U|\leq 2 \consbound$).
By appropriately setting the constant $\tilde c^* $ in the bound on $|U|$, we can ensure that for all $1\leq i\leq |U|$, $|V(W_i)\cap Z'|\geq 3\rho$.

Recall that we are given a graph $H = (U,F)$ with maximum vertex-degree $3$ and that we have denoted $U=\set{u_1,\ldots,u_{|U|}}$.
For $1\leq i\leq |U|$, we think of the graph $W_i$ as representing the vertex $u_i$ of $H$.
For each $1\leq i\leq |U|$, and for each edge $e\in \delta_{H}(u_i)$,
we select an arbitrary subset $Z'_i(e) \subseteq V(W_i) \cap Z'$ of $\rho$ vertices,
such that all resulting sets $\set{Z'_i(e)\mid e\in \delta_H(u_i)}$ of vertices are mutually disjoint.
Let $E_i(e)\subseteq \mset$ be the subset of edges of $\mset$ that have an endpoint in $Z'_i(e)$, so $|E_i(e)|=\rho$.
We let $Z_i(e)$ be the set of vertices of $Z$ that serve as endpoints of the edges in $E_i(e)$.
Notice that all resulting sets $\set{Z_i(e)\mid 1\leq i\leq |U|, e\in \delta_H(u_i)}$ are mutually disjoint, and each of them contains $\rho'$ vertices. 

We apply the algorithm of \Cref{lem: can route large disjoint subsets in expanders} to the graph $G[V_1]$,
together with the parameter $\alpha/2$ and the family
$\set{Z_i(e)\mid 1\leq i\leq |U|, e\in \delta_H(u_i)}$ of vertex subsets, that we order appropriately.

\paragraph{Case 1. The algorithm returns a cut.}
In this case, we obtain a cut $(X, Y)$ in $G[V_1]$ of sparsity less than $\alpha/2$. We will compute a good partition $(V'_1, V'_2)$ of $V$ with $|E(V'_1, V'_2)|~<~|E(V_1, V_2)|$.
We need the following simple observation whose proof appears in Appendix. 
    \begin{observation} \label{obs: connected sparsest cut}
        There is an efficient algorithm, that given a connected graph $G = (V,E)$ and a cut $(X,Y)$ in $G$, produces a cut $(X^*, Y^*)$, whose sparsity is less than or equal to that of $(X,Y)$,
        such that both $G[X^*]$ and  $G[Y^*]$ are connected.
    \end{observation}

We apply \Cref{obs: connected sparsest cut} to graph $G[V_1]$ and cut $(X,Y)$, obtaining a new cut $(X^*,Y^*)$ of sparsity less than $\alpha/2$,
such that both $G[X^*]$ and $G[Y^*]$ are connected.
For convenience, we denote the cut $(X^*,Y^*)$ by $(X,Y)$, and we assume without loss of generality that $|Y|\leq |X|$. 
Notice that $|Y| \leq |V_1|/2 \leq |V|/2$.
Since $G$ is an $\alpha$-expander, $|\delta_G(Y)| \geq \alpha |Y|$ (note that, if this is not the case, then have found a cut $(Y, V \backslash Y)$ of sparsity less than $\alpha$.).

Since $\delta_G(Y)=E(X,Y)\cup E(Y,V_2)$, we get that $|E(Y,V_2)|\geq \alpha |Y|/2$, and $|E(X,Y)|<|E(Y,V_2)|$.
In particular, $E(Y,V_2)\neq \emptyset$.
We now define a new cut $(V'_1,V'_2)$ of $G$, where $V'_2 = V_2 \cup Y$ and $V'_1 = X$.
We claim that $(V'_1,V'_2)$ is a good partition of $V(G)$.
It is immediate to verify that $|V'_1|,|V'_2|\geq n/(4d)$, and that $G[V'_1]=G[X]$ is connected.
Moreover, since $G[Y]$ is connected and $E(Y,V_2)\neq \emptyset$, $G[V'_2]=G[V_2 \cup Y]$ is also connected.
Lastly, we claim that $|E(V'_1,V'_2)| < |E(V_1, V_2)|$.
Indeed, since $|E(X, Y)| < |E(V_2, Y)|$:

\[|E(V'_1, V'_2)| = |E(V_1, V_2)| - |E(V_2, Y)| + |E(Y,X)| < |E(V_1,V_2)|.\]

Therefore, we have computed a good partition $(V'_1, V'_2)$ of $V(G)$, with $|E(V'_1, V'_2)| < |E(V_1, V_2)|$ as required.

\paragraph{Case 2. The algorithm returns paths.}
In this case, we have obtained, for every edge $e=(u_i,u_j)\in F$, 
a path $Q(e)$ in $G[V_1]$, connecting a vertex of $Z_i(e)$ to a vertex of $Z_j(e)$,
such that, with high probability, the paths in $\set{Q(e)\mid e\in F}$ are mutually disjoint.
If the paths in $\set{Q(e)\mid e\in F}$ are not mutually disjoint, the algorithm fails.
We assume from now on that the paths in $\set{Q(e)\mid e\in F}$ are mutually disjoint.
We extend each path $Q(e)$ to include the two edges of $\mset$ that are incident to its endpoints,
so that $Q(e)$ now connects a vertex of $Z'_i(e)$ to a vertex of $Z'_j(e)$. 
 
We are now ready to define the model of $H$ in $G$.
For every $1 \leq i \leq |U|$, we let $f(u_i)=W_i$, and for every edge $e\in F$, we let $f(e)=Q(e)$.
It is immediate to verify that this mapping indeed defines a valid model of $H$ in $G$.
This completes the proof of \Cref{thm: constructive main}.

%% file: cor-proof.tex
\section{\proofof{\Cref{cor: random}}}\label{sec: cor proof}

In this subsection we prove Corollary \ref{cor: random}.
    We use the following result of Krivelevich~\cite{exp_random}:
    \begin{theorem}[Corollary 1 of \cite{exp_random}]
        For every $\epsilon > 0$, there exists $\gamma > 0$, such that for every $n>0$, a random graph $G \sim \gset(n, \frac{1+\epsilon}{n})$ contains an induced bounded-degree $\gamma$-expander $\tilde G$ on at least $\gamma n$ vertices w.h.p.
    \end{theorem}

    Let $G \sim \gset(n,\frac{1+\epsilon}{n})$.
    From the above theorem, w.h.p., there is an induced bounded-degree $\gamma$-expander $\tilde G\subseteq G$ on at least $\gamma n$ vertices, for some $\gamma$ depending only on $\epsilon$.
    From Theorem \ref{thm: general main}, every graph $H$ of size at most $c_\epsilon n/\log n$ is a minor of $\tilde G$, where $c_\epsilon$ is some constant  depending on $\epsilon$ only.
    Corollary \ref{cor: random} now follows.
    \endproofof

%% file: appendix_lower_bounds.tex
\newcommand{\bound}{{20}}

\section{\proofof{\Cref{obs: lower bound}}} \label{sec: lower bound}

Recall that we are given a integer $s$ and a graph $G = (V,E)$ of size $s$.
Assume for now that $2 \leq s < 2^\bound$.
Let $H_G$ be a graph with $s+1$ vertices and $0$ edges.
Notice that the number of vertices in $H_G$ is strictly more than that in $G$, and hence $H_G$ is not a minor of $G$.
The observation now follows since $\bound s / \log s \geq s + 1$.
Thus from now on, we assume that $s \geq 2^\bound$ and hence, $\bound s/ \log s \geq 2^{\bound}$.

We denote by $\mu(G) = |\set{H \> | \> H \text{ is a minor of } G}|$.
For an integer $r$, let $\fset_r$ be the set of all graphs of size at most $r$.
The following two observations now complete the proof of Observation \ref{obs: lower bound}.

\begin{observation}
	$\mu(G) \leq 3^s$.
	\vspace{-1em}
\end{observation}
\begin{proof}
	From the definition of minors, every minor $H$ of $G$ can be identified by a subset $E^{\text{del}}_H \subseteq E$ of deleted edges, a subset $E^{\text{cont}}_H \subseteq E$ of contracted edges and a subset $V^{\text{del}}_H \subseteq V$ of deleted vertices.
	Thus,
	\[\mu(G) \leq 2^{|V|} \cdot 3^{|E|} \leq 3^{|V| + |E|} \leq 3^s.\]
\end{proof}

\begin{observation}
	For every even integer $r \geq 2^{10}$, $|\fset_r| \geq r^{r/10}$.
	\vspace{-1em}
\end{observation}
\begin{proof}
	Let $k = \floor{r^{0.9}}$.
	We lower-bound the number of graphs containing exactly $k$ vertices and exactly $r/2$ edges.
	Notice that, since $r \geq 2^{10}$, $k + r/2 \leq r$.
	For convenience, assume that the set $V^* = \set{1, \ldots, k}$ of vertices and their indices are fixed.	We will first lower-bound the number of vertex-labeled graphs with the set $V^*$ of vertices, that contain exactly $r/2$ edges.
	Since there are only $\binom{k}{2}$ `edge-slots', this number is at least:
	\[ \binom{\binom{k}{2}}{r/2} \geq \binom{r^{1.6}}{r/2} \geq \left( \frac{r^{1.6} - r/2}{r/2}\right)^{r/2} \geq \left(r^{0.6} \right)^{r/2} \geq r^{0.3 r}.\]

	Here, the inequalities hold for all $r \geq 2^{10}$.
	Notice that two graphs $G_1 = (V^*, E_1)$ and $G_2 = (V^*, E_2)$ with labeled vertices are isomorphic to each other iff there is a permutation $\psi$ of the vertices, mapping $E_1$ to $E_2$.
	Thus, the number of non-isomorphic graphs on $k$ vertices and $r/2$ edges is at least:
	\[ \frac{r^{0.3r}}{k!} \geq \frac{r^{0.3r}}{ \left(r^{0.9} \right)!} > \frac{r^{0.3r}}{r^{0.9r^{0.9}}} \geq r^{r^{0.9}(0.3r^{0.1} - 0.9)} \geq r^{r/10}.\]
\end{proof}

We are now ready to complete the proof of Observation \ref{obs: lower bound}.
Assume for contradiction that $G$ contains every graph in the family $\fset^* = \fset_{( \bound s/\log s)}$ as a minor.
Recall that $\bound s/\log s \geq 2^\bound$.
However, from the above two observations, $|\fset^*|\geq (\bound s/\log s)^{\bound s/(10 \log s)}$, while $\mu(G)\leq 3^s$. It is immediate to verify that $|\fset^*|>\mu(G)$, a contradiction.
\endproofof

\section{Proofs Omitted from Section \ref{sec: prelims}}
\subsection{Proof of Observation~\ref{obs: simple partition} }\label{subsec: simple partition proof}
We assume without loss of generality that $x_1\geq x_2\geq\cdots\geq x_r$, and process
the integers in this order. When $x_i$ is processed, we add $i$ to
$A$ if $\sum_{j\in A}x_j\leq \sum_{j\in B}x_j$, and we add it to $B$
otherwise. We claim that at the end of this process, $\sum_{i\in
	A}x_i,\sum_{i\in B}x_i\geq N/4$ must hold. Indeed, 
$1$ is always added to $A$. If $x_1\geq N/4$, then, since $x_1\leq 3N/4$, it is easy to see
that both subsets of integers sum up to at least $N/4$.  Otherwise,
$|\sum_{i\in A}x_i-\sum_{i\in B}x_i|\leq \max_i\set{x_i}\leq x_1\leq
N/4$, and so $\sum_{i\in
	A}x_i,\sum_{i\in B}x_i\geq N/4$.

\subsection{\proofof{\Cref{claim: large expanding subgraph-edges}}} \label{subsec: large expanding subgraph}

Our algorithm iteratively removes edges from $T\setminus E'$, until we obtain a connected component of the resulting graph that is an $\alpha/4$-expander.
We start with $T'=T\setminus E'$ (notice that $T'$ is not necessarily connected).  We also maintain a set $E''$ of edges that we remove from $T'$, initialized to $E''=\emptyset$. While $T'$ is not an $\alpha/4$-expander, let $(X,Y)$ be a cut of sparsity less than $\alpha/4$ in $T'$, that is $|E_{T'}(X,Y)| < \alpha \min{(|X|,|Y|)}/4$. Assume w.l.o.g. that $|X|\geq |Y|$. Update $T'$ to be $T'[X]$, add the edges of $E(X,Y)$ to $E''$, and continue to the next iteration. 

Assume that the algorithm performs $r$ iterations, and for each $1\leq i\leq r$, let $(X_i,Y_i)$ be the cut computed by the algorithm in iteration $i$. Since $|X_i|\geq |Y_i|$, $|Y_i|\leq |V(T')|/2$. At the same time, if we denote $E_i=E''\cap E(X_i,Y_i)$, then $|E_i|< \alpha |Y_i|/4$. Therefore:

\[|E''|=\sum_{i=1}^r|E_i|\leq \alpha\sum_{i=1}^r|Y_i|/4.\]

On the other hand, since $T$ is an expander, the total number of edges leaving each set $Y_i$ in $T$ is at least $\alpha|Y_i|$, and all such edges lie in $E'\cup E''$. Therefore:

\[|E'|+|E''|\geq \alpha\sum_{i=1}^r|Y_i|/2.\]

Combining both bounds, we get that $|E'|\geq \alpha\sum_{i=1}^r|Y_i|/4$.
We get that $\sum_{i=1}^r|Y_i|\leq \frac{4|E'|}{\alpha}$, and therefore $|V(T')|\geq |V(T)|-\frac{4|E'|}{\alpha}$.
\endproofof

\section{\proofof{\Cref{obs: decompose by spanning tree}}} \label{subsec: proof of decompose by spanning tree}
	Let $\tau$ be any spanning tree of $\hG$, rooted at an arbitrary degree-$1$ vertex of $\tau$. We start with $\uset=\emptyset$. Our algorithm performs a number of iterations, where in each iteration we add one new set $U\subseteq V(\hG)$ of vertices to $\uset$, such that $\hG[U]$ is connected and $\floor{|R|/(d r)}\leq |U\cap R|\leq |R|/r$, and we remove the vertices of $U$ from $\tau$. We execute the iterations as long as $|V(\tau)\cap R|\geq \floor{|R|/(d r)}$, after which we terminate the algorithm, and return the current collection $\uset$ of vertex subsets. 
	
	In order to execute an iteration, we let $v$ be the lowest vertex of $\tau$, such that the subtree $\tau_v$ of $\tau$ rooted at $v$ contains at least $\floor{|R|/(d r)}$ vertices of $R$.
	Since the maximum vertex degree in $\hG$ is bounded by $d$, tree $\tau_v$ contains fewer than $d \cdot \floor{|R|/(d r)} \leq |R|/r$ vertices of $R$. We add a new set $U=V(\tau_v)$ of vertices to $\uset$, delete the vertices of $U$ from $\tau$, and continue to the next iteration.
	
	Let $\uset$ be the final collection of vertex subsets obtained at the end of the algorithm. It is immediate to verify that for every set $U\in \uset$, $\hG[U]$ is connected and, from the above discussion, $\floor{|R|/(d r)}\leq  |U\cap R|\leq |R|/r$. Therefore, $|\uset|\geq r$.	
\endproofof

\section{\proofof{\Cref{claim: short paths in expanders}}}\label{subsec: proof of short paths in expanders}

Consider the following sequence of vertex subsets. Let $S_0=Z$, and for all $i>0$, let $S_i$ contain all vertices of $S_{i-1}$, and all neighbors of vertices in $S_{i-1}$. Notice that, if $|S_{i-1}|\leq |V(T)|/2$, then, since $T$ is an $\alpha'$-expander, there are at least $\alpha' |S_{i-1}|$ edges leaving the set $S_{i-1}$, and, since the maximum vertex degree in $T$ is at most $d$, there are at least $\frac{\alpha'|S_{i-1}|}{d}$ vertices that do not belong to $S_{i-1}$, but are neighbors of vertices in $S_{i-1}$. Therefore, $|S_i|\geq |S_{i-1}|\left (1+\frac{\alpha'}{d}\right )$.
We claim that there must be an index $i^*\leq \frac{8d}{\alpha'}\log (n/z)$, such that $|S_{i^*}|> |V(T)|/2$. Indeed, otherwise, we get that for $i=\ceil{\frac{8d}{\alpha'}\log (n/z)}$:

\[|S_{i^*}|\geq |S_0|\left (1+\frac{\alpha'}{d}\right )^{i}\geq z\cdot e^{i\alpha'/(2d)}\geq z\cdot e^{4\log (n/z)}>n/2.\]

Here, the second inequality follows from the fact that $(1+1/x)^{2x}>e$ for all $x>1$.
We construct a similar sequence $S'_0,S'_1,\ldots,$ for $Z'$. Similarly, there is an index $i^{**}\leq  \frac{8d}{\alpha'}\log (n/z')$, such that $S'_{i^{**}}$ contains more than half the vertices of $T$. Therefore, there is a path connecting a vertex of $Z$ to a vertex of $Z'$, whose length is at most $\frac{8d}{\alpha'}(\log (n/z)+\log(n/z'))$. 
\endproofof

%% file: short-paths-routing.tex
\section{Proof of Lemma~\ref{lem: can route large disjoint subsets in expanders}
}\label{sec: routing along short paths}

Recall that we are given a graph $G = (V,E)$, with  $|V|\leq n$ and maximum vertex degree at most $d$, and a parameter $0<\alpha<1$.
We are also given a collection $\set{C_1, \ldots, C_{2 r}}$ of disjoint subsets of $V$, each containing $q = \ceil{cd^2 \log^2n/\alpha^2}$ vertices, for some constant $c$ to be fixed later. 
Our goal is to either find a set $\qset = \set{Q_1, \ldots, Q_{r}}$ of disjoint paths, such that for each $1\leq j\leq r$, path $Q_j$ connects $C_j$ to $C_{j+r}$; or compute a cut $(S, S')$ in $G$ of sparsity less than $\alpha$.

We use a standard definition of multicommodity flow. 
A \emph{flow} $f$ consists of a collection $\pset$ of paths  in $G$, called \emph{flow-paths}, and, for each path $P\in \pset$, an associated flow value $f(P)>0$. The \emph{edge-congestion} of $f$ is the maximum amount of flow passing through any edge, that is, $\max_{e\in E}\set{\sum_{\stackrel{P\in \pset:}{e\in P}}f(P)}$. We say that the flow in $f$ causes \emph{no edge-congestion} iff the edge-congestion due to $f$ is at most $1$. Similarly, the \emph{vertex congestion} of $f$ is the maximum flow passing through any vertex, that is,  $\max_{v\in V}\set{\sum_{\stackrel{P\in \pset:}{v\in P}}f(P)}$. 
If a path $P$ does not lie in $\pset$, then we implicitly set $f(P)=0$. For any pair $s,t\in V$ of vertices, let $\pset(s,t)$ be the set of all paths connecting $s$ to $t$ in $G$. We say that $f$ \emph{transfers $z$ flow units between $s$ and $t$} iff $\sum_{P\in\pset(s,t)} f(P) \geq z$.

The following theorem is a consequence of Theorem 18 from~\cite{LeightonRao}
that we prove after completing the proof of \Cref{lem: can route large disjoint subsets in expanders}.

\begin{theorem}\label{thm: integral routing on expanders}
    There is an efficient randomized algorithm, that,
    given a graph $G=(V,E)$ with $|V|=n$ and maximum vertex degree at most $d$,
    and a parameter $0<\alpha<1$,
    together with a (possibly partial) matching $\mset$ over the vertices of $G$,
    computes one of the following:
    \begin{itemize}
        \item either a collection $\qset'=\set{Q(u,v)\mid (u,v)\in \mset}$ of paths, such that for all $(u,v)\in \mset$, path $Q(u,v)$ connects $u$ to $v$;
        the paths in $\qset'$ with high probability cause vertex-congestion at most $\eta= O(d \log n/\alpha)$,
        and the length of every path in $\qset$ is at most $L= O(d \log n/\alpha)$; or
        \item a cut $(S, S')$ in $G$ of sparsity less than $\alpha$.
    \end{itemize}
\end{theorem}

We are now ready to complete the proof of Lemma~\ref{lem: can route large disjoint subsets in expanders}.
We construct a matching $\mset$ over the vertices of $V$, as follows.
For each $1\leq j\leq r$, we add an arbitrary matching $\mset_j$, containing $q$ edges, between the vertices of $C_{j}$ and the vertices $C_{j + r}$. We then set $\mset=\bigcup_{j=1}^r\mset_j$.
We apply the algorithm from Theorem~\ref{thm: integral routing on expanders} to the graph $G$, parameter $\alpha$ and the matching $\mset$.
If the algorithm returns a cut of sparsity less than $\alpha$, we terminate the algorithm and return the cut. Therefore, we assume from now on that the algorithm returns a set $\qset'$ of paths with the following properties:
\begin{itemize}
    \item For each $j \in [r]$, there is a subset $\qset'_j\subseteq \qset'$ of $q$ paths connecting vertices of $C_j$ to vertices of $C_{j+r}$;
    \item All paths in $\qset'$ have length at most $L=O(d\log n/\alpha)$; and
    \item With high probability, every vertex of $G$ participates in at most $\eta=O(d \log n/\alpha)$  paths of $\qset'$.
\end{itemize}

If the vertex-congestion caused by the paths in $\qset'$ is greater than $\eta$, the algorithm terminates with a failure.
Therefore, we assume from now on that the paths in $\qset'$ cause vertex-congestion  at most $\eta$.
We use the constructive version of the Lovasz Local Lemma by Moser and Tardos~\cite{Moser-Tardos} in order to select one path from each set $\qset_j'$, so that the resulting paths are node-disjoint  with high probability.
The next theorem summarizes the symmetric version of the result of~\cite{Moser-Tardos}.

\begin{theorem}[\cite{Moser-Tardos}]\label{constructive symmetric LLL}
Let  $X$ be a finite set of mutually independent random variables in some probability space. Let $\aset$  be a finite set of bad events determined by these variables. 
 For each event $A\in \aset$, let $\vbl(A)\sse X$ be the unique minimal subset of variables determining $A$, and let $\Gamma(A)\sse \aset$ be a subset of bad events $B$, such that $A\neq B$, but $\vbl(A)\cap \vbl(B)\neq \emptyset$. Assume further that for each $A\in \aset$, $|\Gamma(A)|\leq D$, $\prob{A}\leq p$, and $ep(D+1)\leq 1$. Then there is an efficient randomized algorithm that computes an assignment to the variables of $X$, such that with high probability none of the events in $\aset$ holds.
 \end{theorem}

For each $1\leq i\leq r$, we choose one of its paths $Q_i\in \qset_i$ independently at random. We let $z_i$ be the random variable indicating which path has been chosen. 
For every pair $Q,Q'\in \qset'$ of intersecting paths, such that $Q,Q'$ belong to distinct sets $\qset'_i,\qset'_j$ let $\event(Q,Q')$ be the bad event that both these paths were selected. Notice that the probability of $\event(Q,Q')$ is $1/q^2$. Notice also that $\vbl(\event(Q,Q'))=\set{z_i,z_j}$, where $Q\in \qset'_i,Q'\in \qset'_j$. There are at most $qL\eta$ events $\event(\hat Q,\hat Q')$, with $z_i\in \vbl(\event(Q,Q'))$: set $\qset'_i$ contains $q$ paths; each of these paths has length at most $L$, so there are at most $qL$ vertices that participate in the paths in $\qset'_i$. Each such vertex may be shared by at most $\eta$ other paths. Similarly, there are at most  $qL\eta$ events $\event(\hat Q,\hat Q')$, with $z_j\in \vbl(\event(Q,Q'))$. 
Therefore, $|\Gamma(\event(Q,Q'))|\leq 2qL\eta$. Let $D=2qL\eta$.
It now only remains to show that $(D+1)ep\leq 1$.
Indeed,
\[ (D+1)ep=\frac{O(qL\eta)}{q^2}=\frac{O(L\eta)}{q}=O\left (\frac{d^2\log^2n}{\alpha^2q}\right ).\] 

By choosing the constant $c$ in the definition of $q$ to be large enough, we can ensure that $(D+1)ep\leq 1$ holds.
Using the algorithm from \Cref{constructive symmetric LLL}, we obtain a collection $\qset = \set{Q_1, \ldots, Q_r}$ of paths in $G$,
where for each $j \in [r]$, path $Q_j$ connects a vertex of $C_j$ to a vertex of $C_{j+r}$, and with high probability the resulting paths are disjoint.
This completes the proof of \Cref{lem: can route large disjoint subsets in expanders}, except for the proof of \Cref{thm: integral routing on expanders} that we provide next.

\subsection{\proofof{\Cref{thm: integral routing on expanders}}}
    We use a slight adaptation of Theorem~18 from~\cite{LeightonRao}. 

    \begin{theorem}[Adaptation of Theorem 18 from~\cite{LeightonRao}]\label{thm: Leighton-Rao}
        There is an efficient algorithm, that,
        given a $n$-vertex graph $G$ with maximum vertex degree at most $d$,
        together with a parameter $0 < \alpha <1$ computes one of the following:
        \begin{itemize}
            \item either a flow $f$ in $G$, with every pair of vertices in $G$ transferring
            $\frac{\alpha}{64 n\log n}$ flow units to each other with no edge-congestion,
            such that every flow-path has length at most $\frac{64 d \log n}{\alpha}$; or
            \item a cut $(S, S')$ in $G$ of sparsity less than $\alpha$.
        \end{itemize}
    \end{theorem}

We provide the proof of Theorem~\ref{thm: Leighton-Rao} below, after completing the proof of \Cref{thm: integral routing on expanders} using it.

    We apply \Cref{thm: Leighton-Rao} to the graph $G$ and the parameter $\alpha$.
    If the algorithm returns a cut $(S, S')$ of sparsity less than $\alpha$, then we terminate the algorithm and return this cut. Therefore, we assume from now on that the algorithm returns the flow $f$. Let $f'$ be a flow obtained from $f$ by scaling it up by factor $64 \log n/\alpha$, so that every pair of vertices in $G$ now sends $1/n$ flow units to each other, with total edge-congestion at most $64 \log n/\alpha$. 

    We start by showing that there is a multi-commodity flow $f^*$, where every pair $(u,v)\in \mset$ of vertices sends one flow unit to each other simultaneously, on flow-paths of length at most $128 d\log n/\alpha$, with total vertex-congestion at most $128 d\log n/\alpha$.
    Let $(u,v)\in \mset$ be any pair of vertices.
    The new flow between $u$ and $v$ is defined as follows: $u$ sends $1/n$ flow units to every vertex of $G$, using the flow $f'$,
    and $v$ collects $1/n$ flow units from every vertex of $G$, using the flow $f'$.
    In other words, the flow $f^*$ between $u$ and $v$ is obtained by concatenating all flow-paths in $f'$ originating at $u$ with all flow-paths in $f'$ terminating at $v$.
    It is easy to see then that every flow-path in $f'$ is used at most twice: once by each of its endpoints; all flow-paths in $f^*$ have length at most $128 d\log n/\alpha$; and the total edge-congestion due to flow $f^*$ is at most $128 \log n/\alpha$.
    Since the maximum vertex degree in $G$ is at most $d$, flow $f^*$ causes vertex-congestion at most $128 d\log n/\alpha$.
    
    Next, for every pair $(u,v)\in \mset$, we select one path $Q(u,v)\in \pset(u,v)$ at random, where a path $P\in \pset(u,v)$ is selected with probability $f^*(P)$  -- the amount of flow sent on $P$ by $f^*$.
    We then let $\qset'=\set{Q(u,v)\mid (u,v)\in \mset}$.
    Notice that the length of every path in $\qset'$ is at most $128 d\log n/\alpha$.
    It remains to show that the total vertex-congestion due to paths in $\qset'$ is at most $O(d \log n / \alpha)$ with high probability.
    This is done by standard techniques.
    Consider some vertex $x\in V$.
    We say that the bad event $\event(x)$ happens if more than $8 \cdot 128 d\log n/\alpha$ paths of $\qset$ use the vertex $x$. 
    We use the following variation of the Chernoff bound (see~\cite{measure-concentration}):
    
    \begin{theorem}
        Let $X_1,\ldots,X_n$ be independent random variables taking values in $[0,1]$, let $X=\sum_iX_i$, and
        let $\mu=\expect{X}$. Then for all $t>2e\mu$, 
        $\prob{X>t}\leq 2^{-t}$.
    \end{theorem}
    
    It is easy to see that the expected number of paths in $\qset'$ that contain $x$ is at most $128 d\log n/\alpha$, and so the probability of $\event(x)$ is bounded by $1/n^4$.
    From the Union Bound, the probability that any such event happens for any vertex $x\in V$ is bounded by $1/n^3$.
    Therefore, with high probability, every vertex of $G$ belongs to $2^{10} d \log n/\alpha = O(d \log n / \alpha)$  paths in $\qset'$.
    This finishes the proof of \Cref{thm: integral routing on expanders} except for the proof of \Cref{thm: Leighton-Rao}, that we prove in the next sub-section.
\endproofof

\subsection{Proof of \Cref{thm: Leighton-Rao}}

   \newcommand{\LRconst}{64}

    The proof follows closely that of \cite{LeightonRao}; we provide it here for completeness.
    Recall that we are given a graph $G = (V,E)$ with maximum vertex-degree at most $d$, $|V|=n$ and a parameter $0 < \alpha < 1$.
    We let $L = \LRconst d \log n / \alpha$.
    For every pair $u, v$ of vertices in $V$, let $\pset^{\leq L}(u,v)$ be the set of all paths in $G$ between $u$ and $v$ that contain at most $L$ vertices.
    We employ standard linear program for uniform multicommodity flow:

\begin{eqnarray*}
\mbox{(LP-1)}&	\max\quad\quad f^*&\\
	\mbox{s.t.}&&\\
	&\sum_{P \in \pset^{\leq L}(u,v)}  f(P)\geq f^*& \forall u,v \in V \\
&	\sum_{u,v\in V}\sum_{\stackrel{P \in \pset^{\leq L}(u,v):}{e \in P}} f(P) \leq 1 & \forall e \in E \\
 & f(P) \geq 0 & \forall u,v \in V; \forall P \in \pset^{\leq L}(u,v)
\end{eqnarray*}

    In general, the dual of the standard relaxation of the uniform multicommodity flow problem is the problem of assigning lengths $\ell(e)$ to the edges $e\in E$, so as to minimize $\sum_e\ell(e)$, subject to the constraint that the total sum of all pairwise distances between pairs of vertices is at least $1$, where the distance between pairs of vertices is defined with respect to $\ell$. 
    
    In our setting, given lengths $\ell(e)$ on edges $e\in E$, we need to use $L$-hop bounded distances between vertices, defined as follows:
    for all $u,v \in V$, if  $\pset^{\leq L}(u,v) \neq \emptyset$, then we let:
    \[ D^{\leq L}_{\ell}(u,v) = \min_{P \in \pset^{\leq L}(u,v)} \set{\sum_{e \in P} \ell(e)}; \]
    
    otherwise, we set $D^{\leq L}_{\ell}(u,v) = \infty$.
    The dual of (LP-1) can now be written as follows:

\begin{eqnarray*}
	\mbox{(LP-2)}&\min\quad\quad \sum\limits_{e\in E}\ell(e)&\\
	\mbox{s.t.}&&\\
&	\displaystyle\sum\limits_{u,v \in V} D^{\leq L}_{\ell}(u,v) \geq 1  & \\
	& \ell(e) \geq 0 & \forall e \in E\\
	\end{eqnarray*}

    Even though Linear Programs (LP-1) and (LP-2) are of exponential size, they can be solved efficiently using standard techniques (that is, edge-based flow formulation).
    Let $f^*_{\opt}$ be the value of the optimal solution to (LP-1).
    We let $W^* = \frac{d}{nL} = \frac{\alpha}{\LRconst n \log n}$.
    If $f^*_{\opt} \geq W^*$, then we return the flow $f$ corresponding to the optimal solution of (LP-1); it is immediate to verify that it satisfies all requirements.
    Therefore, we assume from now on that $f^*_{\opt} < W^*$.
    We will provide an efficient algorithm to compute a cut $(S, S')$ in $G$ of sparsity less than $\alpha$.

    Given a length function $\ell : E \mapsto \reals_{\geq 0}$, we denote by $W(\ell) = \sum_{e \in E} \ell(e)$ the total `weight' of $\ell$.
   We need the following definition.
   
   \begin{definition}
   	Given an integer $r$ and a length function $\ell(e)$ on edges $e\in E$, the $r$-hop bounded diameter of $G$ is $\max_{u,v\in V}\set{D^{\leq r}_{\ell}(u,v)}$.
   	\end{definition}

    Consider the optimal solution $\ell_{\opt}:E\rightarrow \reals^+$ to (LP-2).
    Observe that, by the strong duality, the value of the solution $W(\ell_{\opt})  = f^*_{\opt}$, and so $W(\ell_\opt) < W^*$ holds.
    
    We define a new solution $\ell$ to (LP-2) as follows: for each edge $e$, we let $\ell(e)=\ell_{\opt}(e)\cdot \frac{W^*}{W(\ell_{\opt})}$. Since $W^* > W(\ell_{\opt})$, it immediate to verify that we obtain a valid solution to (LP-2), of value $W(\ell)=W^*$.
    Moreover, the constraint governing the sum of pairwise $L$-hop bounded distances is now satisfied with strict inequality:
    \begin{equation} \label{eqn: strinct inequality}
        \sum_{u,v} D^{\leq L}_{\ell}(u,v) > 1.
    \end{equation}
    The lengths $\ell(e)$ on edges are fixed from now on, and we denote $D^{\leq L}_{\ell}$ by $D^{\leq L}$ from now on. We will also use 
    the distance function $D^{\leq L/4}_{\ell}$, that we denote by $D^{\leq L/4}$ from now on.
    
    We use the following lemma. 

    \begin{lemma}[Adaptation of Corollary 20 from \cite{LeightonRao}]\label{cor: LR 20}
        There is an efficient algorithm, that,
        given a graph $G = (V,E)$, a parameter $0 < \alpha < 1$ and any edge length function $\ell : E \mapsto \reals_{\geq 0}$ of total weight $W(\ell) = \sum_{e \in E} \ell(e) \leq \frac{\alpha}{64 n \log n}$,
        returns one of the following:
        \begin{itemize}
            \item 
            either a subset $T \subseteq V$ of at least $\ceil{\frac{2|V|}{3}}$ vertices, such that, for $r=\frac{|E|}{2 n^2 W(\ell)}$, the $r$-hop bounded diameter of $G[T]$ is at most $\frac{1}{2n^2}$; or
            \item a cut $(S, S')$ in $G$ of sparsity less than $\alpha$.
        \end{itemize}
    \end{lemma}

    We complete the proof of Lemma~\ref{cor: LR 20} later, after we complete the proof of \Cref{thm: Leighton-Rao} using it.
    Recall that $W(\ell) = W^* = \frac{\alpha}{64 n \log n}$.
    We apply the algorithm from \Cref{cor: LR 20} to graph $G$, with parameter $\alpha$ and distance function $\ell$.

    If the algorithm returns a cut $(S, S')$ of sparsity less than $\alpha$, we terminate the algorithm and return this cut. Therefore, we assume from now on that the algorithm from \Cref{cor: LR 20} returns a subset $T \subseteq V$ of at least $2|V|/3$ vertices such that
    $G[T]$ has $r$-hop bounded diameter at most $\frac{1}{2n^2}$, where $r=\frac{|E|}{2 n^2 W(\ell)}$.
    Observe that for all $r'>r$, for every pair $u,v$ of vertices, $D^{\leq r'}(u,v)\leq D^{\leq r}(u,v)$. Observe also that: 
    
    \[r=\frac{|E|}{2 n^2 W(\ell)} \leq \frac{\frac{dn}{2}}{2n^2 \frac{d}{nL}} = \frac{L}{4}.\]

Therefore, the $L/4$-hop bounded diameter of $G[T]$ is at most $\frac 1{2n^2}$.

    For convenience, for a subset $S \subseteq V$ of vertices and a vertex $u \in V$, we denote by $D^{\leq L/4}(u, S) := \min_{v \in S} D^{\leq L/4}(u,v)$.
    We use the following lemma.

    \begin{lemma}[Adaptation of Lemma 21 from \cite{LeightonRao}]\label{lem: LR 21}
        There is an efficient algorithm, that,
        given a graph $G = (V,E)$,
        a parameter $0 < \alpha < 1$,
        any edge length function $\ell : E \mapsto \reals_{\geq 0}$,
        a length parameter $L \geq \frac{2 d \ln n}{\alpha}$ and
        a subset $T \subseteq V$ of at least $\ceil{2|V|/3}$ vertices,
        such that $\sum_{v \in V} D^{\leq L}(v,T) > \frac{4 W(\ell)}{\alpha}$,
        returns a cut $(S, S')$ of $V$ with sparsity less than $\alpha$.
    \end{lemma}

We prove Lemma~\ref{lem: LR 21} later, after we complete the proof of \Cref{thm: Leighton-Rao} using it.

    First, we claim that $\sum_{v \in V} D^{\leq L/4}(v, T) > \frac{4 W^*}{\alpha}$.
    Indeed, assume for contradiction otherwise, that is:
    
    \[ \sum_{v \in V} D^{\leq L/4}(v, T) \leq \frac{4 W^*}{\alpha} =\frac{4}{\alpha}\cdot\frac{\alpha}{64n\log n} = \frac{1}{16 n \log n}.\]

    Recall that the $L/4$-hop bounded diameter of $G[T]$ is at most $\frac{1}{2n^2}$.
    From the triangle inequality, for any pair $u, v \in V$ of vertices:

    \[ D^{\leq L}(u,v) \leq D^{\leq L/4}(u,T) + D^{\leq L/4}(v,T) + \frac{1}{2n^2}. \]
    Hence,
    \[ \sum_{u,v \in V} D^{\leq L}(u,v) \leq \sum_{u,v \in V} \left( D^{\leq L/4}(u,T) + D^{\leq L/4}(v,T) + \frac{1}{2n^2} \right) \]
    \[ \indent \leq \frac{1}{2} + 2n \sum_{u \in V}D^{\leq L/4}(u,T)\]
    \[ \indent \leq \frac{1}{2} + 2n \frac{1}{16 n \log n}  = \frac{1}{2} + \frac{1}{8 \log n} < 1,\]

    contradicting the fact that $\ell$ is a valid solution to (LP-2). Therefore, $\sum_{u \in V} D^{L/4}(v, T) > \frac{4 W^*}{\alpha}$ must hold.
    Moreover, notice that $\frac{L}{4} = \frac{16 d \log n}{\alpha} \geq \frac{2 d \ln n}{\alpha}$ holds.
    We now apply the algorithm from \Cref{lem: LR 21} to $G$, with parameters $\alpha$ and $L/4$, edge length function $\ell$ and the subset $T$ of vertices, to 
     obtain a cut $(S,S')$ of $V$ with sparsity less than $\alpha$.
    This completes the proof of \Cref{thm: Leighton-Rao}, except for the proofs of \Cref{cor: LR 20} and \Cref{lem: LR 21} that we provide in the next subsection.

\subsection{Proof of \Cref{cor: LR 20}}

        We start with the following definition:
        \begin{definition}
            Given a graph $G = (V,E)$, a partition of $G$ into components is a collection $\gset = \set{G[V_1], \ldots, G[V_z]}$ of vertex-induced subgraphs such that $\bigcup_{i \in [z]} V_i = V$ and for every $i \neq j$, $V_i \cap V_j = \emptyset$.
        \end{definition}

        We use the following lemma, that we prove later for completeness after completing the proof of \Cref{cor: LR 20} using it.
        \begin{lemma}[Adaptation of Lemma 19 from \cite{LeightonRao}] \label{lem: LR 19}
            There is an efficient algorithm, that,
            given a graph $G = (V,E)$, a parameter $\Delta > 0$, and any edge length function $\ell : E \mapsto \reals_{\geq 0}$,
            partitions $G$ into components
            $\gset = \set{G[V_1], \ldots, G[V_z]}$ such that the following holds:
            \begin{itemize}
                \item For each $G[V_i] \in \gset$, the $r'$-hop bounded diameter of $G[V_i]$ is at most $\Delta$, for $r'=\Delta |E| / W(\ell)$; and
                \item $\sum_{i < j} |E(V_i, V_j)| < 8 W(\ell) \log n/ \Delta$.
            \end{itemize}
        \end{lemma}

        We use \Cref{lem: LR 19} with $\Delta = \frac{1}{2n^2}$ and edge length function $\ell$ to obtain a collection $\gset = \set{G[V_1], \ldots, G[V_z]}$ of components.
        Notice that $r'
        =\frac{\Delta |E|}{W(\ell)} = \frac{|E|}{2n^2 W(\ell)}=r$, so the $r$-hop bounded diameter of each subgrpaph $G[V_i]$ is at most $1/n^2$.
        
        If, for some subgraph $G[V_{i^*}] \in \gset$, $|V_{i^*}| \geq \frac{2|V|}{3}$, then we return $V_{i^*}$.
        Otherwise, we use Observation \ref{obs: simple partition}, to obtain a partition of the graphs in $\gset$ into two subsets, $\gset'$ and $\gset''$, such that, if we let $S=\bigcup_{G_i\in \gset'}V(G_i)$, and $S'=\bigcup_{G_i\in \gset'}V(G_i)$, then $|S|, |S'| \geq |V|/4$ and $|E(S,S')| < \frac{8 W(\ell) \log n}{\Delta} = 16 W(\ell) n^2 \log n$. 
        Therefore, the sparsity of the cut $(S, S')$ is less than:
        \[ \frac{16 W(\ell) n^2 \log n}{n/4} = 64 W(\ell) n \log n  \leq 64 n \log n \cdot \frac{\alpha}{64 n \log n} = \alpha. \]

        This completes the proof of \Cref{cor: LR 20} except for the proof of \Cref{lem: LR 19} that we provide next.

    \proofof{\Cref{lem: LR 19}}
        If $\Delta < \frac{8 W(\ell) \log n}{|E|}$, we output $\gset = \set{G[\set{v}] \> | \> v \in V}$.
        Notice that for each $G[V_i]$, we have $G[V_i] = G[\set{v_i}]$ for some $v_i \in V$.
        Hence, the $r'$-hop bounded diameter of $G[V_i]$ is $0$, and we have $\sum_{i<j} |E(V_i, V_j)| = |E| < \frac{8 W(\ell) \log n}{\Delta}$ as required.
        Therefore, we assume from now on that $\Delta \geq \frac{8 W(\ell) \log n}{|E|} > \frac{8 W(\ell) \ln n}{|E|}$ holds.
        For convenience, we denote $\epsilon := \frac{2 W(\ell) \ln n}{\Delta |E|}$.
        Notice that $\epsilon \leq 1/4$ holds.

        Consider an auxiliary graph $G^{+} = (V^+, E^+)$ obtained from $G$ by replacing each edge $e$ with a path consisting of $\ceil{|E| \ell(e) / W(\ell)}$ edges.
        Notice that $|E^+| \leq 2|E|$.
        For simplicity, we identify the common vertices of $G$ and $G^+$.
        The following observation is now immediate:
        \begin{observation}\label{obs: LR 19 path length}
            For any path of length $\gamma$ in $G^+$, the corresponding path in $G$ has length at most $\frac{W(\ell) \gamma}{|E|}$.
        \end{observation}

        Next, we iteratively partition vertices of $G^+$ into $V_0^+, V_1^+, \ldots,$ and the required partition of $G$ into components will be given by $G[V_1] = G[V_1^+ \cap V], G[V_2] = G[V_2^+ \cap V], \ldots, $.
        We start with $V^+_0 = \emptyset$ and then iterate.
        We now show how to compute $V_{i+1}^+$ given $V_0^+, \ldots, V_i^+$.

        We denote  $V_i^* := V^+ \backslash \bigcup_{j \leq i} V_j^+$.
        If $V \cap V_i^* = \emptyset$, we have computed the desired partition and the algorithm terminates.
        Thus, we assume from now on that there is a vertex $v_{i+1} \in V_i^*$.
        For every integer $j \geq 0$, we denote by $B^{i+1}_{j}$ the subset of vertices $u \in V_i^*$, such that there is some path of length at most $j$ connecting $v_{i+1}$ and $u$ in $G^+[V^*_i]$.

        We let $C_j := \frac{2|E|}{n} + |E_G[B^{i+1}_j]|$ for every integer $j \geq 0$.
        Let $j^*_{i+1}$ be the smallest $j \geq 0$ such that $C_{j+1} < (1+\epsilon)C_j$.
        Notice that some such $j^*_{i+1}$ must exist, since $\epsilon > 0$ and $C_{j+1} = C_j$ for $j \rightarrow \infty$.
        We set $V_{i+1}^+ = B^{i+1}_{j^*_{i+1}}$ and proceed to the next iteration.
        The following observation is now immediate:
        \begin{observation}\label{obs: LR 19}
            For every index $i > 0$, $V_i^+ \cap V \neq \emptyset$ and $|E(V_i^+, V_i^*)| < \epsilon \left( \frac{2|E|}{n} + \left|  E[V_i^+] \right|\right)$.
        \end{observation}
        \begin{proof}
            Notice that for every index $i > 0$ and $j \geq 0$, we have $v_i \in B_j^i$.
            Thus, $v_i \in V_i^+ \cap V$, and hence $V_i^+ \cap V \neq \emptyset$.
            From our construction, we have
            \[\frac{2|E|}{n} + \left| E[B_{j_i^* + 1}^i] \right| < (1+\epsilon) \left( \frac{2|E|}{n} + \left| E[B_{j_i^*}^i] \right| \right). \]
            
            Equivalently:
            
            \[ \left|  E[B_{j_i^* + 1}^i] \right| - \left| E[B_{j_i^*}^i] \right| < \epsilon \left( \frac{2|E|}{n} + \left|  E[B_{j_i^*}^i] \right|\right).\]
            
            Therefore,
            \[  |E(V_i^+, V_i^*)| \leq \left|  E[B_{j_i^* + 1}^i] \right| - \left| E[B_{j_i^*}^i] \right| < \epsilon \left( \frac{2|E|}{n} + \left|  E[B_{j_i^*}^i] \right|\right) = \epsilon \left( \frac{2|E|}{n} + \left|  E[V_i^+] \right|\right).\]
        \end{proof}

        The following two claims will complete the proof of \Cref{lem: LR 19}.

        \begin{claim}
            $\sum_{i<j} |E(V_i, V_j)| < \frac{8 W(\ell) \log n}{\Delta}$.
        \end{claim}
        \begin{proof}
            \[\sum_{i<j} |E(V_i, V_j)| = \sum_{i>0} \left| E \left( V_i, \bigcup_{j> i} V_j \right) \right| \leq \sum_{i>0} |E(V_i^+, V_i^*)| < \sum_{i > 0} \epsilon \left( \frac{2|E|}{n} + |E(|V_i^+|)| \right)\]
            \[ \indent \leq \epsilon \left( 2|E| + |E^+| \right) \leq 4|E|\epsilon  = \frac{8 W(\ell) \ln n}{\Delta} < \frac{8 W(\ell) \log n}{\Delta} .\]
            Here, the second inequality follows from \Cref{obs: LR 19} and the penultimate inequality follows from the fact that $|E^+| \leq 2|E|$.
        \end{proof}

        \begin{claim}
            For each $G[V_i]$, the $r'$-hop bounded diameter of $G[V_i]$ is at most $\Delta$, for $r' = \frac{\Delta |E|}{W(\ell)}$.
        \end{claim}
        \begin{proof}
            We claim that it suffices to show that, for each $G[V_i]$, the diameter of $G^+[V^+_i]$ is at most $r'= \frac{\Delta |E|}{W(\ell)}$.
            Indeed, if this is the case, \Cref{obs: LR 19 path length} implies that the $r'$-hop bounded diameter of $G[V_i]$ is at most $\frac{W(\ell) r'}{|E|} = \Delta$.
            Notice that, in order to show that the diameter of $G^+[V_i^+]$ is at most $r'$, it suffices to show that $j^*_i \leq \frac{r'}{2} = \frac{\Delta |E|}{2 W(\ell)}$.
            Fix any index $i$ and the corresponding graph $G^+[V_i^+]$. 
            If $j_i^* \neq 0$, we must have:
            \[ 2|E| \geq |E^+| \geq |E(V_i^+)| > (1+\epsilon)^{j^*_i} \frac{2|E|}{n}.\]
            
            Therefore, 
            $ (1+\epsilon)^{j^*_i} < n$ must hold, and so: 
            \[ j^*_i < \frac{\ln n}{\epsilon} = \frac{\Delta |E|}{2 W(\ell)}.\]
            (We have used the fact that $\epsilon < 1/4$).
        \end{proof}
    \endproofof

\subsection{Proof of \Cref{lem: LR 21}}    
        Similarly to the proof of \Cref{lem: LR 21}, consider an auxiliary graph $G^{+} = (V^+, E^+)$ obtained from $G$ by replacing each edge $e$ with a path consisting of $\ceil{|E| \ell(e) / W(\ell)}$ edges.
        Notice that $|E^+| \leq 2|E|$.
        For simplicity, we identify the common vertices of $G$ and $G^+$.
        Given a subset $S \subseteq V(G^+)$ of vertices, we denote by $N(S)$ the set of all vertices $v\in V(G^+)$ such that $v\not\in S$, but $v$ has a neighbor in $S$.
        
        Next, we iteratively partition the vertices of $G^+$ into layers, $V^+_0,V^+_1,\ldots$, and for each $i\geq 0$, we define the corresponding graph $G^+_i=G^+[V^+_i]$, as follows. We start with $V^+_0=T$, $G^+_0 = G^+[T]$ and then iterate. We now show how to compute $V^+_{i+1}$ and $G^+_{i+1}$, given  $V^+_i$ and $G^+_i$, assuming that $V^+_i\neq V^+$ (otherwise, the algorithm terminates). 
        
        Let $E_i := \delta_{G^+}(V^+_i)$ and $C_i := |E_i|$. We partition $E_i$ into two subsets: set $E'_i$ containing all edges $(u,v)$ with $u\in V^+_i$, such that $v$ is a vertex of the original graph $G$; and set $E''_i$ containing all remaining edges. Let $C'_i=|E_i'|$, and let $C''_i=|E''_i|$.
        We distinguish between the following two cases:
        \begin{itemize}
            \item \textbf{Case 1: $C'_i \geq C_i/2$}.
                    In this case, we let $V^+_{i+1}$ contain all vertices of $V^+_i \cup N(V^+_i)$.
                    We also set $G^+_{i+1}=G^+[V^+_{i+1}]$.
                    Notice that in this case, $|E[G^+_{i+1}]\setminus E[G^+_i]|\geq C_i$.

            \item \textbf{Case 2: $C''_i > C_i/2$}.
                    In this case, we let $V^+_{i+1}$ only contain the vertices of $V^+_i$, and those vertices of $N(V^+_i)$ that do not lie in the original graph $G$, that is:
                    
                    \[V^+_{i+1}=V^+_i\cup (N(V^+_i)\setminus V(G)).\] 
                    
                    As before, we set $G^+_{i+1}=G^+[V^+_{i+1}]$. Notice that in this case, $E[G^+_{i+1}]\setminus E[G^+_i]$ contains all edges of $E''_i$, and so $|E[G^+_{i+1}]\setminus E[G^+_i]|\geq C''_i> C_i/2$.
        \end{itemize}

        From the above discussion we obtain the following observation:
        \begin{observation} \label{obs: LR edges grow}
            For each level $i$, $|E(G^+_{i+1})\setminus E(G^+_i)|\geq \frac{C_i}{2}$, and in particular $\sum_iC_i\leq 2|E^+|$.
        \end{observation}

        For each level $i$, let $n_i = |V(G) \backslash V^+_i|$ -- the number of vertices of the original graph $G$ that do not lie in $V^+_i$.
        Recall that $|T| \geq \ceil{2|V|/3}$, and so for all $i$, $n_i \leq |V|/3 \leq |V|/2$.
        Moreover, $C_i = |\delta_{G^+}(V^+_i)| \geq |\delta_G(V \cap V^+_i)|$.
        
        If, for any level $i$, $C_i<\alpha n_i$, then we return the cut $(V \cap V^+_i, V \backslash V^+_i)$; it is immediate to see that its sparsity is less than $\alpha$. Therefore, we assume from now on, that for all $i$, $C_i\geq \alpha n_i$. 
        We will reach a contradiction by showing that $\sum_{v \in V} D^{\leq L}(v,T) \leq \frac{4 W(\ell)}{\alpha}$ must hold.
        In order to do so, we use the following two claims.
        
        \begin{claim}\label{claim: bound Case 1}
        	The number of indices $i$ for which Case $1$ is invoked is at most $L$.
        	\end{claim}
        
        \begin{proof}
        	Let $i$ be an index for which Case $1$ is invoked, so $C'_i \geq C_i/2$. Recall that we have assumed that $C_i\geq\alpha n_i$. 
        Since the maximum vertex-degree of $G$ is bounded by $d$, the number of new vertices of $V$ that are added to $V^+_{i+1}$ is at least $\frac{C'_i}{d}\geq \frac{\alpha n_i}{2d} $. Therefore, $n_{i+1} \leq n_i(1- \frac{\alpha }{2d})$, and the total number of indices $i$ in which Case 1 is invoked must be bounded by  $\frac{2d \ln n}{\alpha} \leq L$.
        \end{proof}

        \begin{claim}\label{claim: bound nis}
        	$\sum_i n_i\leq \frac{4|E|}{\alpha}$.
        \end{claim}
        \begin{proof}
            Recall that we have assumed $C_i \geq \alpha n_i$ for all $i$.
            Thus, 
            \[ \sum_i n_i \leq \sum_i \frac{C_i}{\alpha} = \frac{\sum_{i} C_i}{\alpha} \leq \frac{2 |E^+|}{\alpha} \leq \frac{4 |E|}{\alpha}.\]
            Here, the second-last inequality follows from \Cref{obs: LR edges grow} and the last inequality follows from the fact that $|E^+| \leq 2|E|$.
        \end{proof}

        For each vertex $v \in V \backslash T$, let $i_v$ be the unique index, such that $v \in V(G^+_i)$ and $v \not \in V(G^+_{i-1})$.
        For the remaining vertices $v \in T$, we set $i_v = 0$. Notice that $v$ must be connected by an edge to a vertex $u$ with $i_u<i_v$. Therefore, we can construct a path $P^+_v=(v_0,v_1,\ldots,v_r)$ in $G^+$, where $v_0\in T$, $v_r=v$, and for all $1\leq j\leq r$, $i_{v_{j-1}}<i_{v_j}$. 
        
        Let $P_v$ be the path corresponding to $P^+_v$ in the original graph $G$. Since we invoke Case $1$ at most $L$ times, it is easy to verify that $P_v$ contains at most $L$ edges. 
        Moreover:

        \[D^{\leq L}(v, T) \leq \sum_{e \in P_v} \ell(e) \leq \sum_{e \in P_v} \frac{W(\ell)}{|E|} \ceil{\frac{|E| \ell(e)}{W(\ell)}} = \frac{W(\ell)}{|E|} |E(P^+_v)| \leq i_v \frac{W(\ell)}{|E|}.\]

        Altogether:

        \[\sum_v D^{\leq L}(v, T) \leq \frac{W(\ell)}{|E|} \sum_v i_v = \frac{W(\ell)}{|E|} \sum_i n_i \leq \frac{4W(\ell)}{\alpha}, \]
        where the last inequality follows from Claim~\ref{claim: bound nis}. This contradicts the assumption that $\sum_v D^{\leq L}(v, T) > \frac{4W(\ell)}{\alpha}$,

        completing the proof of \Cref{lem: LR 21}.

%% file: appendix-sec7.tex
\section{Proofs Omitted from Section~\ref{sec: pos to poe}}\label{sec: proofs for sec 7}

\subsection{Proof of Claim~\ref{claim: large expanding subgraph cheeger}}\label{subsec: efficient expander fixing}

The proof is very similar to the proof of Claim~\ref{claim: large expanding subgraph-edges}.
The algorithm iteratively removes edges from $G\setminus E'$, until we obtain a connected component of the resulting graph that is an $\Omega\left(\frac{\alpha^2}{d}\right)$-expander.
We start with $G'=G\setminus E'$ (notice that $G'$ is not necessarily connected).  We also maintain a set $E''$ of edges that we remove from $G'$, initialized to $E''=\emptyset$. 
We then perform a number of iterations. In every iteration, we apply Theorem~\ref{thm: spectral} to $G'$, and obtain a cut $(X,Y)$ in $G'$. If $|E_{G'}(X,Y)| \geq \alpha \cdot\min{(|X|,|Y|)}/4$, then, from Theorem~\ref{thm: spectral}, $G'$ is an $\Omega\left(\frac{\alpha^2}{d}\right)$-expander. We terminate the algorithm and return $G'$. We later show that $|V(G')|\geq |V| - \frac{4|E'|}{\alpha}$. Assume now that  $|E_{G'}(X,Y)| < \alpha \cdot \min{(|X|,|Y|)}/4$, and assume w.l.o.g. that $|X|\geq |Y|$. Update $G'$ to be $G'[X]$, add the edges of $E(X,Y)$ to $E''$, and continue to the next iteration. 
Clearly, at the end of the algorithm, we obtain a graph $G'$ that is an $\Omega\left(\frac{\alpha^2}{d}\right)$-expander. It only remains to show that $|V(G')|\geq |V| - \frac{4|E'|}{\alpha}$. The remainder of the analysis is identical to the analysis of Claim~\ref{claim: large expanding subgraph-edges}.

Assume that the algorithm performs $r$ iterations, and for each $1\leq i\leq r$, let $(X_i,Y_i)$ be the cut computed by the algorithm in iteration $i$. Since $|X_i|\geq |Y_i|$, $|Y_i|\leq |V(G)|/2$. At the same time, if we denote $E_i=E''\cap E(X_i,Y_i)$, then $|E_i|< \alpha |Y_i|/4$. Therefore:

\[|E''|=\sum_{i=1}^r|E_i|< \alpha\sum_{i=1}^r|Y_i|/4.\]

On the other hand, since $G$ is an $\alpha$-expander, the total number of edges leaving each set $Y_i$ in $G$ is at least $\alpha|Y_i|$, and all such edges lie in $E'\cup E''$. Therefore:

\[|E'|+|E''|\geq \alpha\sum_{i=1}^r|Y_i|/2.\]

Combining both bounds, we get that $|E'|\geq \alpha\sum_{i=1}^r|Y_i|/4$, and so $\sum_{i=1}^r|Y_i|\leq \frac{4|E'|}{\alpha}$. 
Therefore, $|V(G')|=|V|-\sum_{i=1}^r|Y_i|\geq |V|-\frac{4|E'|}{\alpha}$.

\subsection{Proof of Claim~\ref{claim: flow in expander}}
We can compute the largest-cardinality set of disjoint paths connecting vertices of $A$ to vertices of $B$  in $G$ using standard maximum $s$--$t$ flow computation and the integrality of flow. Therefore, it is sufficient to show that there exists a set of $\ceil{\alpha z/d}$ disjoint paths connecting $A$ to $B$ in $G$.

Assume otherwise. Then, from Menger's theorem, there is a set $Z$ of fewer than $\alpha z/d$ vertices in $G$, such that $G\setminus Z$ contains no path from a vertex of $A\setminus Z$ to a vertex of $B\setminus Z$. Let $E'$ be the set of all edges of $G$ incident to the vertices of $Z$. Since the maximum vertex degree in $G$   is at most $d$, $|E'|<\alpha z$.
Therefore, graph $G\setminus E'$ contains no path connecting a vertex of $A$ to a vertex of $B$. Let $X$ be the union of all connected components of $G\setminus E'$ containing the vertices of $A$, and let $Y=V(G)\setminus X$. Then $|E(X,Y)|\leq |E'|<\alpha z\leq \alpha\cdot\min\set{|X|,|Y|}$, contradicting the fact that $G$ is an $\alpha$-expander.

\subsection{Proof of Theorem \ref{thm: exp to wl}} \label{subsec: exp to wl}
The main tool that we use for the proof of Theorem~\ref{thm: exp to wl} is the following theorem, whose proof appeared in~\cite{CC_gmt}; we include the proof here for completeness.

\begin{theorem}[Restatement of Theorem A.4 in~\cite{CC_gmt}]\label{thm: path grouping}
  There is an efficient algorithm, that, given a graph $G$ with maximum vertex degree at most $d$, an integer $q\geq 1$, and a set $\pset$ of at least $16dq$ disjoint paths in $G$, computes a subset $\pset'\subseteq \pset$ of at least $|\pset|/2$ paths, and a collection $\cset$ of
  disjoint connected subgraphs of $G$, such that each path $P\in
  \pset'$ is completely contained in some subgraph $C\in \cset$, and
  each such subgraph contains at least $q$ and at most $4d q$
  paths in $\pset$.
\end{theorem}

\begin{proof}
Starting from $G$, we construct a new graph $H$, by contracting every path $P\in \pset$ into a super-node $u_P$. Let $U=\set{u_P\mid P\in \pset}$ be the resulting set of super-nodes. Let $\tau$ be any spanning tree of $H$, rooted at an arbitrary vertex $r$. Given a vertex $v\in V(\tau)$, let $\tau_v$ be
  the sub-tree of $\tau$ rooted at $v$. Let $J'_v\subseteq V(G)$ be the set of all vertices of
  $\tau_v$ that  belong to the original graph $G$ (that is, they are not super-nodes), and let $J''_v$ be the set of all vertices of $G$ that lie on paths $P\in \pset$ with $u_P\in \tau_v$. We then let $J_v=J'_v\cup J''_v$. We also denote $G_v=G[J_v]$; observe that it must be a connected graph. Over the course of the algorithm, we will delete some vertices from $\tau$. The notation $\tau_v$ and $G_v$ is always computed with respect to the most current tree
  $\tau$. We start with $\cset=\emptyset,\pset'=\emptyset$, and then iterate.

  Each iteration is performed as follows. If $q\leq |V(\tau)\cap U|\leq 4d q$, then we add the graph $G_r$ corresponding to the root $r$ of $\tau$ to $\cset$, and terminate
  the algorithm. If $|V(\tau)\cap U|<q$, then we also terminate the
  algorithm (we will show later that $|\pset'|\geq |\pset/2|$ at this point). Otherwise, let $v$ be the lowest
  vertex of $\tau$ with $|\tau_v\cap U|\geq q$. If $v\not \in U$, then,
  since the degree of every vertex in $G$ is at most $d$, $|\tau_v\cap
  U|\leq d q$. We add $G_v$ to $\cset$, and all paths in
  $\set{P\mid u_P\in \tau_v}$ to $\pset'$. We then delete all vertices of
  $\tau_v$ from $\tau$, and continue to the next iteration.

  Assume now that $v=u_P$ for some path $P\in \pset$. If $|\tau_v\cap
  U|\leq 4d q$, then we add $G_v$ to $\cset$, and all paths in
  $\set{P'\mid u_{P'}\in \tau_v}$ to $\pset'$ and continue to the next
  iteration. So we assume that $|\tau_v\cap U|> 4d q$.

  Let $v_1,\ldots,v_z$ be the children of $v$ in $\tau$.  Build a new
  tree $\tau'$ as follows. Start with the path $P$, and add the vertices
  $v_1,\ldots,v_z$ to $\tau'$. For each $1\leq i\leq z$, let
  $(x_i,y_i)\in E(G)$ be any edge connecting some vertex $x_i\in V(P)$ to
  some vertex $y_i\in V(G_{v_i})$; such an edge must exist from the definition
  of $G_{v_i}$ and $\tau$. Add the edge $(v_i,x_i)$ to $\tau'$. Therefore,
  $\tau'$ is the union of the path $P$, and a number of disjoint stars
  whose centers lie on the path $P$, and whose leaves are the vertices
  $v_1,\ldots,v_z$. The degree of every vertex of $P$ is at most
  $d$. We define the \emph{weight} of the vertex $v_i$ as the number
  of the paths in $\pset$ contained in $G_{v_i}$ (equivalently, it is $|U\cap \tau_{v_i}|$). Recall that the weight
  of each vertex $v_i$ is at most $q$, by the choice of $v$. For each
  vertex $x\in P$, the weight of $x$ is the total weight of its
  children in $\tau'$. Recall that the the total weight of all vertices
  of $P$ is at least $4d q$, and the weight of every vertex is at
  most $d q$. We partition $P$ into a number of disjoint segments
  $\Sigma=(\sigma_1,\ldots,\sigma_{\ell})$ of weight at least $q$ and
  at most $4dq$ each, as follows. Start with $\Sigma=\emptyset$,
  and then iterate. If the total weight of the vertices of $P$ is at
  most $4d q$, we build a single segment, containing the whole
  path. Otherwise, find the shortest segment $\sigma$ starting from
  the first vertex of $P$, whose weight is at least $q$. Since the
  weight of every vertex is at most $d q$, the weight of $\sigma$
  is at most $2d q$. We then add $\sigma$ to $\Sigma$, delete it
  from $P$ and continue. Consider the final set $\Sigma$ of
  segments. For each segment $\sigma$, we add a new graph
  $C_{\sigma}$ to $\cset$. Graph $C_\sigma$ consists of the union of
  $\sigma$, the graphs $G_{v_i}$ for each $v_i$ that is connected to a
  vertex of $\sigma$ with an edge in $\tau'$, and the corresponding edge
  $(x_i,y_i)$. Clearly, $C_{\sigma}$ is a connected subgraph of
  $G$, containing at least $q$ and at most $4d q$ paths of
  $\pset$. We add all those paths to $\pset'$, delete all vertices of
  $\tau_v$ from $\tau$, and continue to the next iteration. We note that path $P$ itself is not added to $\pset'$, but all paths $P'$ with $u_{P'}\in V(\tau_v)$ are added to $\pset'$.

  At the end of this procedure, we obtain a collection $\pset'$ of
  paths, and a collection $\cset$ of disjoint connected subgraphs of
  $G$, such that each path $P\in \pset'$ is contained in some $C\in
  \cset$, and each $C\in \cset$ contains at least $q$ and at most
  $4d q$ paths from $\pset'$. It now remains to show that
  $|\pset'|\geq |\pset|/2$. We discard at most $q$ paths in the last
  iteration of the algorithm. Additionally, when $v=u_P$ is processed,
  if $|\tau_v\cap U|> 4d q$, then path $P$ is also discarded, but at
  least $4d q$ paths are added to $\pset'$. Therefore, overall,
  $|\pset'|\geq |\pset|-\frac{|\pset|}{4d q+1}-q\geq |\pset|/2$,
  since $|\pset|\geq 16dq$.
\end{proof}

We now turn to prove Theorem~\ref{thm: exp to wl}. Recall that we are given an $\alpha$-\posexp System $\Sigma = (\sset, \mset, A_1, B_3)$ of width $w$ and length $3$, where $0<\alpha <1$, and the corresponding graph $G_{\Sigma}$ has maximum vertex degree at most $d$. Our goal is to compute subsets $\hat A_1\subseteq A_1,\hat B_3\subseteq B_3$ of $\Omega(\alpha^2 w/d^3)$ vertices each, such that $\hat A_1\cup \hat B_3$ is well-linked in $G_{\Sigma}$. Notice that we can assume w.l.o.g. that $w\geq 256 d^3/\alpha^2$, as otherwise it is sufficient that each set $\hat A_1,\hat B_3$ contains a single vertex, which is trivial to ensure.

We apply Claim~\ref{claim: flow in expander} to graph $S_1$, together with the sets $A_1,B_1$ of vertices, to compute a set $\pset_1$ of $\ceil{\alpha w/d}$ node-disjoint paths in $S_1$, connecting vertices of $A_1$ to vertices of $B_1$. We then set $q=\floor{16d/\alpha}$, and use Theorem~\ref{thm: path grouping}, 
to compute a subset $\pset'_1\subseteq \pset_1$ of at least $|\pset_1|/2\geq \alpha w/(2d)$ paths, and a collection $\cset$ of
  disjoint connected subgraphs of $S_1$, such that each path $P\in
  \pset'_1$ is completely contained in some subgraph $C\in \cset$, and
  each such subgraph contains at least $q$ and at most $4d q$ paths of $\pset'_1$. 
  (Note that from our assumption that $w\geq 256 d^3/\alpha^2$, $|\pset_1|\geq 16dq$).
  Clearly, $|\cset|\geq \frac{|\pset_1'|}{4dq}\geq \frac{\alpha^2 w}{256d^3}$. We select one representative path $P\in \pset'_1$ from each subgraph $C\in \cset$, so that $P\subseteq C$, and we let $\pset^*_1\subseteq \pset'_1$ be the resulting set of paths. We are now ready to define the set $\hat A_1\subseteq A_1$ of vertices: set $\hat A_1$ contains, for every path $P\in \pset^*_1$, the endpoint of $P$ that lies in $A_1$. Note that $|\hat A_1|=|\pset^*_1|=|\cset|\geq  \frac{\alpha^2 w}{256d^3}$. For convenience, for every vertex $a\in \hat A_1$, we denote by $P_a\in \pset^*_1$ the unique path originating at $a$, and we denote by $C_a\in \cset$ the unique subgraph of $S_1$ containing $P_a$.

We select a subset $\hat B_3\subseteq B_3$ of at least $ \frac{\alpha^2 w}{256d^3}$ vertices similarly, by running the same algorithm in $S_3$.
The set of paths obtained as the outcome of Theorem~\ref{thm: path grouping} is denoted by $\pset_3'$, and the set of connected subgraphs of $S_3$ by $\cset'$. We also denote by $\pset^*_3\subseteq \pset'_3$ the set of representative paths that we select from each subgraph of $\cset'$.
 For every vertex $b\in \hat B_3$, we denote by $P_b\in \pset^*_b$ the unique path originating at $b$, and we denote by $C_b\in \cset'$ the unique subgraph containing $P_b$.
 
 It remains to show that $\hat A_1\cup \hat B_3$ is well-linked in $G_{\Sigma}$. We show this using the same arguments as in ~\cite{CC_gmt}.
Let $X,Y\subseteq \hat A_1\cup \hat B_3$ be two equal-cardinality sets of vertices. We need to show that there is a set $\qset$ of $|X|=|Y|$ disjoint paths connecting them in $G_{\Sigma}$, such that the paths in $\qset$ are internally disjoint from $\hat A_1\cup \hat B_3$. We define a new subgraph $H\subseteq G_{\Sigma}$ as follows: graph $H$ is the union of the graph $S_2$ and the matchings $\mset_1$ and $\mset_2$; additionally, for every vertex $v\in X\cup Y$, we add the graph $C_v$ to $H$. It is now enough to show that  there is a set $\qset$ of $|X|=|Y|$ disjoint paths connecting $X$ to $Y$ in $H$; such paths are guaranteed to be internally disjoint from $\hat A_1\cup \hat B_3$. From the integrality of flow, it is sufficient to show a flow $F$ in $H$, where every vertex in $X$ sends one flow unit, every vertex in $Y$ receives one flow unit, and every vertex of $H$ carries at most one flow unit. We now construct such a flow.
This flow will be
a concatenation of three flows, $F_1,F_2,F_3$.

We start by defining the flows $F_1$ and $F_3$. 
Consider some vertex $v\in X\cup Y$, and assume w.l.o.g. that $v\in \hat A_1$. We select an arbitrary subset $U_v\subseteq B_1$ of $q=\floor {16d/\alpha}$ vertices that serve as endpoints of paths $P\in \pset_1'$ that are contained in $C_v$. Since $C_v$ is a connected graph, vertex $v$ can send $1/q$ flow units to every vertex in $U_v$ simultaneously, inside the graph $C_v$, so that the flow on every vertex is at most $1$. We denote the resulting flow by $F^v$.

We obtain the flow $F_1$ by taking the union of all flows $F^v$ for $v\in X$, and we obtain the flow $F_3$ by taking the union of all flows $F^v$ for $v\in Y$ (we reverse the direction of the flow $F^v$ in the latter case). 

Let $R_1=\bigcup_{v\in X}U_v$, and let $R_2=\bigcup_{v\in Y}U_v$. Note that  $R_1\cup R_2\subseteq B_1\cup A_3$. For every vertex $x\in R_1\cup R_2$ that lies in $B_1$, we let $x'$ be the vertex of $A_2$, that is connected to $x$ by an edge of $\mset_1$. Similarly,  for every vertex $x\in R_1\cup R_2$ that lies in $A_3$, we let $x'$ be the vertex of $B_2$, that connects to $x$ by an edge of $\mset_2$. Let $R_1'=\set{x'\mid x\in R_1}$ and $R_2'=\set{x'\mid x\in R_2}$. Note that $R_1',R_2'$ are disjoint sets of vertices in $S_2$. Since graph $S_2$ is an $\alpha$-expander, there is a flow $F_2'$ in $S_2$, where every vertex in $R_1'$ sends one flow unit, every vertex in $R_2'$ sends one flow unit, and every edge carries at most $1/\alpha$ flow units. Scaling this flow down by factor $q=\floor{16d/\alpha}$, we obtain a new flow $F_2$ in $S_2$, where every vertex of $R_1'$ sends $1/ {q}$ flow units, every vertex of $R_2'$ receives $1/ {q}$ flow units, and every vertex of $S_2$ carries at most one flow unit.

The final flow $F$ is obtained by concatenating the flows $F_1,F_2$ and $F_3$, and sending $1/{q}$ flow units on every edge of $\mset_1\cup \mset_2$ that is incident to a vertex of $R_1\cup R_2$. The flow in $F$ guarantees that every vertex of $X$ sends one flow unit, every vertex in $Y$ receives one flow unit, and every vertex of $G_{\Sigma}$ carries at most one flow unit.

\section{Proof of Observation~\ref{obs: connected sparsest cut}}

	We start with the cut $(X,Y)$ and perform a number of iterations. In every iteration, we modify the cut $(X,Y)$ so that the number of connected components in $G\setminus E(X,Y)$ strictly decreases, while ensuring that the cut sparsity does not increase.
We now describe the execution of an iteration. Let $(X,Y)$ be the current cut. Let $\cset_X$ and $\cset_Y$ be the sets of all connected components of $G[X]$ and $G[Y]$ respectively. If $|\cset_X| = |\cset_Y| = 1$, then we return the cut $(X,Y)$, and terminate the algorithm. We assume from now on that this is not the case.

 Assume w.l.o.g. that $|X|\leq |Y|$. 
Let $\rho_X := \frac{|E(X,Y)|}{|X|}$ and $\rho_Y := \frac{|E(X,Y)|}{|Y|}$.
We consider the following two cases.

	\paragraph{Case 1: } The first case happens when $|\cset_X| > 1$.
		Recall that $|E(X,Y)| = \rho_X |X|$.
		Thus, there is a connected component $C \in \cset_X$ such that $|E(C, Y)| \geq \rho_X |C|$.
		Consider a new partition $(X', Y')$, obtained by setting $X' = X \backslash C$ and $Y' = Y \cup C$.
		Notice that the number of connected components in $G\setminus E(X',Y')$ decreases by at least one.
		The sparsity of the new cut is:
		\[ \frac{|E(X', Y')|}{\min\set{|X'|, |Y'|}} = \frac{|E(X', Y')|}{|X'|} = \frac{|E(X,Y)| - |E(C,Y)|}{|X| - |C|} \leq \frac{\rho_X |X| - \rho_X |C|}{|X| - |C|} = \rho_X.\]

		\paragraph{Case 2:} If Case 1 does not happen, then $|\cset_Y| > 1$ must hold.
		As before, there is a connected component $C \in \cset_Y$ such that $|E(C,Y)| \geq \rho_Y |C|$.
		Consider the new partition $(X', Y')$ by setting $X' = X \cup C$ and $Y' = Y \backslash C$.
		Notice that the number of connected components in $G\setminus E(X',Y')$ decreases by at least one.
		In order to bound the sparsity of the new cut, we consider two cases.
		If $|X'| \geq |Y'|$, then the sparsity of the new cut is 
		\[  \frac{|E(X', Y')|}{|Y'|} = \frac{|E(X,Y)| - |E(C,Y)|}{|Y| - |C|} \leq \frac{\rho_Y|Y| - \rho_Y|C|}{|Y| - |C|} = \rho_Y \leq \rho_X.\]
		Otherwise, the sparsity of the new cut is 
		\[ \frac{|E(X', Y')|}{|X'|} = \frac{|E(X,Y)| - |E(C,Y)|}{|X|+|C|} < \frac{|E(X,Y)|}{|X|} = \rho_X.\]
		
It is immediate to verify that the algorithm is efficient, and that it produces the cut $(X^*,Y^*)$ with the required properties.